\documentclass[12pt,onecolumn]{article}
\usepackage[left=2cm,right=2cm,top=3cm,bottom=3cm]{geometry}
\usepackage[utf8]{inputenc}
\usepackage[english]{babel}
\usepackage{hyperref}
\usepackage{paralist}
\usepackage{graphicx}
\usepackage{authblk}
\usepackage{natbib}
\usepackage{enumitem}
\usepackage{color}
\usepackage{multicol}
\usepackage{soul}

\newcommand{\citeA}[1]{\citet{#1}}
\renewcommand{\cite}[1]{\citep{#1}}
\newcommand{\authors}[1]{}
\newcommand{\affiliation}[2]{\affil[#1]{#2}}
\newcommand{\acknowledgments}{\section*{Acknowledgments}}
\newcommand{\new}{}
\begin{document}
%%%%%%%%%%%%%%%%%%%%%%%%
\title{A study of daytime convective vortices and turbulence in the martian Planetary Boundary Layer
based on half-a-year of InSight atmospheric measurements and Large-Eddy Simulations} 

\author[1,2]{A. Spiga}
\author[3]{N. Murdoch}
\author[4]{R. Lorenz}
\author[1]{F. Forget}
\author[5]{C. Newman}
\author[6]{S. Rodriguez}
\author[7]{J. Pla-Garcia}
\author[7]{D. Vi{\'u}dez Moreiras}
\author[8]{D. Banfield}
\author[6]{C. Perrin}
\author[9]{N. T. Mueller}
\author[10]{M. Lemmon}
\author[1]{E. Millour}
\author[11]{W.B. Banerdt}

%%%%
%%%%
\authors{A. Spiga\affil{1,2}, 
N. Murdoch\affil{3},
R. Lorenz\affil{4}, 
F. Forget\affil{1},
C. Newman\affil{5},
S. Rodriguez\affil{6},
J. Pla-Garcia\affil{7},
D. Vi{\'u}dez Moreiras\affil{7},
D. Banfield\affil{8},
C. Perrin\affil{6},
N. T. Mueller\affil{9},
M. Lemmon\affil{10},
E. Millour\affil{1},
W.B. Banerdt\affil{11}}
%%%%
\affiliation{1}{Laboratoire de M\'et\'eorologie Dynamique / Institut Pierre-Simon Laplace (LMD/IPSL), Sorbonne Universit\'e, Centre National de la Recherche Scientifique (CNRS), \'Ecole Polytechnique, \'Ecole Normale Sup\'erieure (ENS), Campus Pierre et Marie Curie BC99, 4 place Jussieu 75005 Paris, France}
%%%%
\affiliation{2}{Institut Universitaire de France (IUF), 1 rue Descartes, 75005 Paris, France}
%%%%
\affiliation{3}{Institut Supérieur de l'Aéronautique et de l'Espace (ISAE-SUPAERO), 10 Avenue Edouard Belin, 31400 Toulouse, France}
%%%
\affiliation{4}{Johns Hopkins Applied Physics Laboratory, 11100 Johns Hopkins Road, Laurel, MD 20723, USA}
%%%%
\affiliation{5}{Aeolis Research, 333 N Dobson Road, Unit 5, Chandler AZ 85224-4412, USA}
%%%%
\affiliation{6}{Université de Paris, Institut de physique du globe de Paris, CNRS, F-75005 Paris, France}
%%%%
\affiliation{7}{Centro de Astrobiología (CSIC-INTA), 28850 Torrejón de Ardoz, Madrid, Spain}
%%%%
\affiliation{8}{Cornell University, Cornell Center for Astrophysics and Planetary Science, Ithaca, NY, 14853, USA}
%%%%
\affiliation{9}{German Aerospace Center (DLR), Institute of Planetary Research, Rutherfordstr. 2, 12489 Berlin, Germany}
%%%%
\affiliation{10}{Space Science Institute, 4765 Walnut Street, Suite B, Boulder, CO 80301, USA}
%%%%
\affiliation{11}{Jet Propulsion Laboratory, California Institute of Technology, Pasadena, CA 91109, USA}

\date{Version: \today}
\maketitle
\newpage
\newcommand{\dakar}{Here we use the high-sensitivity continuous 
pressure, wind, temperature measurements
in the first 400 sols of InSight operations
(from northern late winter 
to midsummer)
to analyze wind gusts, convective cells 
and vortices in Mars' daytime PBL.} 

\begin{abstract}
Studying the atmospheric 
Planetary Boundary Layer (PBL)
is crucial to understand 
the climate of a planet.
The meteorological measurements
by the instruments onboard InSight
at a latitude of 4.5$^{\circ}$N
make a uniquely rich dataset
to study the 
active turbulent dynamics
of the daytime PBL on Mars.
%%%
\new{\dakar}
%%%
We compare InSight measurements
to turbulence-resolving
Large-Eddy Simulations (LES).
The daytime PBL turbulence at
the InSight landing site is
very active,
with clearly identified
signatures of convective cells
and a vast population of
6000 recorded 
vortex encounters, adequately
represented by a power-law 
with a 3.4 exponent.
%in agreement 
%with LES results.
While the daily variability
of vortex encounters
at 
%the 
InSight 
%landing site 
can be explained
by the statistical
nature of turbulence,
the seasonal variability
is positively correlated
with ambient wind speed,
which is supported by LES.
However, wind gustiness
is positively correlated 
to surface temperature
rather than ambient wind speed
and sensible heat flux,
confirming the radiative
control of the daytime martian PBL;
and fewer convective
vortices are forming in LES
when the background wind
is doubled.
Thus, the long-term
seasonal variability
of vortex encounters 
at the InSight landing
site is mainly controlled by the
advection of convective vortices
by ambient wind speed.
Typical tracks
followed by vortices forming
in the LES show
a similar distribution
in direction and length
as orbital imagery.
%of the InSight region.
\end{abstract}

\section*{Plain Language Summary} 
InSight is a lander sent to 
the surface of Mars
with a weather station capable,
like never before,
to measure 
pressure,
temperature and winds
continuously and
at high cadence.
We use this InSight atmospheric data set
acquired over half a Martian year,
along with computer simulations,
to study the intense turbulence 
that develops
in the daytime hours on Mars.
InSight detect
periodic variations 
in the measurements
of the weather station,
corresponding to 
air motions
driven by convection.
We also detect a large
population of 6000 whirlwinds
passing close to the InSight
lander and causing
the pressure at the weather
station to suddenly drop.
The number of those 
whirlwind encounters
vary from day to day,
because of the random 
turbulence,
and on a seasonal basis,
because of
the varying ambient wind
that transports the whirlwinds
towards InSight.
Unlike the population 
of whirlwinds,
the strength of wind gusts
follow the
ground temperature
varying with season.
%We think part of those properties
%illustrate key differences
%between turbulence on Mars
%and on the Earth.
Whirlwinds also leave
graffiti-like 
dark tracks at the surface
of Mars that can be imaged
by satellites
in the InSight region
and reproduced by
our numerical simulations.

\newpage\setcounter{tocdepth}{3}\tableofcontents
%%%%%%%%%%%%%%%%%%%%%%%%
\newpage

\section{Introduction}

Mars is a cold desert; 
yet its near-surface atmosphere,
the so-called Planetary Boundary Layer (PBL),
is prone to 
strong turbulent motions in the daytime
(\citeA{Petr:11} and references therein).
Daytime turbulent motions
in the thin martian atmosphere
include
spectacular vortices
that may appear as
\emph{dust devils} if they raise sufficient dust,
strong updrafts at the borders of convective cells, and powerful wind gusts.
These motions result in a mixing of 
heat,
momentum,
dust particles
and chemical species
over altitudes of several kilometers 
above the surface, making PBL processes
on Mars a crucial step
to understand the meteorology
and climate.
The Martian PBL also
exhibits interesting differences
with the terrestrial PBL,
notably 
a strong control on
the daytime PBL turbulence
by the 
near-surface atmospheric
absorption of
surface infrared emission
\cite{Habe:93pbl,Savi:99,Spig:10bl}.

Phenomena related to daytime 
turbulence on Mars cause 
pressure, wind, and temperature 
to fluctuate at timescales 
shorter than a Martian hour 
(defined as 1/24$^{\small\textrm{th}}$
of a martian day or sol).
Such signatures have been recorded
in the \emph{in situ} meteorological measurements of 
Viking \cite{Hess:77,Till:94},
Pathfinder \cite{Scho:97,Lars:02},
Phoenix \cite{Elle:10,Hols:10},
Spirit and Opportunity \cite{Smit:06}, and
Curiosity \cite{Stea:16,Kaha:16,Ordo:18},
as is summarized in the review by \citeA{Mart:17}.
Turbulence-resolving numerical modeling
referred to as Large-Eddy Simulations (LES)
can help to make sense of
the PBL events
arising in 
time series obtained 
by single-station measurements
(e.g., pressure drops, wind gusts,
quasi-periodic temperature fluctuations)
in the broader context
of convective turbulence in the PBL
(see \citeA{Spig:16ssr} for a review).

The instrumentation implemented on the InSight spacecraft, 
which landed on Mars on the flat plains of Elysium Planitia 
(4.5$^{\circ}$N 135.6$^{\circ}$E)
on November 28$^{\small\textrm{th}}$ 2018,
is particularly suitable to conduct 
studies of PBL turbulence \cite{Spig:18insight,Banf:20,Bane:19nat}.
The pressure sensor is characterized by
its unprecedented sensitivity
and high-frequency acquisition \cite{Banf:18}.
The wind and temperature measurements,
albeit similar to the ones performed on board
Curiosity \cite{Gome:12},
benefit for the first time from
the simultaneous use
of two booms facing in opposite directions
-- inadvertent destruction of 
one of the Curiosity wind sensors 
by flying debris at landing 
produced observational biases 
which made only winds coming
from certain directions 
reliably measurable
\cite{Newm:17}, making
the Curiosity wind 
retrieval challenging
\cite{Viud:19}.
Another key novel characteristic of InSight's \textit{in situ } meteorological observations 
is that measurements of pressure, temperature, and wind are made continuously \cite{Spig:18insight},
as they are needed to constrain
the atmosphere-induced seismic noise
\cite{Murd:17,Kend:17,Garc:20} at all times in order to assess how much of the seismic
signal corresponds to the
activity in the interior of Mars.
Those direct atmospheric
measurements are complemented
by surface brightness 
temperature sensing \cite{Muel:20},
color imaging \cite{Maki:18},
and, for the first time 
at the surface of Mars,
seismic measurements \cite{Logn:19nat}.
Furthermore, 
solar array currents 
can also be used
for atmospheric investigations
\cite{Lore:20}.

\begin{figure}[h!]
\begin{center}
\includegraphics[width=\textwidth]{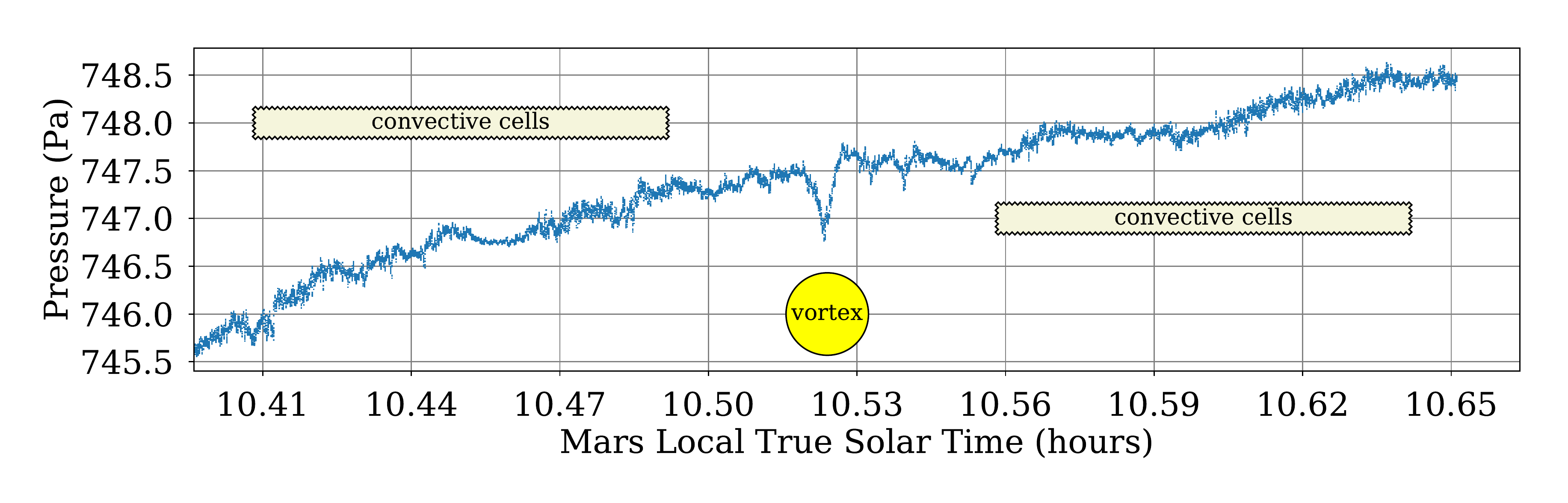}
\caption{ \label{fig:firstpbl}
\textit{ The first pressure measurements on board InSight
on sol 4 ($L_s = 298^{\circ}$, November 30th 2018)
directly shows
how daytime 
convective 
turbulence
in the PBL
leaves
distinctive
signatures in
the pressure 
time series
(the time axis
is the Mars local
true solar time
as described in 
section~\ref{sec:obs}).
An encounter with
a convective vortex
caused a sudden drop
in pressure and the 
convective cells
led to quasi-periodic
fluctuations of pressure
with a period 
of about 100 seconds.}}
\end{center}
\end{figure}

The potential of InSight 
to study the daytime PBL turbulence
was actually unveiled the very first time the pressure sensor was switched
on for a 900 second test, at about 10:30 Mars local time on InSight sol 4. 
\newcommand{\rabat}{Figure~\ref{fig:firstpbl} shows the occurrence of
a sudden, short-duration pressure drop of amplitude 0.6~Pa,
characteristic of a convective vortex, 
in the midst of longer, 100-second-period fluctuations of pressure of amplitude 0.1~Pa, characteristic of convective cells.} \new{\rabat}

The goal of the present study is to use InSight measurements in
the first half year of operations
in order to explore the
atmospheric
PBL dynamics in the
daytime, especially convective cells
and vortices, and to propose a preliminary
assessment of 
the seasonal variability thereof
and the vortex population statistics.
This paper focuses on
PBL turbulent structures
and
convective vortices, 
regardless of whether
they carry dust particles or not.
\newcommand{\washington}{The question
as to whether the identified
vortices in this study are
carrying a sufficient
amount of dust particles
to be visible as dust devils
in InSight's cameras
is left for future studies;
as is described in \citeA{Banf:20},
despite extensive imaging campaigns,
no dust devil were observed by InSight
(this extends to the sols 0-400
spanned by the current study).} \new{\washington}
The diagnostics drawn from InSight
observations are compared
to turbulence-resolving 
LES using the model described in \citeA{Spig:10bl}.

The topics
related to
PBL dynamics
left out of
the current study
are detailed
in other papers
of this issue:
notable individual dust devil events
\cite{Lore:20catalog},
seismic signatures of vortices 
(Murdoch et al., in revision, \citeA{Kend:20}),
orbital observations of vortex tracks
\cite{Perr:20}
and aeolian science with InSight
(Charalambous et al.
and Baker et al., 
in revision for this issue).
As is shown in \citeA{Banf:20},
InSight has a great potential too
for studies of the nighttime,
shear-driven turbulence
associated with the nocturnal
inversion.
This topic is out of the scope
of the present paper, which
focuses on the daytime PBL dynamics,
but will be developed in
future papers.

\section{Methods}

\subsection{InSight observations \label{sec:obs}}

This study includes
observations acquired
in the first 400 sols
of Insight operations
at the surface of Mars.
To indicate seasons
on Mars, 
the Mars-Sun angle, referred to as the areocentric solar longitude~$L_s$ in degrees ($^{\circ}$), is used
with the standard convention
that 0$^{\circ}$ corresponds 
to northern spring equinox.
InSight landing on November 26th, 2018 corresponds to InSight sol 0
and $L_s = 295^{\circ}$
(northern late winter).
The last sol considered in this paper, sol 400, corresponds to $L_s = 134^{\circ}$
(northern mid-summer).
As far as local time
is concerned, in order to permit
the analysis on seasonal timescales
and the comparison of InSight
observations with models,
we use the sundial-equivalent
Mars Local True Solar Time (LTST)
in which noon corresponds precisely
to the moment when the sun 
crosses the meridian.

The characteristics of the InSight
instruments relevant for atmospheric
science are summarized 
in \citeA{Spig:18insight}
and the \textit{Methods} section
of \citeA{Banf:20}.
The current paper uses
measurements from the pressure, temperature and wind
sensors on board InSight,
which in addition to the magnetometer
form the
Auxiliary Payload Sensor Suite (APSS).
Details of the calibration of
the pressure sensor and
the Temperature and Winds
for INSight (TWINS) sensors
can be found in \citeA{Banf:18}.
\newcommand{\losangeles}{APSS measurements
of pressure, wind, temperature
are continuously performed
-- except during brief, random anomalies
of APSS electronics that
cause measurements to stop
during a couple of days,
and the solar conjunction
between sols 269 and 283
that prevented data transmission
from Mars to Earth.} \new{\losangeles}

\newcommand{\noisep}{mPa~Hz$^{-1/2}$~}
The Insight pressure measurements
are routinely carried out at 10 Hz
with a noise level of 10 \noisep from 0.1-1~Hz 
\newcommand{\berlin}{(typical noise at 1 Hz
is thus of the order 10$^{-2}$~Pa)} \new{\berlin}
rising to 50 \noisep at 0.01~Hz.
\newcommand{\istanbul}{
Downlink limitations 
caused the InSight pressure datasets
to be downsampled
at 2 Hz in the sol intervals
14--167 and 262--269.} \new{\istanbul}
This is 
significantly 
higher frequencies
and
lower noise levels
than the previous pressure sensors
sent to Mars \cite{Mart:17}
and is appropriate to
study expected daytime turbulent
signatures: 
gusts, 
vortices, 
and cells 
\cite{Spig:18insight}.

\newcommand{\prague}{
As is described in 
the \emph{Methods} section of \citeA{Banf:20},
the inclusion of 
an inlet tubing 
often does not prevent the pressure signal
above 2 Hz to be correlated with wind speed
-- pointing towards either 
a loss of effectiveness of the pressure inlet,
or mechanical or electrical noise
in the pressure sensor.
Thus, we consider herein only the pressure signal
at frequencies $\leq$~2~Hz:
before our pressure drop search
is performed, the signal
is smoothed using a window of
0.5~s -- i.e. pressure
is low-pass-filtered with 
a 2 Hz cutoff.
This is appropriate for the
search of vortex-induced 
pressure drops of duration
above 1 second,
which makes the majority of events
according to existing studies,
see e.g. \citeA{Murp:16}.} \new{\prague}

The TWINS sensor booms
are similar to those
on board 
the Curiosity rover 
\cite{Gome:12}.
The booms are located
on the InSight platform,
facing outward in opposite
directions over the two solar
panels.
Their altitude from the surface
is respectively 121.5 cm and 111.5 cm
for the west and east booms 
\cite{Banf:20}.
Wind and air temperature 
are acquired 
at a frequency of 
1 Hz and 
an accuracy of $\sim$~1~m~s$^{-1}$ 
for wind speed, 
22.5$^{\circ}$ for wind direction, 
and 5 K for temperature.
\newcommand{\athenes}{
Owing to downlink limitations,
the InSight TWINS dataset
is available with
a frequency of 
0.1 Hz on sol intervals
32--182 and 262--269,
0.5 Hz on sol intervals
14--30, 183--230, 284--292,
and 1 Hz on sol intervals
4--10, 231--261, 293--400.} \new{\athenes}
\newcommand{\oleron}{The wind 
measurements
are obtained from the 
TWINS booms' 
sensor acquisitions
using look-up tables
built on 
wind-tunnel 
calibration 
experiments
and correcting from
the influence of
the lander platform elements 
using computational fluid 
dynamics simulations
\cite{Banf:18}.
The measurements from
the boom facing the prevailing wind
are preferentially selected.} 
\new{\oleron}
\newcommand{\lecaire}{In practice,
the TWINS booms' wind measurements 
may be discarded because of
sensor saturation
(usually when wind speed is high,
typically wind gusts $>$~20 m~s$^{-1}$)
and/or low Reynolds number
(usually when wind speed is low,
typically below 2.8~m~s$^{-1}$,
see \citeA{Banf:20})
and/or unfavorable wind direction
(e.g., perpendicular
to both booms).
The actual frequency of
TWINS wind and temperature
measurements is thus
typically $0.1-0.3$~Hz.} \new{\lecaire}

The use of atmospheric temperatures
retrieved by InSight 
deserves particular care. 
\newcommand{\tokyo}{In daytime, the differences in air temperatures 
measured by the two booms can be large (Figure~\ref{fig:analyzet}).
In sustained wind conditions, 
thermal contamination 
by the lander elements
(deck and solar panels)
perturbs the
air temperature measurements
of the TWINS boom facing 
the incoming ambient wind \cite{Banf:20}.
As is reported in \citeA{Viud:20},
during the dust storm from sol 40 to sol 60,
temperature measurements by the west-facing boom are clearly anomalous;
an increase in daytime air temperature is observed
rather than the decrease expected in dustier conditions
and correctly detected by the east-facing boom.
This difference can be related to the 
southeasterly wind direction at that time:
before the atmospheric flow reaches the 
west-facing boom,
enhanced convective heat transfer
between the lander and the boom
causes the measured air temperature
to be strongly overestimated by 15~K 
compared to the east-facing boom.
A 10-K daytime excess in the 
temperature measurements of the
west-facing boom,
compared to the east-facing boom,
is also found repeatedly
from sol 160 to sol 400
characterized by steady
southeasterly winds.
This bias is stable
in this 240-sol interval.
In weak wind conditions
(sol 60 to 160),
the two TWINS booms yield
consistent diagnostics
for temperature,
although with higher
uncertainties
associated with
smaller Reynolds number
(similarly to wind
measurements, see
\citeA{Banf:20}).
Discrepancies between
the two booms' measured
air temperature remain
within the 5~K limit,
which is the sensor uncertainty.
\begin{figure}[h!]
\begin{center}
\includegraphics[width=0.9\textwidth]{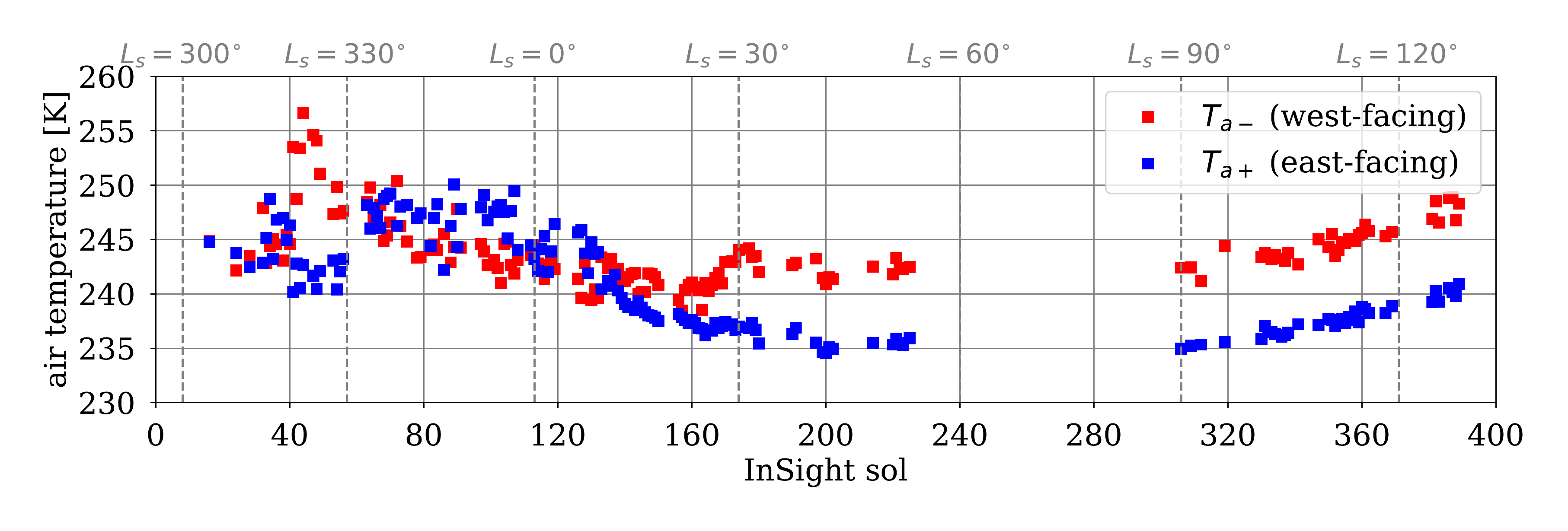}
\includegraphics[width=0.9\textwidth]{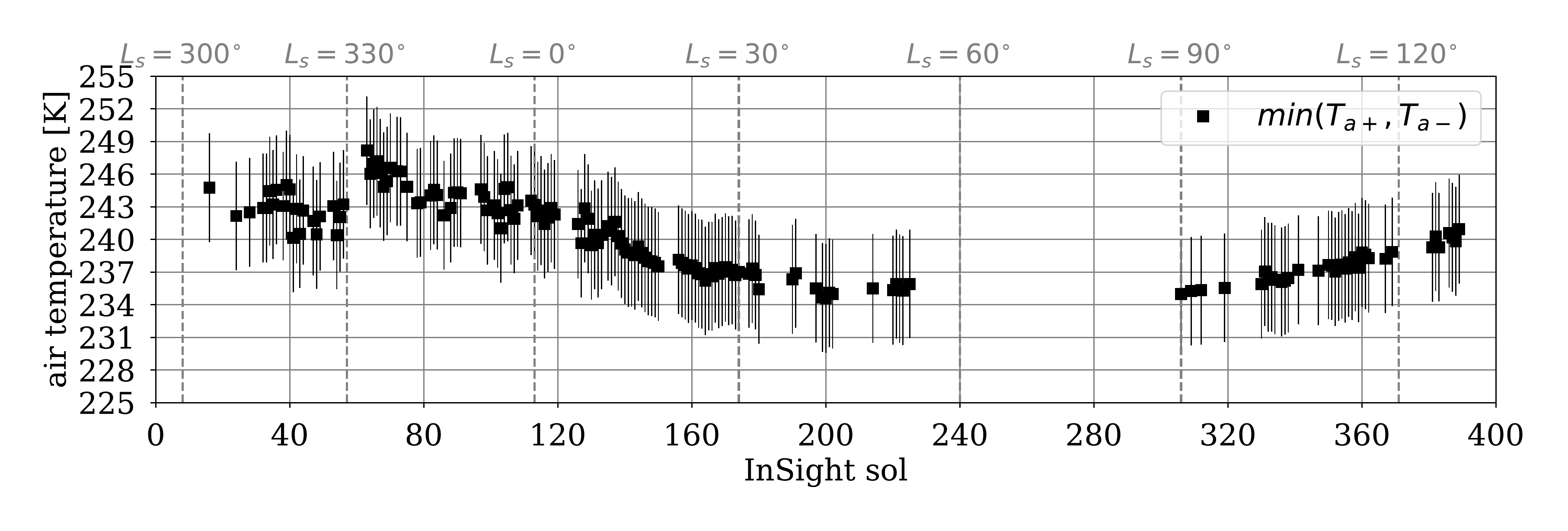}
\includegraphics[width=0.9\textwidth]{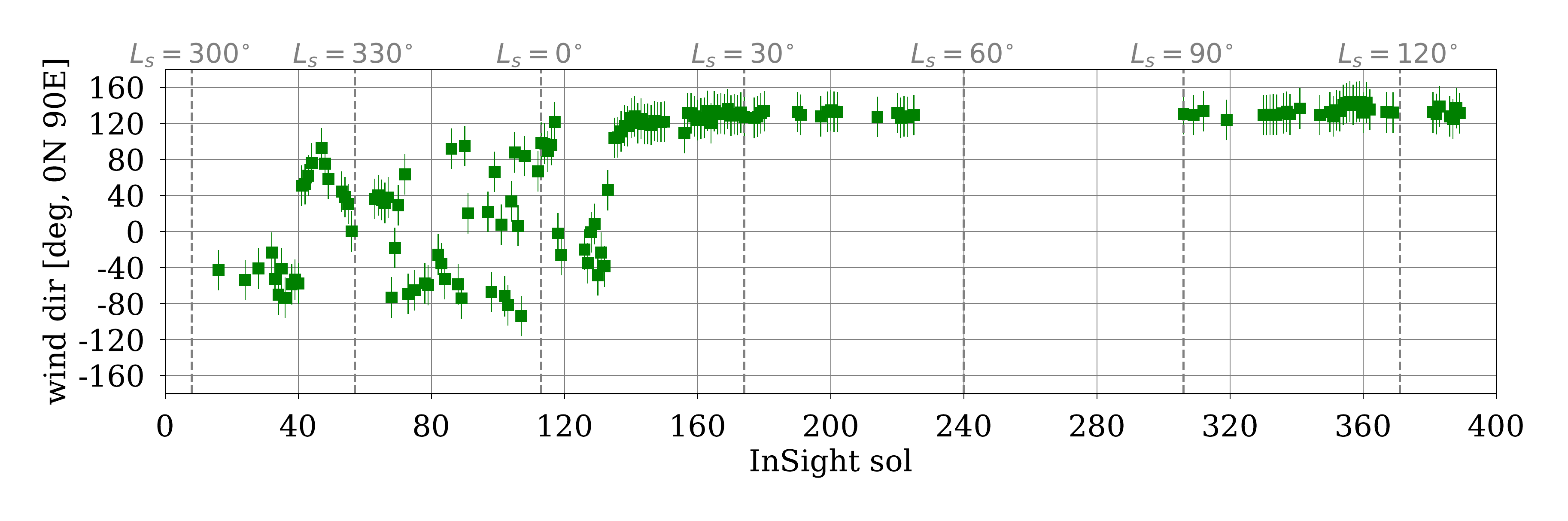}
\caption{ \label{fig:analyzet}
\textit{Typical measurements
of TWINS air temperature
(top panel for each separate boom, 
middle panel for air temperature estimated
by taking the minimum value of both booms)
and wind direction
(bottom, obtained as indicated in the text)
are shown for the first 400 sols
of Insight operations.
This figure is obtained
with a similar methodology
as what is described in 
greater detail in Figure~\ref{fig:season}.}
}
\end{center}
\end{figure}
} \new{\tokyo}

A strategy to mitigate those effects
in the daytime is to consider,
at each time, the minimum of 
the two temperature values
deduced from each TWINS booms.
In what follows,
what is named ``air temperature'' and denoted $T_a$ 
refers to this quantity.
\newcommand{\sydney}{This is 
the simplest 
and most efficient
method to date
to select the boom performing
the most reliable 
air temperature measurements.} 
\new{\sydney}
In what follows, we use this TWINS air temperature
mostly to compute the surface-to-atmosphere
temperature gradient.
\newcommand{\auckland}{Our 
conclusions still stand
if another
air-temperature 
estimate 
is considered
(such as the maximum
of the two temperatures
from each of the booms,
or an average of those
temperatures, see
Figure~\ref{fig:season})
since only the relative
seasonal variations
of atmospheric
temperatures
(and related quantities,
such as surface-to-atmosphere
temperature gradients)
are analyzed.} \new{\auckland}

Surface (i.e., ground) 
brightness temperature
measurements are performed 
by the
HP$^3$ radiometer 
on board the InSight
lander.
The details on this sensor
calibration are
described in \citeA{Muel:20}.
The HP$^3$ radiometer 
sensors measure 
the surface brightness temperature 
in three spectral bands
(8-14~$\mu$m,
8-10~$\mu$m,
15-19~$\mu$m)
at two different spots
relative to the InSight lander,
named the close spot 
and the far spot
\cite{Spoh:18}.
In what follows, we only use 
the surface brightness
temperature retrieved in the
far spot that, contrary to
the close spot, is devoid
of lander contamination
(shadowing, thermal effects).
\newcommand{\sofia}{The solar panel shadows pass 
through the near spot during 
the northern winter 
and can result in up to 20~K 
cooler temperatures in the afternoon 
and a up to 10~K cooler daily average temperature.} \new{\sofia}
Considering the 
larger calibration uncertainties
of the two spectral bands 
8-10~$\mu$m and 15-19~$\mu$m
\cite{Muel:20},
in what follows 
surface brightness temperature
is based on the sole 
8-14~$\mu$m spectral band.
\newcommand{\villejuif}{Daytime conditions are the most
favorable for surface temperature
retrievals which uncertainty
is better than 1.5 K between local
times 10:00 and 15:30.} \new{\villejuif}

\subsection{Vortex detection method \label{sec:methodovortex}}

Convective vortices developing in
the martian PBL, and passing closely
enough to the InSight lander,
manifest as sudden
pressure drops
\newcommand{\lyon}{which 
amplitude ranges from 0.1 Pa
to 10 Pa and
duration ranges 
from several seconds
to several tens of seconds}\new{\lyon}.
This is
the most distinctive signature of those
vortices
in the atmospheric sensors
at the surface
(see \citeA{Murp:16}
for a review).
Frequent -- albeit not
systematic -- wind direction reversals
are also associated
with those encounters,
as well as
an increase of wind speed.

\begin{figure}[p!]
\begin{center}
\includegraphics[width=0.48\textwidth]{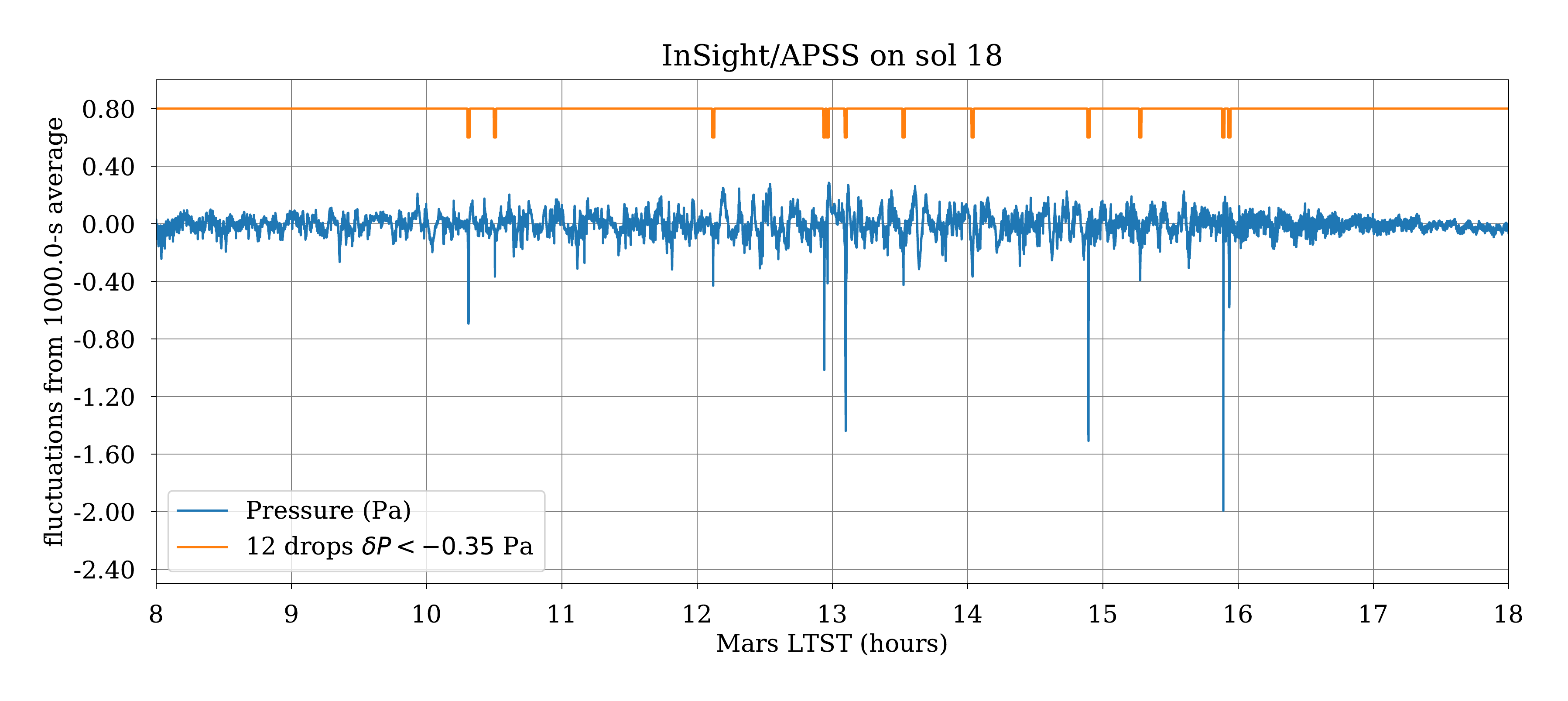}
\includegraphics[width=0.48\textwidth]{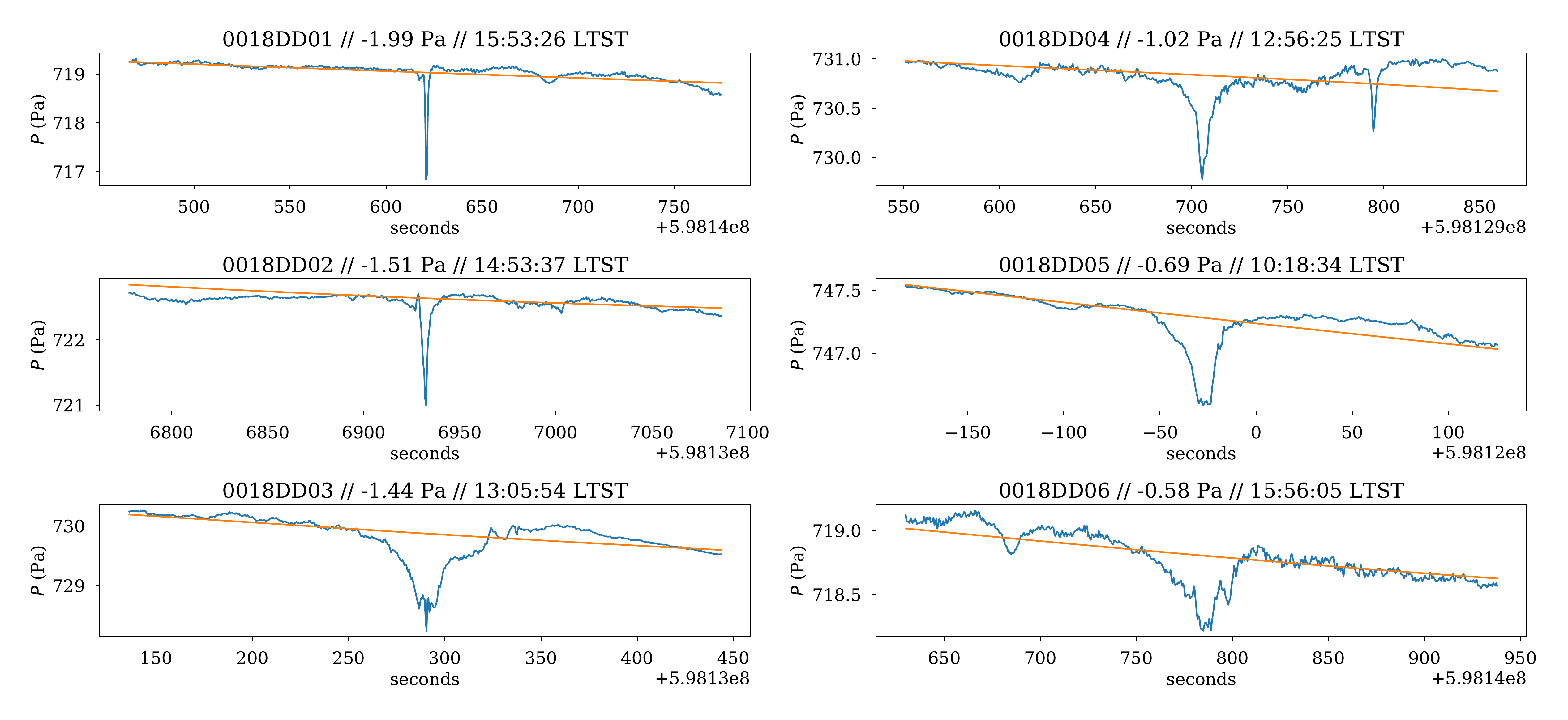}
\vskip 0.3cm
\includegraphics[width=0.48\textwidth]{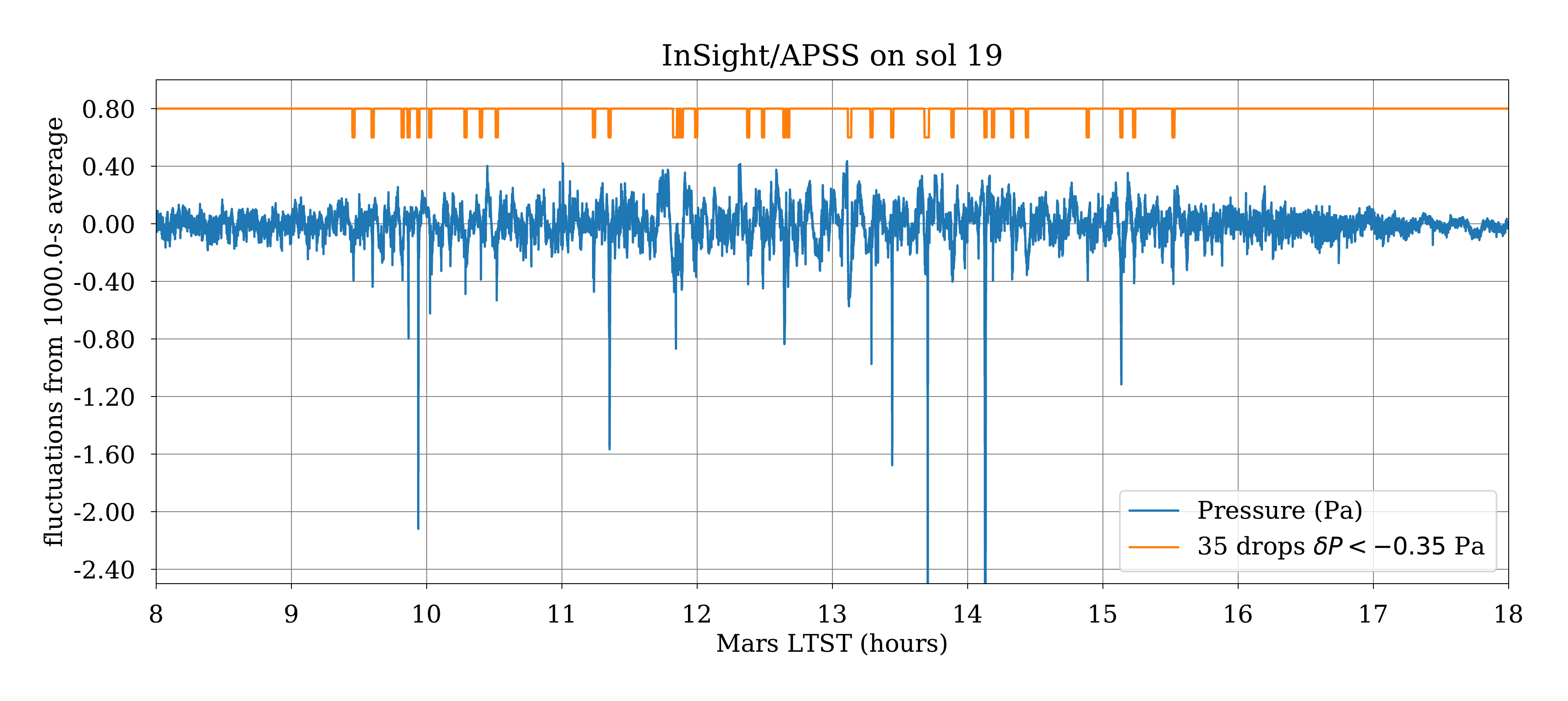}
\includegraphics[width=0.48\textwidth]{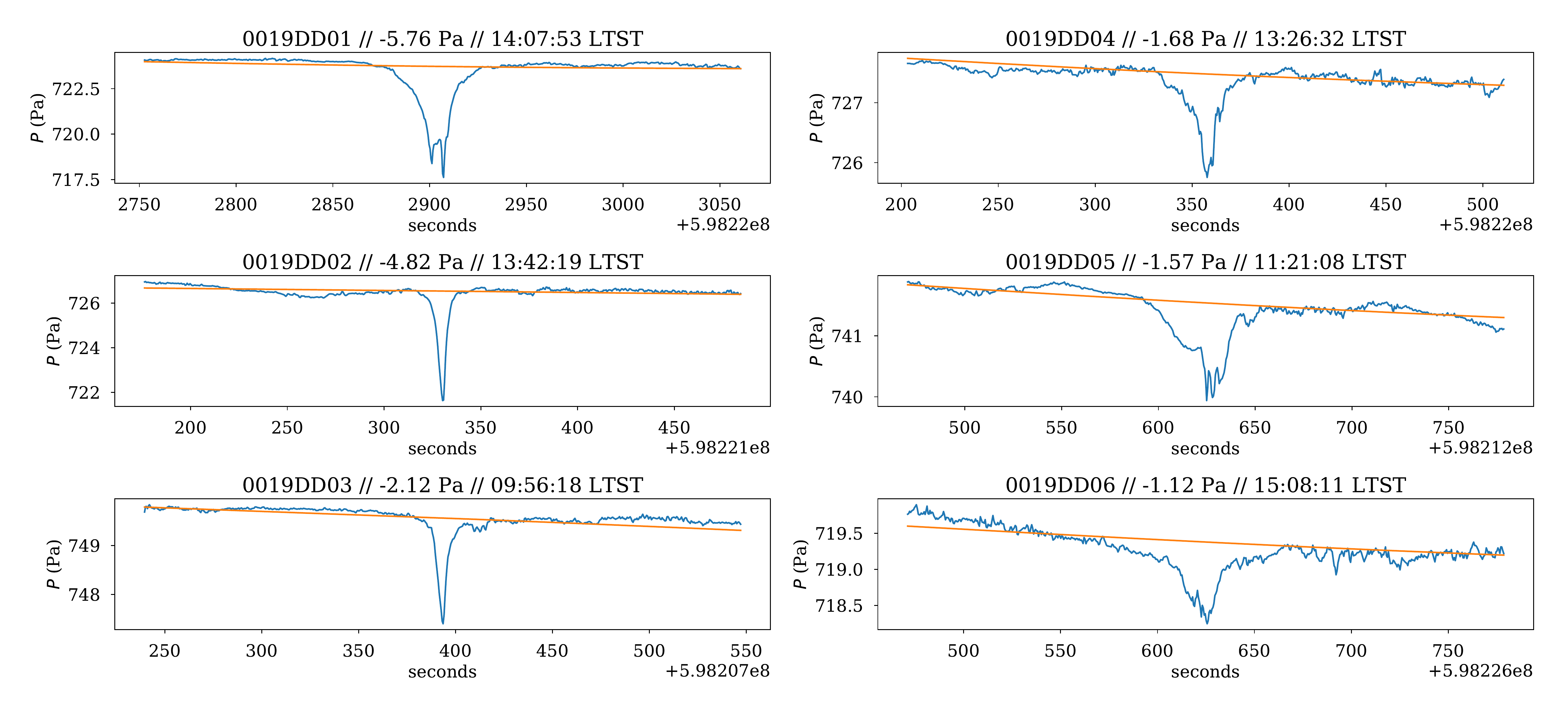}
\vskip 0.3cm
\includegraphics[width=0.48\textwidth]{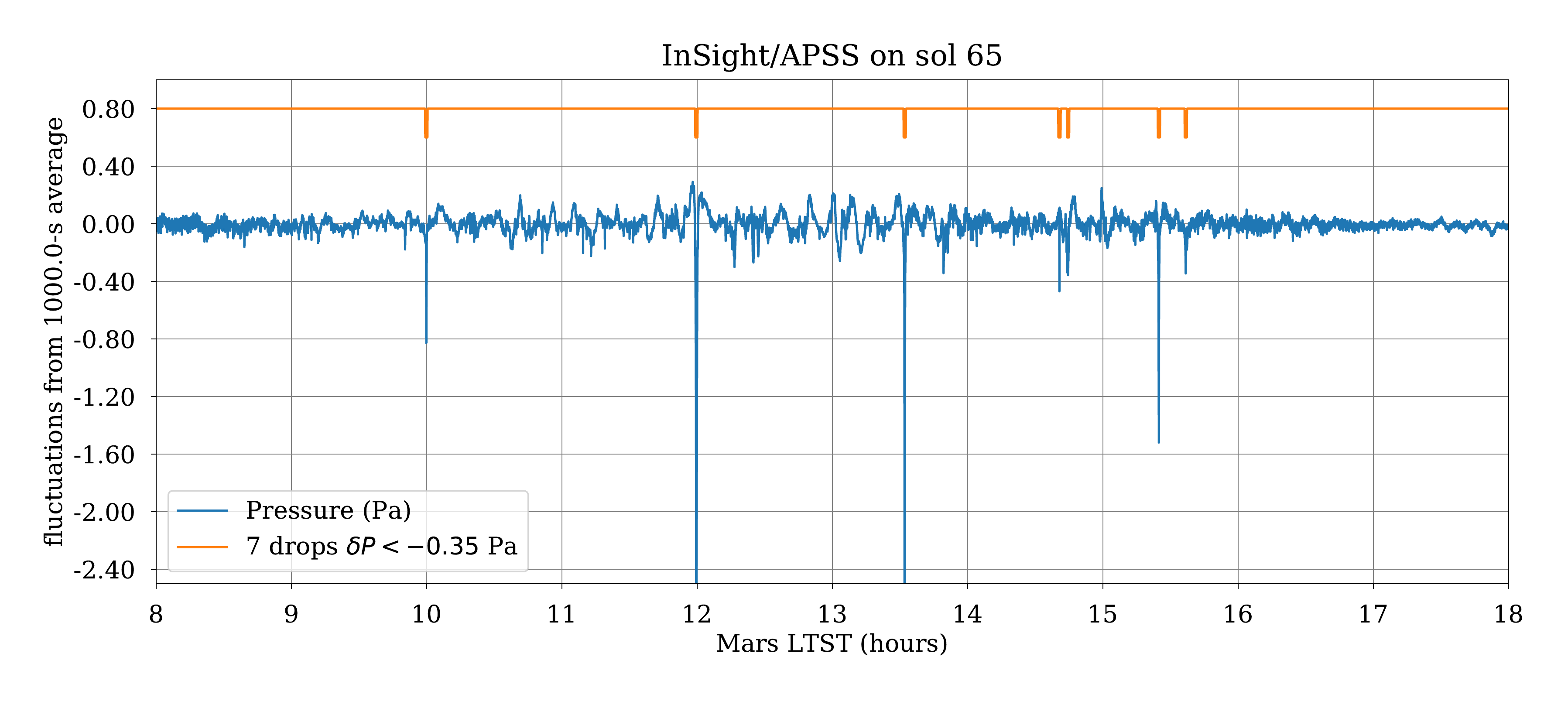}
\includegraphics[width=0.48\textwidth]{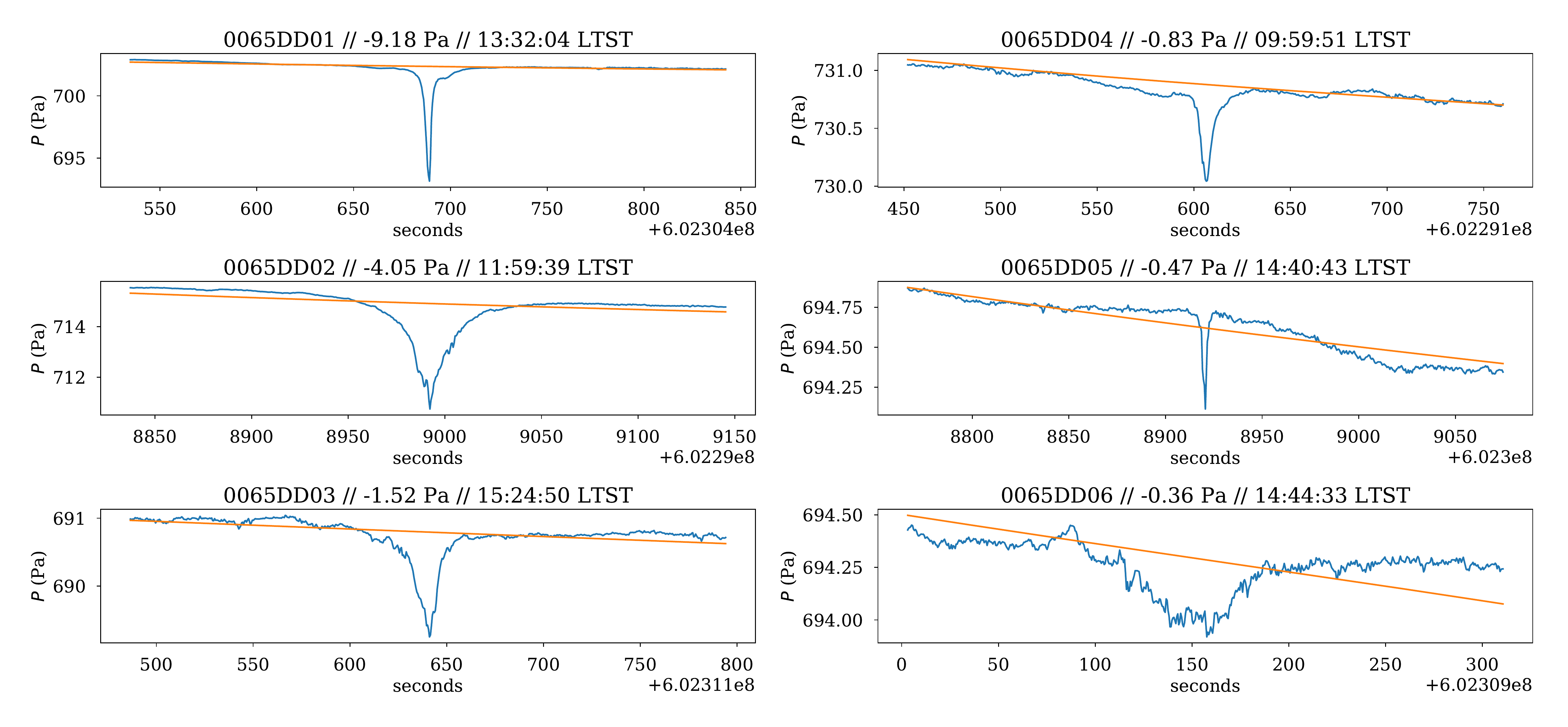}
\vskip 0.3cm
\includegraphics[width=0.48\textwidth]{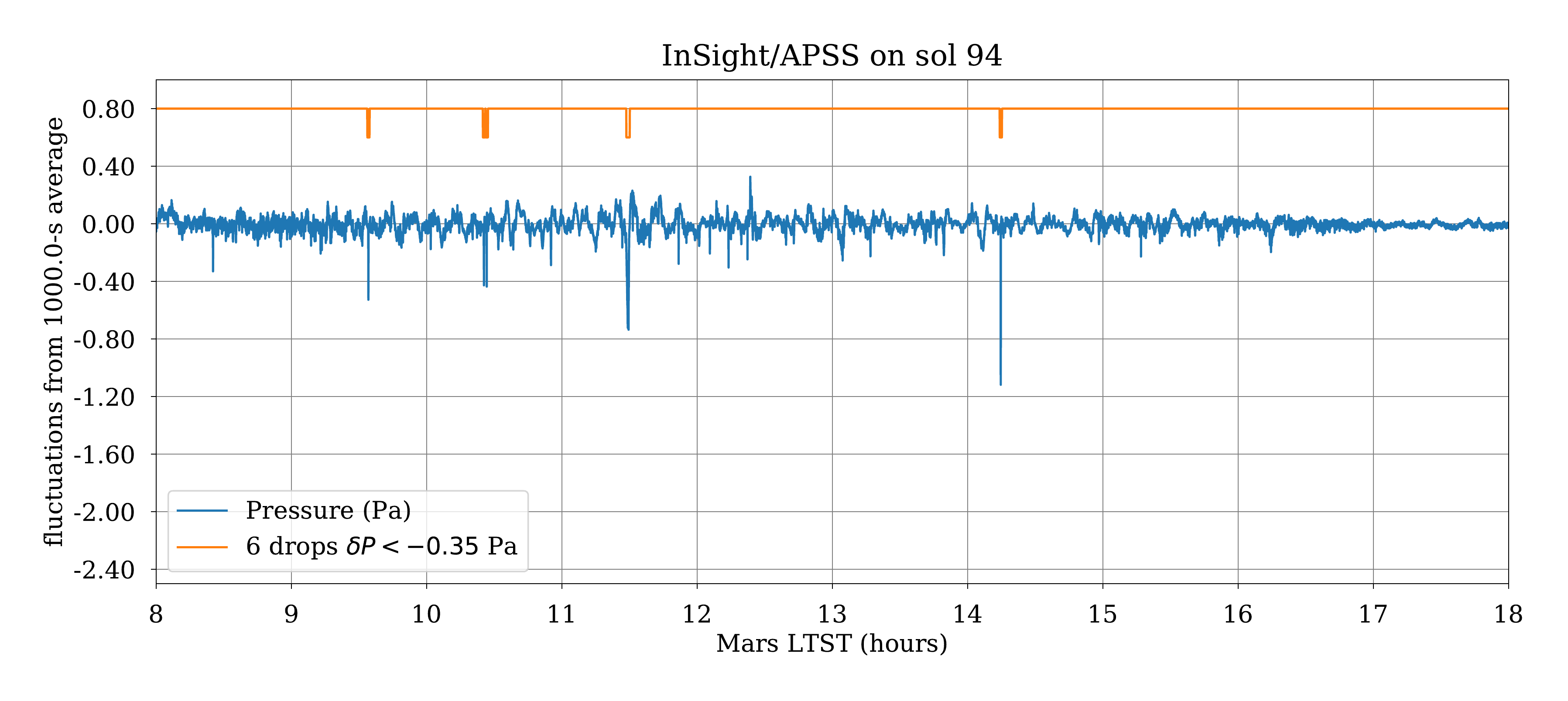}
\includegraphics[width=0.48\textwidth]{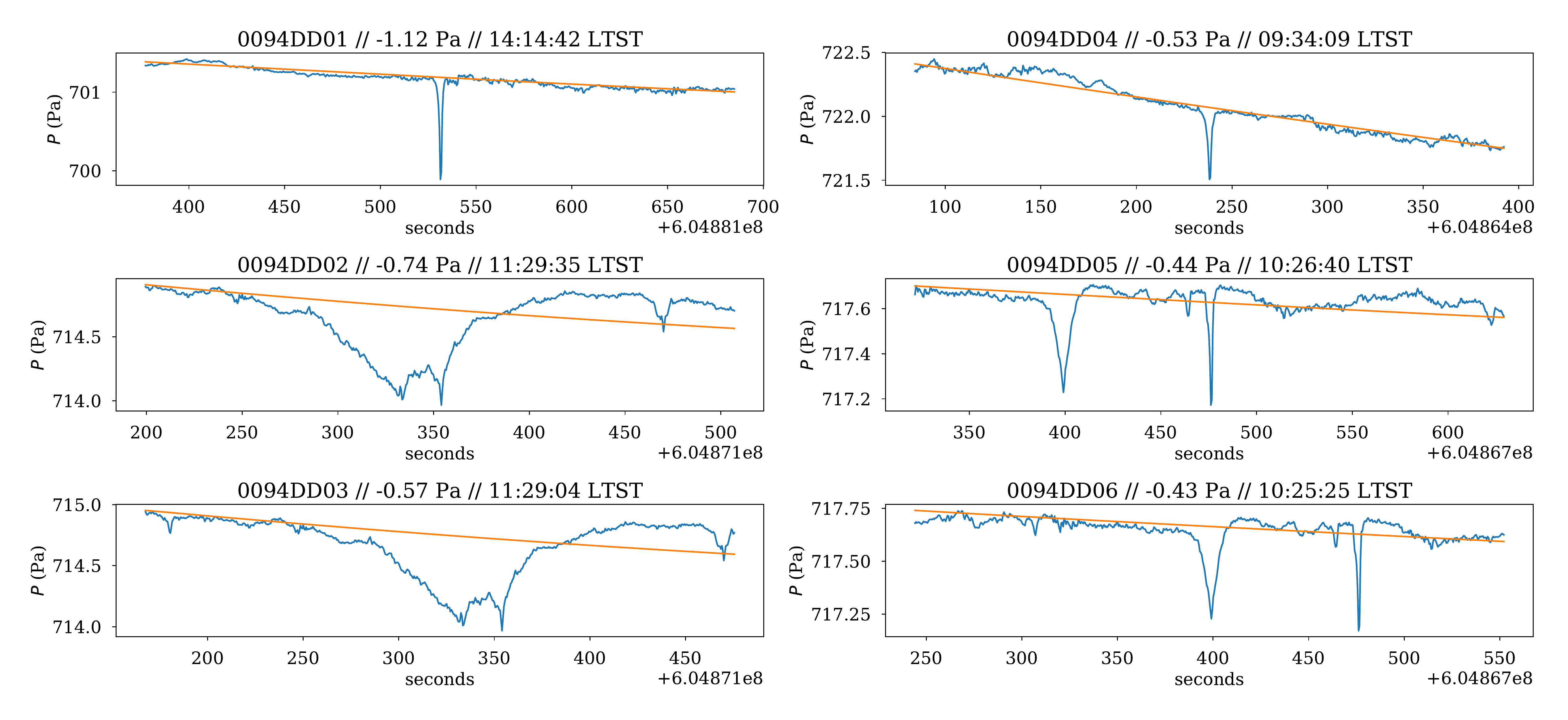}
\vskip 0.3cm
\includegraphics[width=0.48\textwidth]{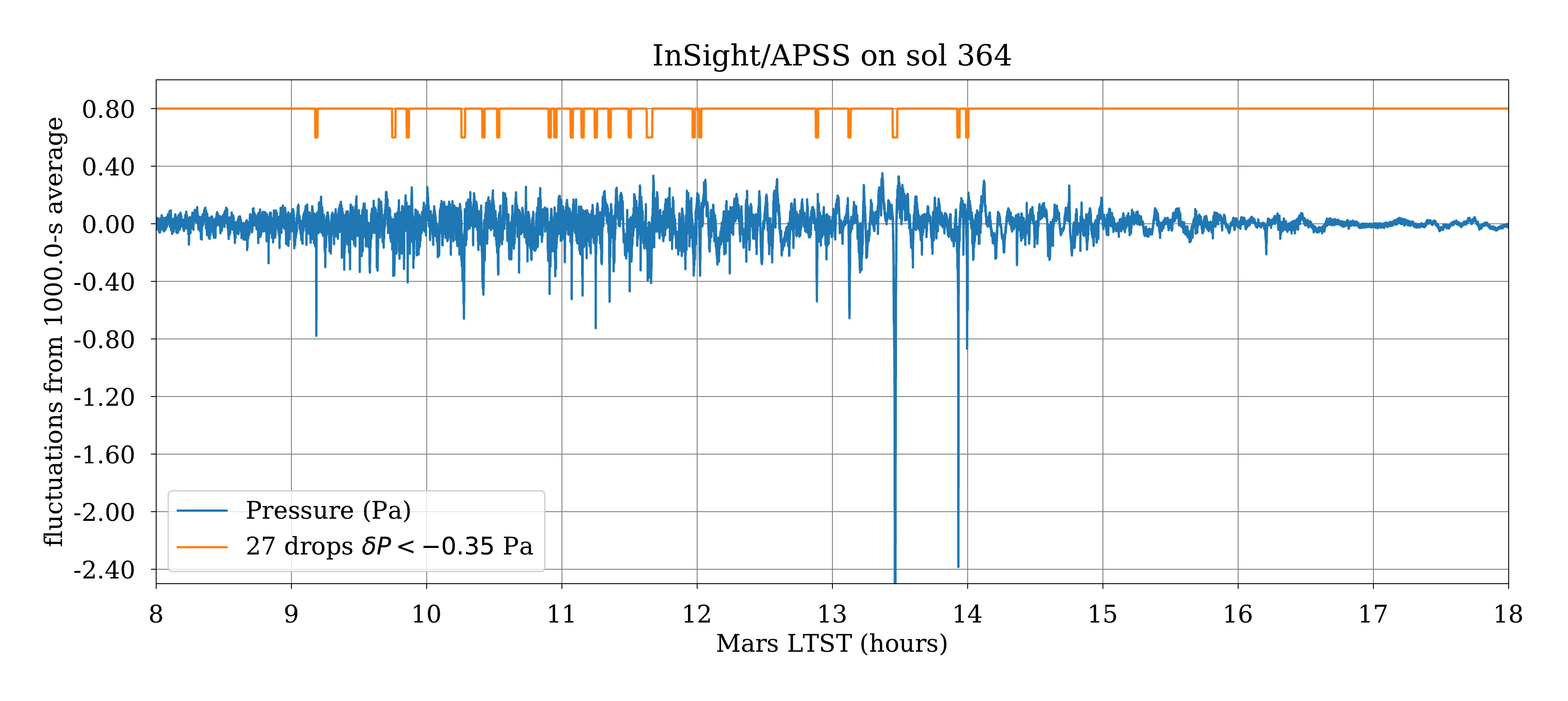}
\includegraphics[width=0.48\textwidth]{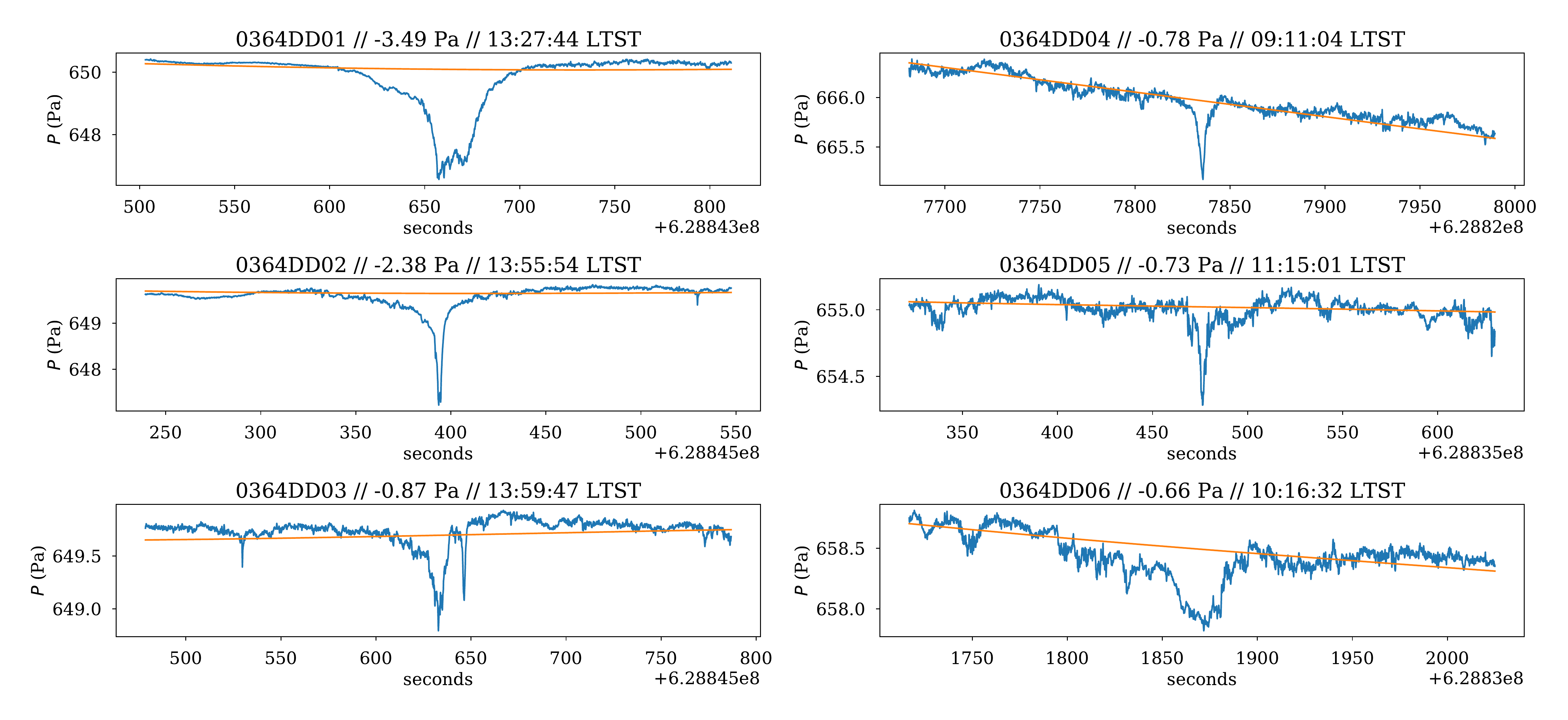}
\caption{ \textit{
Typical examples of detecting 
vortex-induced
pressure drops in the InSight
time series are
shown for 
sols 18, 19, 65, 94, 364
(from top to bottom).
The left plots show,
in blue, the InSight 
daytime pressure
measurements detrended
by subtracting the
signal smoothed with a
1000-s Hanning window
applied on the whole sol
and, in orange,
ticks for detected
pressure drops.
The right plots
feature a subpanel
plot for each of 
the deepest drops
detected on each sol:
the blue lines are 
the InSight
pressure measurements
and
the orange lines are
the smoothed signal.}  
\label{fig:vortices_examples}}
\end{center}
\end{figure}

The method of detecting vortex pressure
drops in the time series of
InSight is slightly different
than most published studies
(e.g., \citeA{Kaha:16}).
Those existing studies
adopt a method
detecting locally 
the drop of pressure
from the ambient pressure
measured just before 
the passage of a vortex.
Here we adopt a method
detecting globally
pressure drops over
the record of pressure
of a full day.
This simply appeared
as more efficient
and straightforward
in the (unprecedentedly continuous)
pressure records obtained on board the InSight lander.
\newcommand{\toronto}{
We do not complement our 
vortex detection by
analysis of the duration
of vortex event --
this is discussed 
in the paper by
Murdoch et al. (in revision).
} \new{\toronto}
An example of five typical
InSight sols with vortex detection
using pressure time series 
is shown in
Figure~\ref{fig:vortices_examples}.

\new{Our detection algorithm
for vortex-induced pressure drop
is set as follows.}
Firstly, 
the InSight pressure signal over a complete sol
is detrended from the diurnal
cycle of pressure
by subtracting a 1000-second window
average from the signal;
then a search of the 
minima of pressure is
performed
between 
LTST
08:00 and 17:00 
(which covers the local time 
of occurrence of 
drop events, see section~\ref{sec:statvortex}), 
starting from the
deepest pressure drops
and gradually
removing the detected
drops from the signal.
\newcommand{\lisbon}{This removal 
is realized
on a window $\pm$~30 seconds
around the pressure
local minimum, meaning
that one limitation of
the method is that double-dip
pressure drops 
are counted as one event,
unless the two local pressure
minimum are occurring
more than 60~s apart
(as is the case for instance
for sols 18 \& 94 in 
Figure~\ref{fig:vortices_examples}).
Using a removal window $\pm$~50 seconds
rather than $\pm$~30 seconds
yields a population
of detected pressure drops
about 10\% less abundant,
yet similar results
on statistics
and seasonality.} \new{\lisbon}

\newcommand{\paris}{How
do we know that a detected 
pressure minimum
correspond to a convective
vortex event?
Compared to vortex-induced
signatures, the variations of pressure 
associated with convective cells
develop on longer timescales
(several hundreds seconds)
and convective cells
induce alternating
pressure highs and lows,
the cell-induced lows 
being less deep than 
the vortex-induced drops
\cite{Lore:12,Spig:12gi}.
This distinction between
daytime PBL vortices and cells
is illustrated, for instance,
in the observed pressure signal
shown in Figure~\ref{fig:firstpbl}
and the modeling results
in Figure~\ref{fig:les}.
In the literature,
a method to select
vortex-induced 
pressure drops
over 
cell-induced 
pressure lows
is to select
only pressure drops
deeper than 
a certain threshold.
A threshold value of
-0.3~Pa is usually
adopted
\cite{Elle:10},
although for
pressure sensors
with a higher noise level, %than InSight,
a conservative
-0.5~Pa limit
is used \cite{Kaha:16}.}
\new{\paris}

\newcommand{\madrid}{We can take advantage
of the continuous InSight
pressure data
to discuss the
choice of the threshold
value for
vortex detection in
pressure time series.
We examined the distribution
of daytime pressure fluctuations 
from the 1000-second window 
smoothed signal, separating
positive and negative
perturbations
(a typical example is shown
in Figure~\ref{fig:histopress}).
A good threshold to
discriminate between
convective cells and vortices
corresponds to the
value for which
the distributions
of positive and negative
perturbations differ
significantly,
i.e. negative pressure perturbations
are more abundant than positive
counterparts hence cannot be
attributed to convective cells.
Figure~\ref{fig:histopress}
illustrates that
the threshold value of
-0.3 Pa employed in the
literature is acceptable,
yet adopting a threshold
of -0.35 Pa is a more
conservative choice
for which pressure
minimum can be more
unambiguously attributed
to convective vortices.
We built two catalogs
using the two distinct
thresholds and
found that the results 
discussed 
in this paper
on drop statistics and
seasonal variability
are not 
significantly altered by
the choice of
threshold.
In what follows,
results obtained with
the most conservative
catalog (using
a -0.35 Pa
threshold value)
are shown.
\begin{figure}[h!]
\begin{center}
\includegraphics[width=0.7\textwidth]{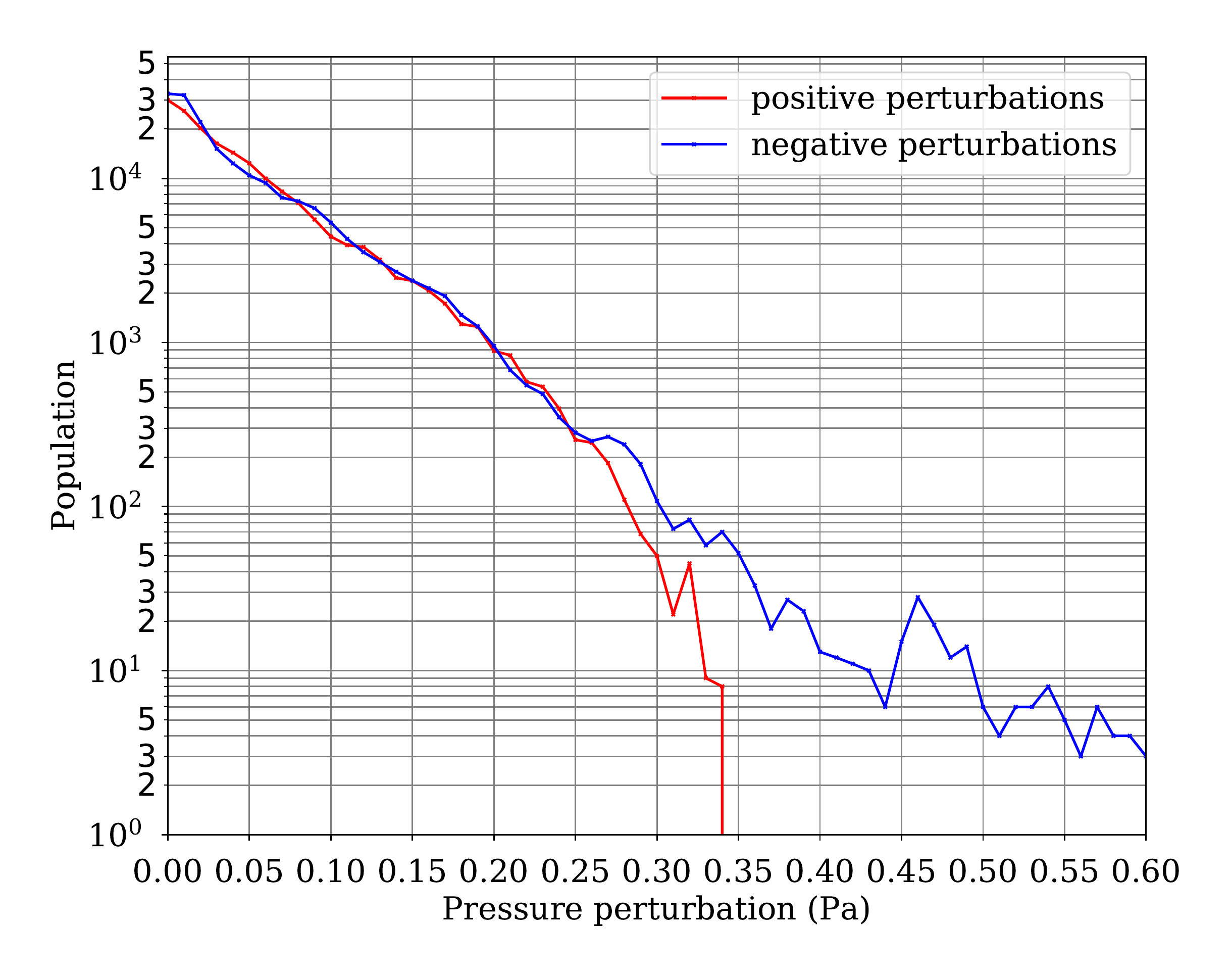}
\caption{ \textit{Histogram of absolute values of pressure perturbations from the 1000-s smoothed signal, with positive perturbations in red and negative perturbations in blue. The analysis of sol 319 is shown as a typical example -- a similar figure for other sols is obtained. The left part, where equivalent populations of positive and negative perturbations are found, corresponds to convective-cell-induced signatures. The right part, where only negative perturbations are found, corresponds to convective-vortex-induced signatures. The threshold adopted in the literature for vortex-induced detection on Mars is 0.3 Pa; we adopt here a more conservative estimate of 0.35 Pa.}  
\label{fig:histopress}}
\end{center}
\end{figure}
}
\new{\madrid}

Detrending with~500-s to 2000-s 
smoothing 
windows was tested
and the 1000-s window
was finally selected
\new{as an optimum
for vortex detection}.
\newcommand{\rome}{A shorter window of 500-s
tends to make 
vortex-induced
pressure
drops to become part
of the smoothed signal
and no longer considered
as perturbations by the algorithm:
as a result, 
long-lasting weak
pressure drops
are not detected
and
the other
pressure drops
are underestimated.
We found that about
20\% pressure drops
detected with the 1000-s
window are left undetected
with the 500-s window.} \new{\rome}
\newcommand{\naples}{
A longer window of 2000~s 
causes long-period fluctuations of pressure
attributed to convective cells
to be almost systematically included
in the perturbation (detrended) signal
rather than the smoothed signal;
as a result, more than 30\%
of detected pressure drops
are related to convective cells
rather than convective vortices.} \new{\naples}

\subsection{Large-Eddy Simulations (LES) \label{sec:methodo_les}}

The results obtained from the InSight measurements 
are compared with turbulence-resolving Large-Eddy Simulations 
(LES).
The principle of LES
is to run a hydrodynamical solver 
of the Navier-Stokes equations
at fine enough spatial resolution
-- on Mars, 
several tens of meters --
to resolve the largest
turbulent eddies in the
daytime PBL, responsible
for most of the transport
of heat and momentum 
there
\cite{Toig:03,Mich:04,Spig:10bl}.
Such computationally-expensive
simulations are usually performed
following the idealized setting
of an infinite flat plain
through doubly-periodic 
boundary conditions.
The turbulent eddies
resolved by LES
include the 
convective cells, gusts, and
vortices developing
in the daytime 
PBL -- only the very-small-scale 
``local'' turbulence is not
resolved by LES.

Here we use the model described
in \citeA{Spig:09}
and \citeA{Spig:10bl}
which couples the 
Weather Research
and Forecast (WRF) 
hydrodynamical solver
\cite{Skam:08},
run at high spatial
and temporal
resolutions 
typical of LES \cite{Moen:07},
to the physical parameterizations,
notably radiative transfer,
developed for Mars at
the Laboratoire de 
M{\'e}t{\'e}orologie
Dynamique (LMD, 
see e.g. \citeA{Forg:99}
and \citeA{Made:11}).

Large-Eddy Simulations performed for this
study dedicated to InSight
extend those developed as pre-landing
investigations in 
\citeA{Kend:17},
\citeA{Murd:17},
and \citeA{Spig:18insight}.
The first two papers used 
LES with a resolution of 50~m 
and the third paper presented 
LES with a resolution of 10~m.
Both are appropriate
to resolve convective cells, 
provided the horizontal 
domain is sufficiently large
to include several convective cells
so as to avoid boundary effects
\cite{Maso:89,Mich:04}.
However, as far as vortices
are concerned, the 50-m configuration
only allows the largest vortices
to be resolved and
the 10-m configuration 
is too computationally
expensive to be 
run on the 
whole local time period
in which vortex activity
takes place.
Moreover, our objective in this study
is to 
perform several LES
runs in order to
explore the sensitivity
of vortex activity
to local time,
seasonal conditions
and ambient wind speed, which
makes the 10-m-resolution
approach untractable for
this purpose.

We thus carry out
in the present study 
LES with a spatial resolution of 25~m
(using an integration timestep of 1/4 second),
\newcommand{\amsterdam}{
hence resolving vortices
of diameters above 50 meters.} \new{\amsterdam}.
The horizontal domain extends over
$481 \times 481$ grid points
in the horizontal directions, which
makes the total extent of the
simulation domain 
$12$~km~$\times$~$12$~km.
The top of the model is set at 10~km altitude
(about twice the expected 
PBL depth) 
with 241
vertical levels.
The surface temperature 
calculations in the model
use a thermal inertia
of~180~J~m$^{-2}$~K$^{-1}$~s$^{-1/2}$
and an albedo of~0.16,
corresponding to
the conditions
encountered at the InSight
landing site, 
\newcommand{\marseille}{based on the 
HP$^3$ radiometer far spot
measurements
considered to be representative
of regional average conditions} \new{\marseille}
\cite{Golo:19nat}.
Radiative transfer computations
assume the longitude and latitude
of the InSight landing site
for the whole LES horizontal domain.

LES runs are initialized 
with a vertical temperature
profile set to be uniformly
similar at all model grid points
and extracted 
at the relevant
season and location
from Global Climate
Model 
simulations
(GCM, \citeA{Forg:99,Mill:15}).
\newcommand{\jakarta}{A random (noise) 
perturbation of
0.1~K amplitude is added
to the initial temperature field 
to break its symmetry 
and help trigger convective motion.} \new{\jakarta}
The LES integrations
are started at 07:00 local time (LTST)
and the diurnal evolution
of incoming sunlight and temperature
profile in the PBL are computed online during
the LES integrations
by the radiative transfer scheme
\newcommand{\lecap}{(visible column dust opacity
considered in the model is 0.8)} \new{\lecap}.
An uniform and constant profile
of ambient wind speed~$V$ 
(positive in the $x$ direction)
is prescribed
in the model. 
Surface friction
and turbulence alter
this prescribed profile 
during the LES integrations,
so that the value~$V$ of prescribed 
ambient wind represents
wind conditions in the
free atmosphere above the PBL;
the value of ambient wind speed
is about~$V/2$
at the
height relevant for InSight 
comparisons (1.165~m, see section~\ref{sec:obs}).
\newcommand{\seoul}{The
two values 10 and 20 m~s$^{-1}$
of ambient wind speed~$V$
prescribed in LES
thus correspond
to the near-surface 
ambient wind conditions
encountered in the distinct
sequences identified from
Figure~\ref{fig:season}
and discussed in 
section~\ref{sec:envseason}.} \new{\seoul}

\section{Vortex population and statistics \label{sec:statvortex}}

\subsection{General population}

\newcommand{\ibiza}{the total number of included sols} 

\begin{figure}[h!]
\begin{center}
\includegraphics[width=0.9\textwidth]{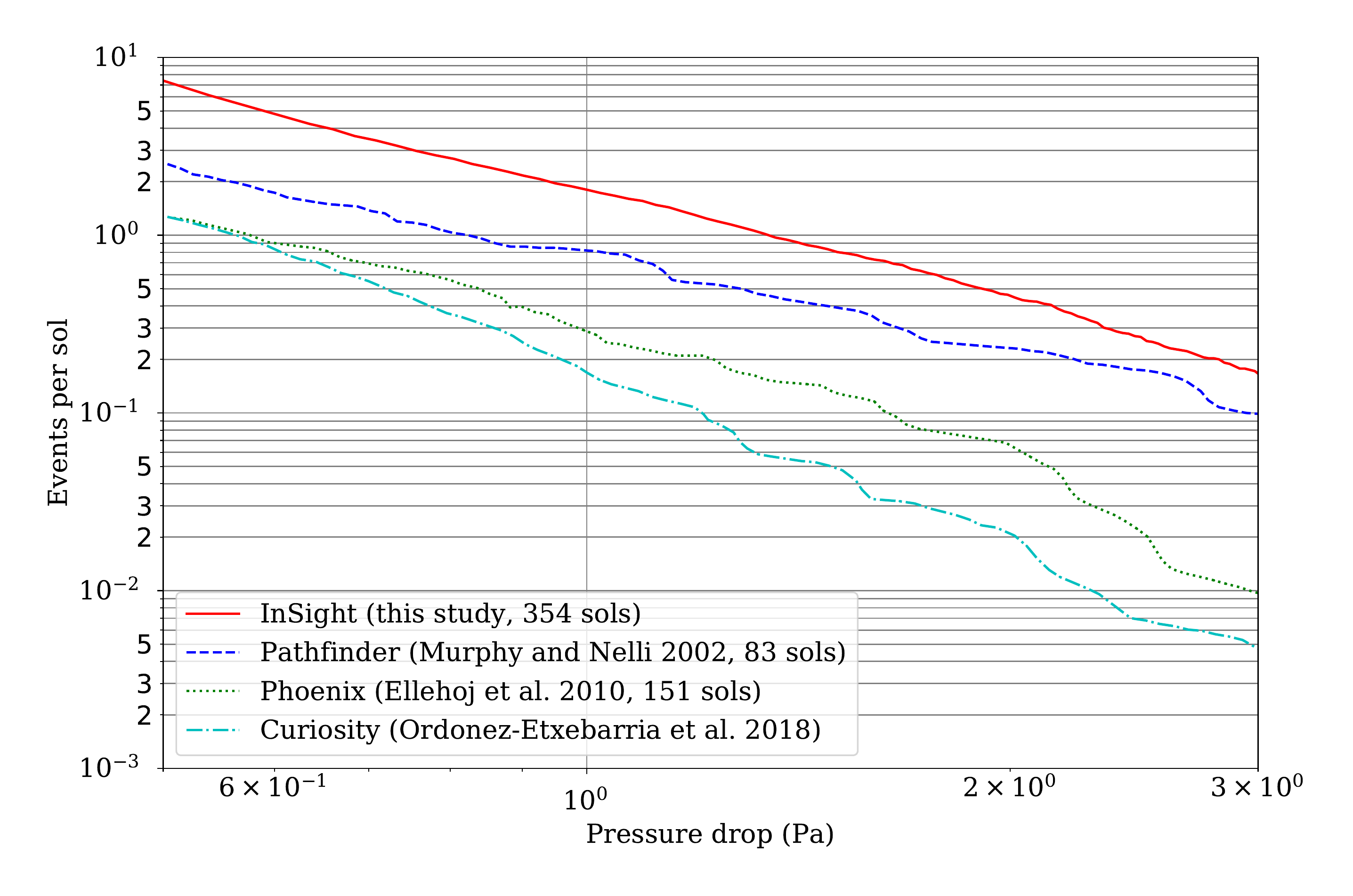}
\caption{ \textit{This logarithmic
diagram shows
the cumulative distribution of detected pressure
drops per sol in the InSight time series
(sols 0-400, 
from northern late winter~$L_s$ = 295$^{\circ}$ 
to midsummer~$L_s$ = 134$^{\circ}$),
normalized by 
\new{\ibiza}.
A total of 354 InSight sols, with
uninterrupted pressure measurements 
in the daytime hours, are included 
to obtain the full red line,
drawn with a bin size 
of 1.7$\times$10$^{-2}$ Pa.
The equivalent statistics 
from other landers are included
for the sake of comparison:
Pathfinder (northern summer
$L_s = 142--183^{\circ}$ at latitude~19$^{\circ}$N) 
with the blue dashed line \cite{Murp:02},
Phoenix (northern spring and summer
$L_s = 77--148^{\circ}$ at latitude~68$^{\circ}$N)
with the green dotted line
\cite{Elle:10},
Curiosity (a full Martian year is included
at latitude 4.6$^{\circ}$S)
with the cyan dash-dotted line
\cite{Ordo:18}.} }
\label{fig:vortices_distrib}
\end{center}
\end{figure}

Convective vortices are known 
to be ubiquitous on Mars \cite{Fent:16ssr}, 
yet InSight appeared as 
a particularly active site 
for convective vortices.
This has been demonstrated with 200 sols
of observations by InSight in \citeA{Banf:20}.
Figure~\ref{fig:vortices_distrib}
confirms, after
400 sols of InSight
observations, that 
the InSight lander operates in
a location prone to numerous
vortex encounters compared 
to previous missions equipped with
a pressure sensor:
Pathfinder \cite{Murp:02},
Phoenix \cite{Elle:10},
Curiosity \cite{Kaha:16,Ordo:18}.
\newcommand{\mercure}{Considering
the 1-Pa pressure drops as a proxy for the total number of vortex events
per sol,
the InSight lander
experiences 
from northern late winter to midsummer
ten times more convective
vortices than the near-equatorial Curiosity 
lander did on average all year long
and about twice as many as the 
tropical Pathfinder lander did
in the northern summer season.
This is all the more striking since, 
in the first 400 sols considered here for analysis,
the annual peak of surface temperature 
has not been reached yet at the near-equatorial site of InSight
(see section~\ref{sec:envseason}).} \new{\mercure}

A total of about 6000 vortex-induced pressure drop events
deeper than 0.35~Pa
are detected between sol 0 and sol 400
of InSight operations.
The strongest detected pressure
drop is 9.2 Pa
\cite{Banf:20,Lore:20},
which is the deepest vortex-induced pressure drop detected to date on Mars.
The sample
of vortex detections
shown in 
Figure~\ref{fig:vortices_examples} 
for 
five typical sols
illustrates the strong 
diurnal, 
daily, and seasonal variability of detected vortex encounters at the InSight landing site.

\subsection{Local time}

\begin{figure}[h!]
\begin{center}
\includegraphics[width=0.7\textwidth]{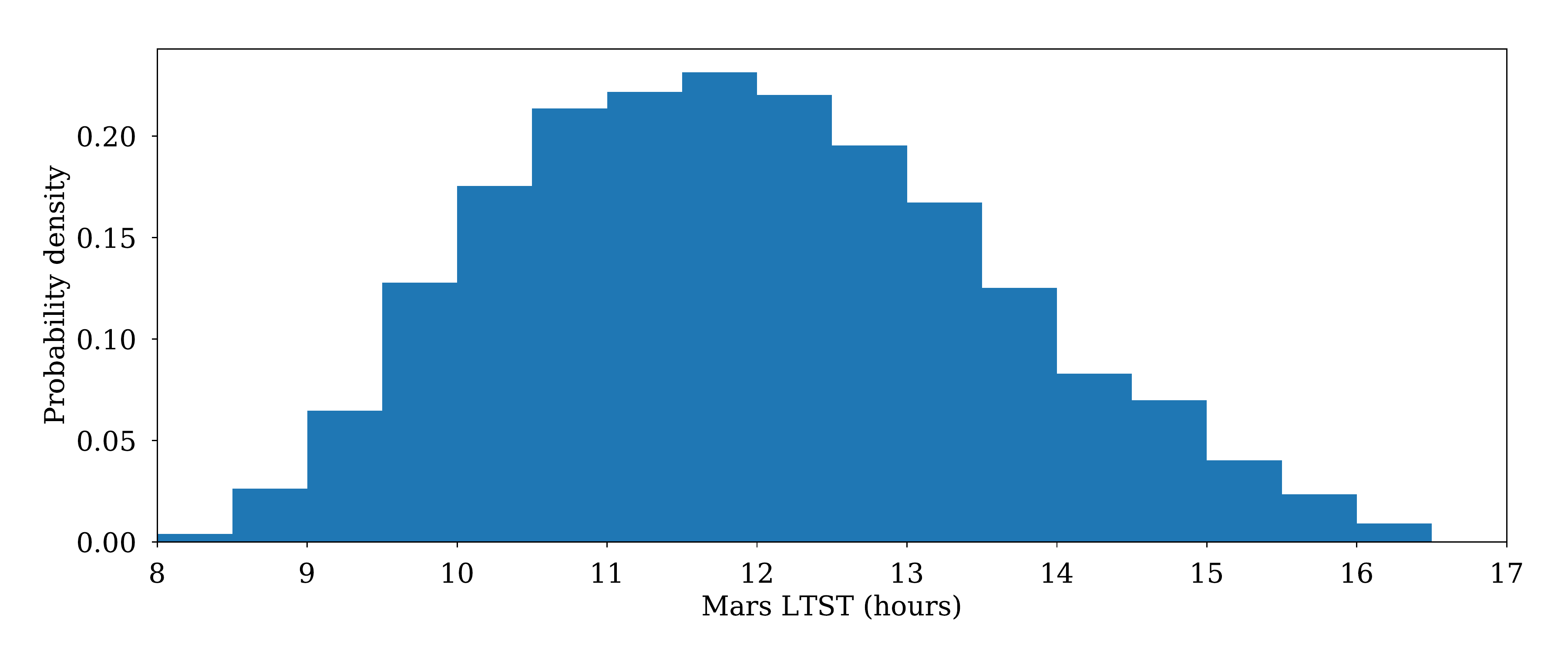}
\includegraphics[width=0.7\textwidth]{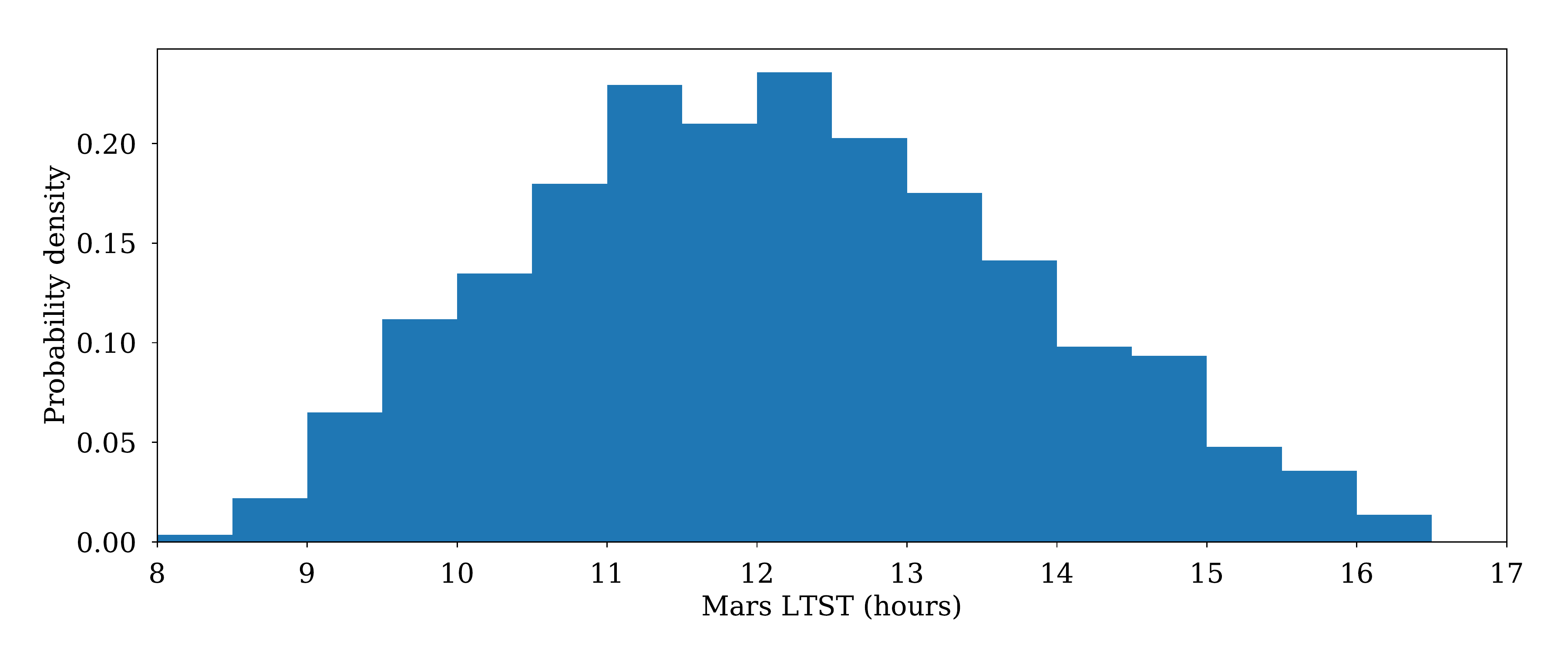}
\caption{ \textit{The vortex-induced
pressure drops
detected in the first 400 sols of
InSight operations are 
gathered here in histogram plots
with bins spanning half-an-hour
intervals of Local True Solar Time.
The top plot includes all detected
pressure drops 
(with a threshold of 0.35~Pa)
while the bottom plot
includes only the pressure
drops deeper than 0.5~Pa.} 
\label{fig:vortices_lt}}
\end{center}
\end{figure}

The local time of occurrence
of convective vortices
at the InSight landing
site is between LTST 08:00 and 17:00
(Figure~\ref{fig:vortices_lt}).
The latest vortex-induced
pressure drop 
deeper than 0.35 Pa
detected at
the InSight landing site
in the first 400 sols
of operations
is at LTST 16:29. %16:38.
No detection prior to LTST 08:00 
is obtained
in the first 400 sols of Insight operations.

Overall, at the InSight landing site,
the vortex activity is high
between local times LTST 
10:00 and 14:00.
As is shown in 
Figure~\ref{fig:vortices_lt},
the peak of activity
for pressure drops 
is between 11:00 and 12:00
LTST, with
an extent towards 12:30
for pressure drops
stronger than 0.5~Pa.
The mean of the
distribution is
found at
respectively
11:57 LTST
and 12:09 LTST;
the standard deviation
of the two distributions
shown in e
Figure~\ref{fig:vortices_lt}
is 1.6 hour.
This is, apparently,
an earlier peak than
expected 
from studies
based on missions other
than InSight
which exhibit 
a maximum occurrence of vortex-induced
pressure drops
around noon
\cite{Murp:16,Kaha:16}.
However, \citeA{Ordo:18}
(their Figure 12)
found that the distribution
of daytime pressure drops
detected by Curiosity
peaked between
11:00 and 12:00 LTST
when considering only
local spring and summer,
which are also the seasons
covered by the present
study addressing
the first half a year
of InSight operations.
In the InSight data,
there is also a tendency of 
the peak of vortex activity to 
occur earlier 
in the summer season,
by about half an hour LTST.
In local summer,
\citeA{Newm:19} found
that
the vortex encounters 
detected by Curiosity
exhibited a double-peak
structure at LTST
10:00 - 11:00
and
13:00 - 14:00
(see also simulations
by \citeA{Chap:17}).
This double-peak structure
is not found during
the first northern summer
season experienced by InSight.

\newcommand{\denali}{No error bars are included
in the histograms when this
power-law fit is performed;
instead, the 
$3\,\sigma$ error
on the power-law exponent
(estimated from the covariance matrix)
is indicated in the plot legend.} 

\begin{figure}[h!]
\begin{center}
\includegraphics[width=0.99\textwidth]{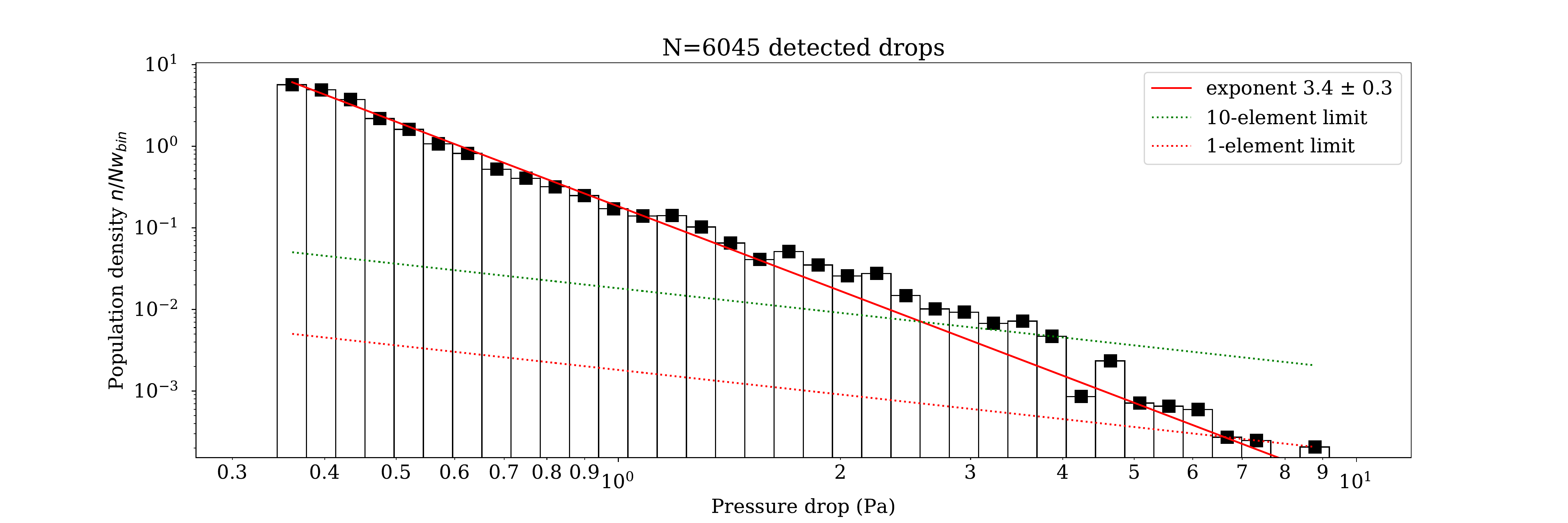}
\includegraphics[width=0.99\textwidth]{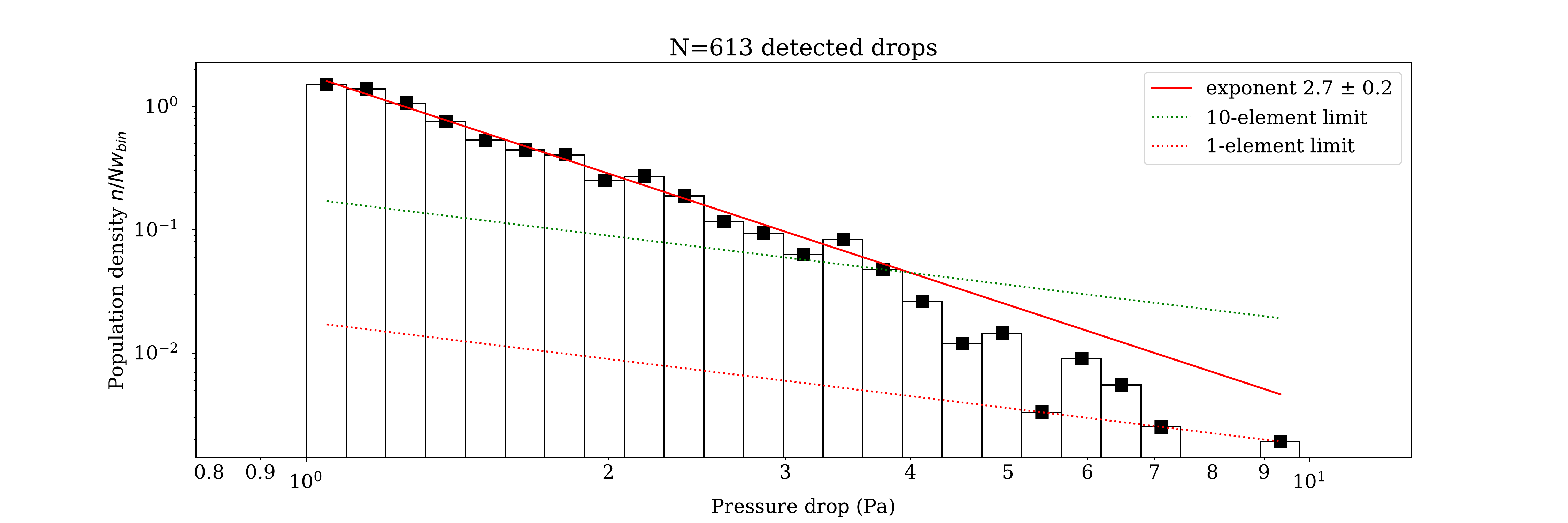}
\caption{ \textit{The population of
detected pressure drops 
normalized by bin sizes (widths)
is shown here
on a logarithmic histogram
with logarithmically-scaled 
bin sizes,
following e.g. \citeA{Lore:11}.
The top plot includes all
vortex-induced pressure drops 
detected by InSight
in the first 400 sols of
operations.
The bottom plot shows
only vortices with
pressure drops deeper than 1 Pa.
An optimal power-law fit
of the distribution,
obtained by a non-linear 
Levenberg-Marquardt
least-squares approach,
is shown as a red line
with the optimum exponent
shown in the legend.
\new{\denali}
The red and green
dotted lines
indicate
the normalized
population 
(value on
the $y$-axis)
corresponding
to respectively
one vortex and
ten vortices
in each respective
bins in the $x$-axis.}  
\label{fig:vortices_distrib_fit}}
\end{center}
\end{figure}

\subsection{Statistical distribution of pressure drops \label{sec:statdrop}}

A question discussed at length 
in the existing literature
\cite{Lore:11,Jack:15,Kurg:19}
is whether the distribution
of pressure drops
caused by convective vortices
(Figure~\ref{fig:vortices_distrib})
follows a power law or not,
and what is the exponent
of this power law.
Clearly, the rich InSight dataset
permits the exploration of
this question 
with an interesting
new statistical perspective,
given the large population
of detected vortex events. 
The upper panel of Figure~\ref{fig:vortices_distrib_fit}
shows a normalized log-log
distribution 
(with logarithmic-sized bins)
of all the vortices
detected in the first 400 sols
of InSight operations.
A power-law
distribution would appear
as a linear
trend in this diagram.
Normalized distribution
means that the 
number of events
per bin is divided by the
bin widths \cite{Lore:11}, which allows the
differential distribution
in pressure drops to be retrieved
\cite{Kaha:16}.

The optimal fit we
obtain in 
Figure~\ref{fig:vortices_distrib_fit} (upper panel), 
with a non-linear 
Levenberg-Marquardt
least-squares
approach,
suggests that
the observed distribution
of pressure drops at the InSight
landing site
is well represented by
a power-law distribution 
with a 3.4~$\pm$~0.3 exponent.
The fit is particularly
good for vortices having
pressure drops
between 0.3
and $\sim$1.5 Pa.
This is reasonably close
to the exponent 3.7 - 3.8
found for Curiosity observations
\cite{Stea:16,Kurg:19}
and to the exponent
of 3 - 3.5 obtained from
corrected Phoenix
observations
\cite{Jack:18}.
Here we caution the reader
that we
did not attempt
to perform a statistical
analysis on the choice of
function to fit
the pressure-drop
population.
We adopted the power law
as a means to 
compare the distributions
obtained 
by InSight observations
versus
other measurements
and numerical simulations,
but our analysis does
not rule out
other possible
functions to fit
the distribution.

For pressure drops deeper
than 1.5 Pa,
the 3.4-exponent 
power law appears to
underestimate the
number of events
actually detected by InSight.
For that particular
population, a power law
with an exponent of 2.7~$\pm$~0.2
provides a better fit,
as is shown in 
the lower panel of
Figure~\ref{fig:vortices_distrib_fit}.
We did not identify
problems or biases 
in our detection
method that would
explain why the
deepest pressure drops
might be systematically
overestimated,
as would be implied
by Figure~\ref{fig:vortices_distrib_fit} 
(top panel) if we assume
that the 3.4-exponent
power law is the reality.
This power-law slope break
might be 
due to the fact that the 
total number
of the deepest detected drops 
is not sufficient to draw
statistically-meaningful
conclusions about
power-law exponents.
This possibility is supported
by the fact that the 
drop distribution did follow
a 2.6-exponent power law 
when we considered the statistics
of all pressure drops deeper than 0.3 Pa
after only 40 sols of InSight operations.

\section{Daytime turbulence and seasonal variability observed by InSight}

\subsection{Environmental conditions and PBL forcings \label{sec:envseason}}

\newcommand{\ottawa}{The sky optical depth 
in the visible,
obtained from InSight cameras
(see section 3.3.2
of \citeA{Spig:18insight}
and section 2.1
of \citeA{Viud:20}),
is shown in the bottom panel.}

\newcommand{\sun}{The uncertainties indicated
in section~\ref{sec:obs}
are reported as error bars on
the figures.}

\begin{figure}[p!]
\begin{center}
\includegraphics[width=0.75\textwidth]{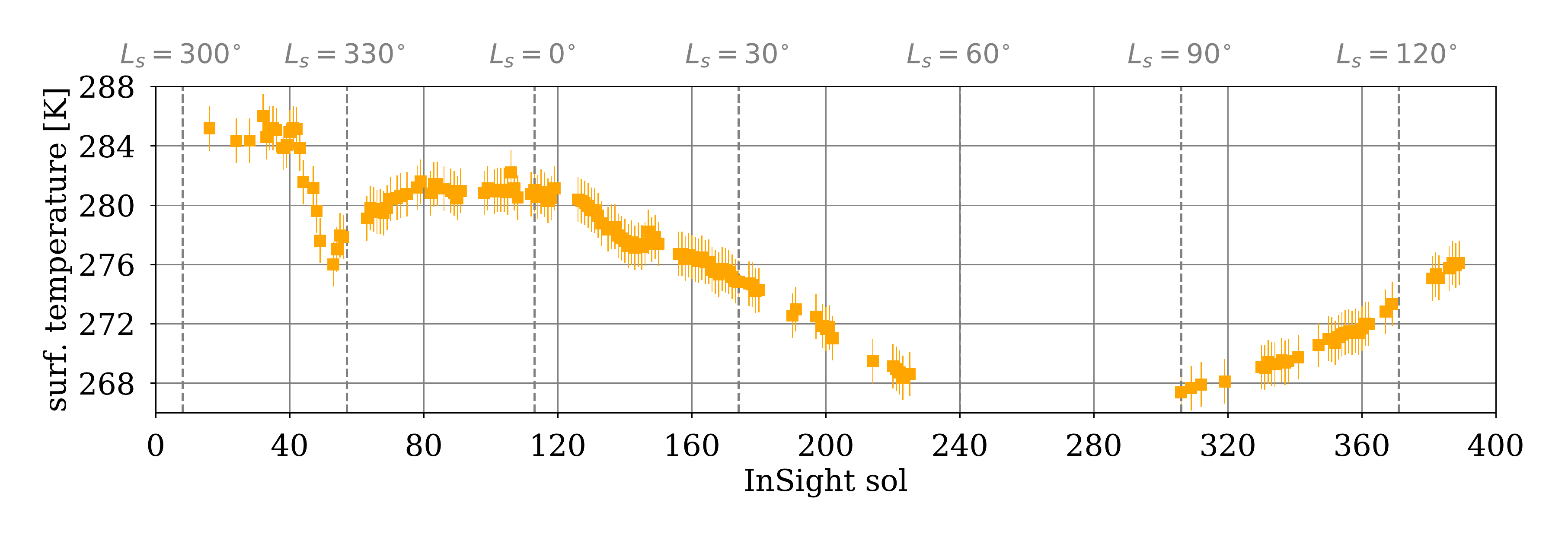}
\includegraphics[width=0.75\textwidth]{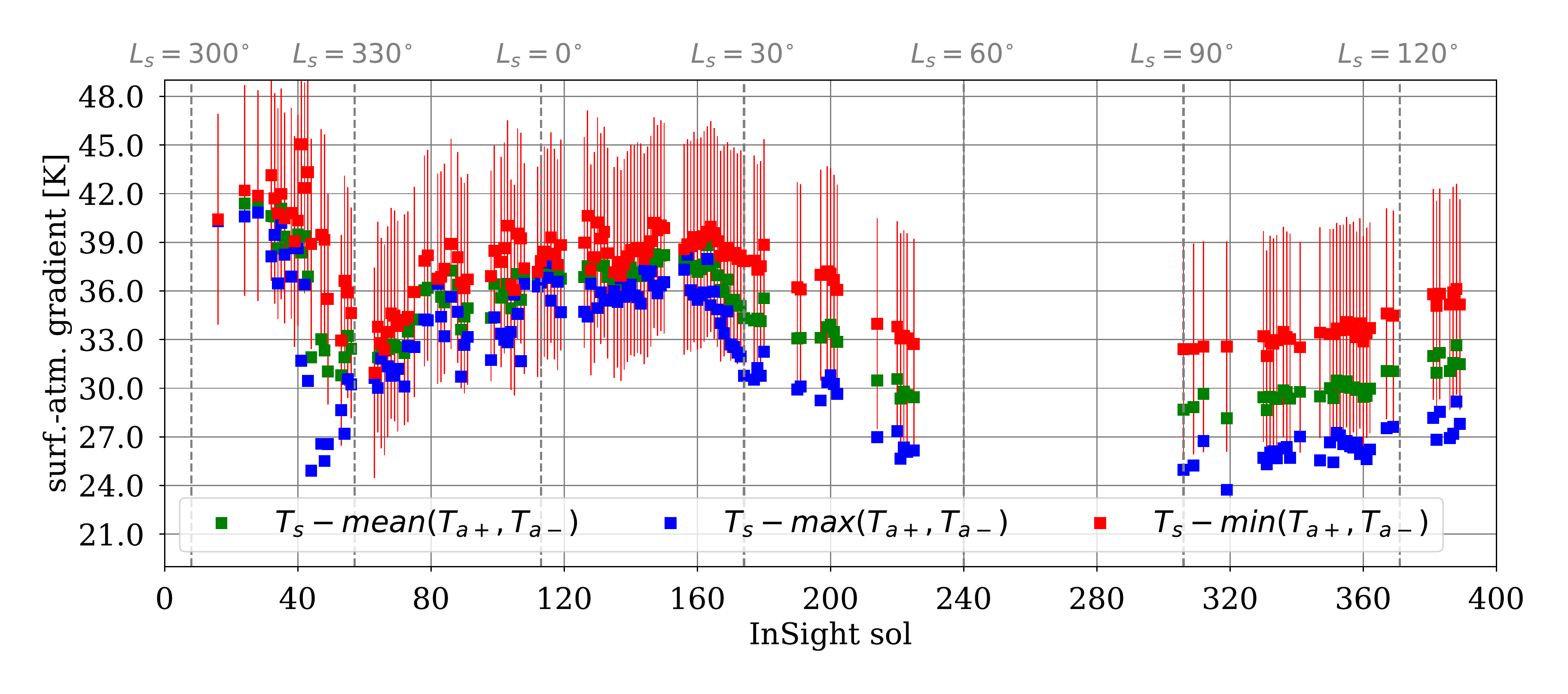}
\includegraphics[width=0.75\textwidth]{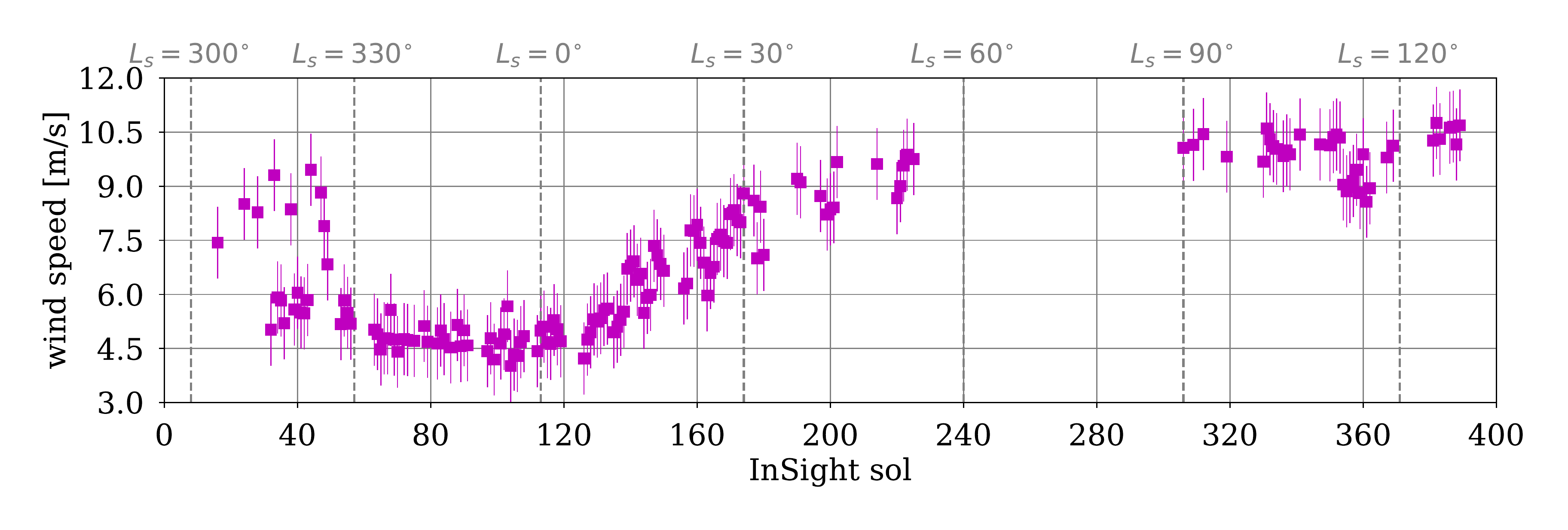}
\includegraphics[width=0.75\textwidth]{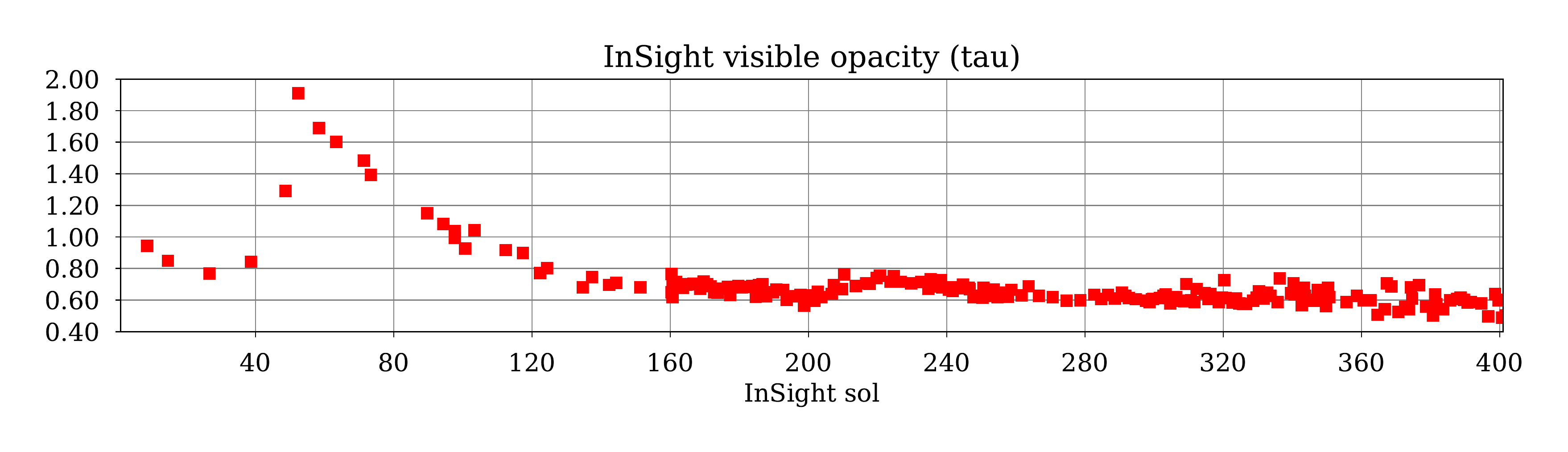}
\caption{ \textit{The seasonal evolution
of daytime
surface temperature (first panel from top),
surface-to-atmosphere temperature gradient (second panel from top),
and ambient wind speed (third panel from top),
are shown for the first 400 sols
of Insight operations.
Each point is an average 
of the indicated quantity
performed for each sol 
in the local time
interval 11:00 - 14:00
(Local True Solar Time).
\new{\sun}
Atmospheric wind and temperature 
measurements by APSS/TWINS
data
are available for more sols
than shown here; 
the three top diagrams only
show the sols for which
sufficient HP$^3$ radiometer data 
for 
surface brightness 
temperature 
measurements were collected
in the local time interval considered.
\new{\ottawa}} \label{fig:season}}
\end{center}
\end{figure}

The seasonal evolution of the conditions relevant for turbulence in the daytime PBL
is summarized in Figure~\ref{fig:season}.

Surface temperature~$T_s$ behaves as expected from the seasonal evolution at the equator, given InSight values of albedo and thermal inertia typical of Martian bare soil. A seasonal decrease of surface temperature is observed from northern winter ($L_s=300^{\circ}$) to northern summer ($L_s=90^{\circ}$), then surface temperatures rise again, pointing towards an expected seasonal peak at northern fall equinox ($L_s=180^{\circ}$) as predicted, e.g., from the Mars Climate Database \cite{Mill:15}. \newcommand{\lapaz}{This behavior could appear as counter-intuitive, however the behaviour of surface temperature observed at the InSight landing site in Figure~\ref{fig:season} is explained by the near-equatorial position of the spacecraft.} \new{\lapaz}

A notable dip of daytime surface temperature -- departing from the sine-shaped seasonal evolution -- occurred from sol 40 to sol 80 and corresponds to a large regional dust storm outside the InSight landing site region that doubled the dust optical depth in the InSight landing site region \cite{Banf:20,Viud:20} (see Figure~\ref{fig:season} lowermost panel). As a result, at the InSight landing site, the incoming sunlight reaching the Martian surface is
lower as a result of enhanced
absorption and scattering by the additional dust particles present in the atmosphere, hence the dip in daytime surface temperature.

What drives PBL convection in daytime is the near-surface convective instability that we could diagnose
by computing the surface-atmosphere gradient~$T_s - T_a$. 
The seasonal evolution of this gradient
follows to first order the seasonal evolution
of surface temperature; yet the impact of the local dust
storm from sol 40 to sol 80 appears 
more prominent than it is
on the surface temperature signal (Figure~\ref{fig:season}). 
The surface-atmosphere gradient also stays quite
high on sol 150 while the surface temperature has
started its seasonal decrease.

Another important control on PBL convection is
the near-surface ambient wind speed~$V$.
Discussing the 
physical mechanisms
underlying the
seasonal evolution
of large-scale wind speeds
is out of the scope of the present paper
(see \citeA{Spig:18insight},
\citeA{Banf:20}).
Suffice to say here that
the high daytime wind speeds 
in northern winter
(beginning of the InSight mission,
$L_s = 300-330^{\circ}$)
and northern spring to summer
($L_s = 60-120^{\circ}$),
and the decrease
in late northern winter
correspond to the transition
between two annual wind regimes
driven by a combination
of large-scale 
(Hadley cells)
and regional
(western boundary currents)
circulations.

From the seasonal evolution
of surface-atmosphere temperature gradient
and
ambient wind speed,
three sequences
in the first half year of InSight 
can be drawn -- the indicated
season references the northern hemisphere.

\begin{enumerate}

    \item \textit{Early mission} (late \new{northern} winter, sols 0 to 40, $Ls=300-330^{\circ}$). This sequence is characterized by both high surface-atmosphere temperature gradient, mostly as a result of surface temperature being high, and high ambient wind speed in the northern winter ``windy'' season. 

    \item \textit{Dust storm and spring} (early \new{northern} spring, sols 40 to 160, $Ls=330-30^{\circ}$). Following the rise of dust opacity at the InSight landing site caused by a regional dust storm that started on sol 40 (\citeA{Banf:20}, Viúdez-Moreiras et al., this issue), both the surface-atmosphere temperature gradient and the ambient wind speed decrease. 
    The decrease in surface-atmosphere temperature gradient is significant (from about 45~K to about 32~K) but, even in those regional dust storm conditions, near-surface temperature gradients on Mars remain  super-adiabatic.
    The behavior of the wind speed is more subtle and less clearly related to the regional dust storm than temperature.
    Wind speed actually remains high at the beginning of the regional dust storm from sol 40 to sol 45.
    Then, the decrease in wind speed starting at sol 50 at $L_s = 326^{\circ}$ is predicted as a normal seasonal evolution by the pre-landing LMD GCM simulations in \citeA{Spig:18insight} even with no regional dust storm at this season (see their Figure 8).
    Indeed, the transition from northern winter solstice to spring equinox cause the wind speed to decrease, as a prelude to the seasonal reversal of the Hadley circulation closer to northern summer solstice.
    This difference of evolution between temperature and wind speed is also clear in the aftermath of the regional dust storm.
    When the dust opacity returns to levels seen at the beginning of the mission (around sol 100), both surface temperature and surface-atmosphere temperature gradient has risen again to close to pre-storm values, to follow the sine-shaped long-term seasonal variations; conversely, ambient wind speed remains low, in agreement with the seasonal evolution predicted by models (see \citeA{Spig:18insight} and also Baker et al. in revision for this issue).

    \item \textit{Aphelion season} (from \new{northern} mid-spring to summer, sols 160 to 400, $Ls=30-120^{\circ}$). Starting from sol 160, both the surface temperature and the surface-to-atmosphere temperature gradient decrease dramatically (-10~K), while at the same time the ambient wind speed rises by almost a factor 2 to reach values slightly larger than in the \textit{Early mission} sequence. This sequence is interesting for the seasonal evolution of turbulence, since it combines a wind speed equivalent to the \textit{Early mission} sequence
    but surface temperature conditions 30~K colder than during this earlier sequence.
    Note that there is a gap in the range $Ls=60-90^{\circ}$, due to a combination of HP$^3$ radiometer troubleshooting and solar conjunction, but the pressure, temperature, and wind measurements available in this range shows that the atmospheric conditions are equivalent to those before and after the data gap.
    
\end{enumerate}

\subsection{Seasonal evolution of turbulence \label{sec:season_vortex}}

\subsubsection{Convective vortices \label{sec:season_vortex_spec}}

\begin{figure}[p!]
\begin{center}
\includegraphics[width=0.99\textwidth]{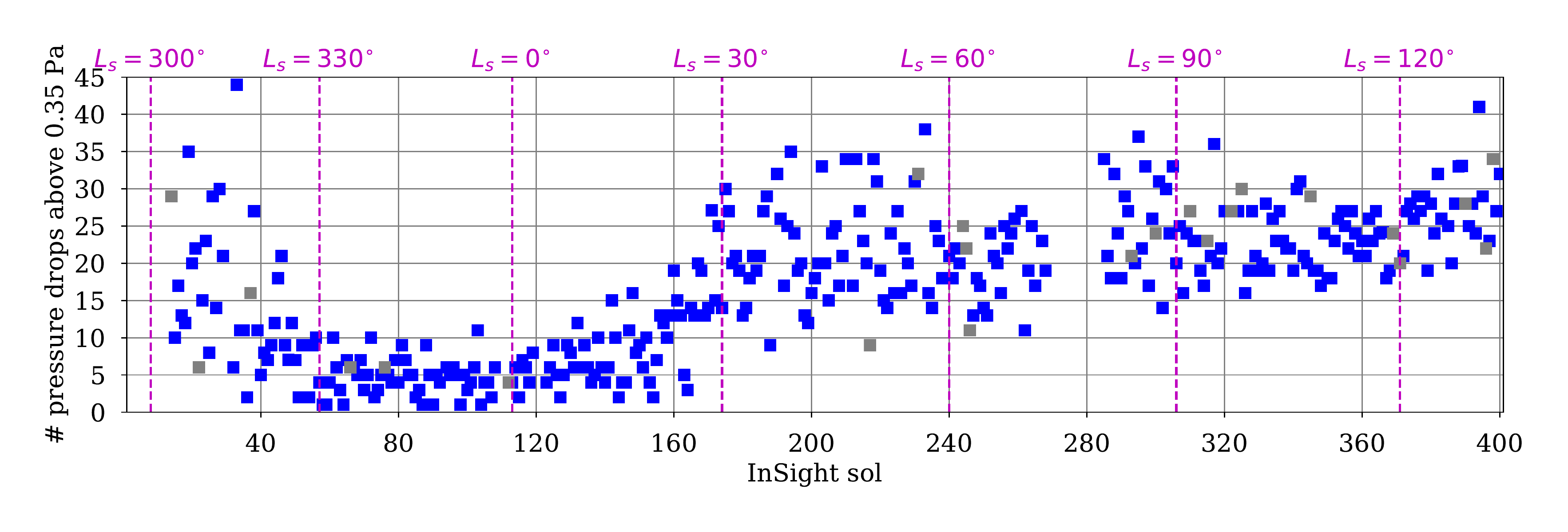}
\includegraphics[width=0.99\textwidth]{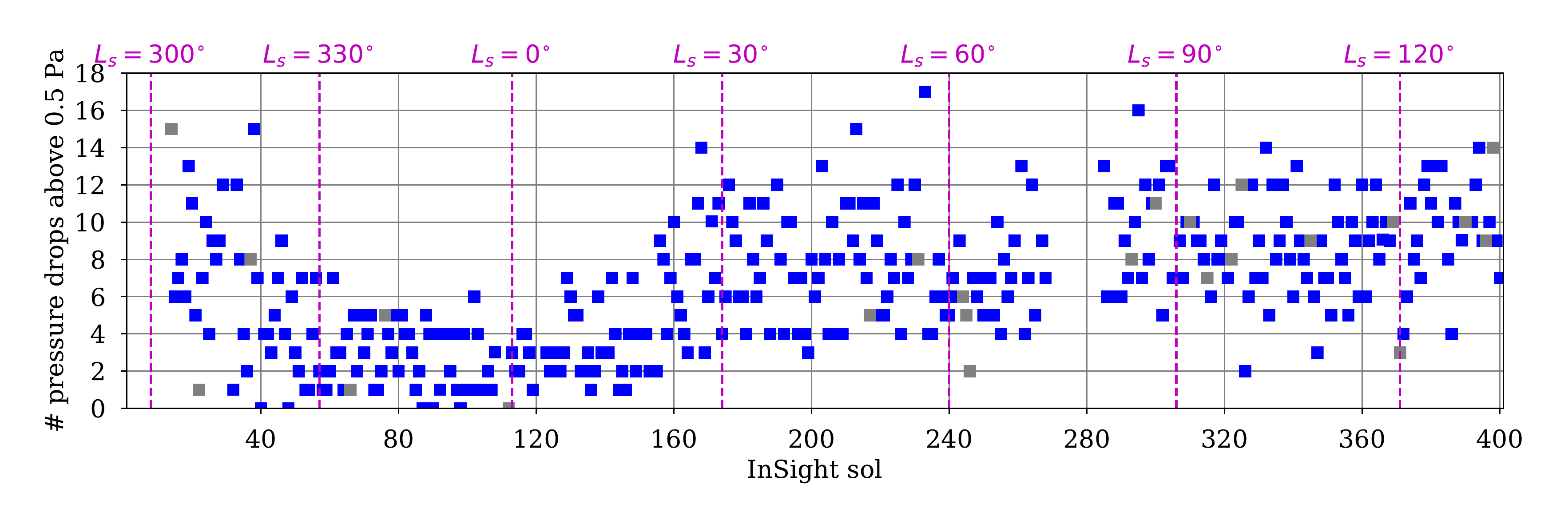}
\includegraphics[width=0.99\textwidth]{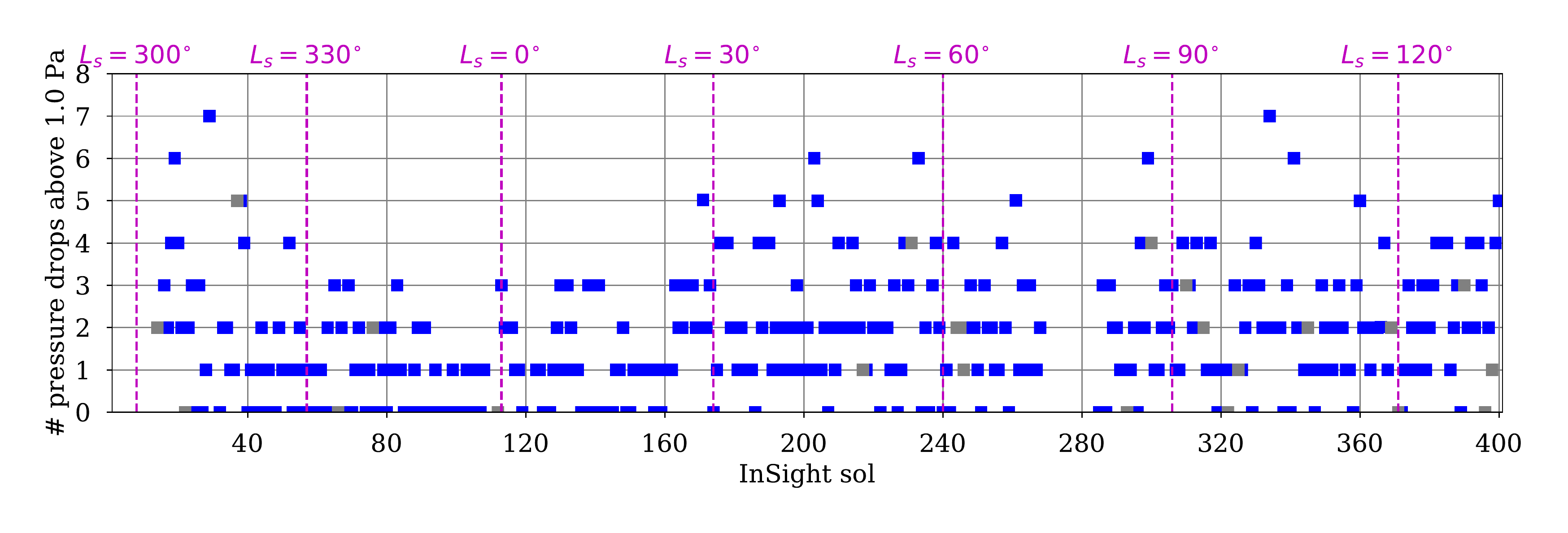}
\caption{ \textit{The number of 
vortex encounters
detected in each sol of InSight operations are
shown here for retrieved pressure drops 
above 0.3 Pa (top), 
0.5 Pa (middle), 1 Pa (bottom).
The blue squares correspond to
InSight sols with complete daytime coverage
in the local time interval 08:00-17:00.
The gray squares correspond to 
incomplete InSight sols,
having gaps several hours long
in the local time interval 08:00-17:00.
In those particular cases, the
number of vortex encounters
is obtained by considering
the number of detected pressure drops
in the covered local times
and correcting for local time gaps
using a Gaussian diurnal distribution
approximating the observed 
local-time distribution 
in Figure~\ref{fig:vortices_lt}.} 
\label{fig:vortices}}
\end{center}
\end{figure}

Figure~\ref{fig:vortices}
shows the seasonal evolution
of the number of convective-vortex 
pressure drops detected at the
InSight landing site.
\newcommand{\rio}{
In the literature,
a quantity
named the ``Dust Devil Activity'',
combining the sensible heat flux
with PBL depth
\cite{Renn:98,Newm:17},
is used to 
relate ambient
conditions to vortex activity.
It is difficult to use
this diagnostic with
InSight data since
PBL depth estimates are not
robust enough
(see section~\ref{sec:ccell})
and, as is explained above,
sensible heat flux
is a degenerate diagnostic.
We use instead
more direct diagnostics
to interpret the seasonal
evolution 
of detected vortices:
surface temperature,
surface-to-atmosphere gradient 
and ambient wind speed.} \new{\rio}

Figure~\ref{fig:vortices} clearly
indicates,
over half a year of
InSight observations,
a much clearer correlation
of the number of detected vortices
with the ambient wind speed
than with the surface-to-atmosphere
gradient (or surface temperature) -- 
whether the total
population of vortices
or the population of deepest-drop
vortices are considered.
The vortex activity at the InSight
landing site is as intense
in the \textit{Aphelion season} sequence
as in the \textit{Early mission} sequence,
despite a significant drop
in surface-to-atmosphere gradient.
In the \textit{Dust storm and spring}
sequence, the vortex activity also 
closely follows the evolution of
ambient wind speed and, 
in the latest stages of this sequence,
rises while the 
surface-to-atmosphere gradient
is dropping significantly as a 
result of seasonal evolution.
We note that past studies 
also reported an increase in vortex
detections in frontal conditions
when the ambient
wind speed
was likely to
be significantly
higher
\cite{Elle:10,Stea:16,Kaha:16}.
\newcommand{\santiago}{
The strong correlation 
between InSight vortex detections
and ambient wind speed
is confirmed quantitatively
in Figure~\ref{fig:correlation}:
the Pearson correlation coefficient
is close to 0.9 (positive linear correlation),
while vortex detections
are not correlated to 
surface-air temperature gradient,
and weakly anti-correlated
to surface temperature
(at odds with physical 
expectations given the surface
forcing of daytime PBL convection).
\begin{figure}[h!]
\begin{center}
\includegraphics[width=0.6\textwidth]{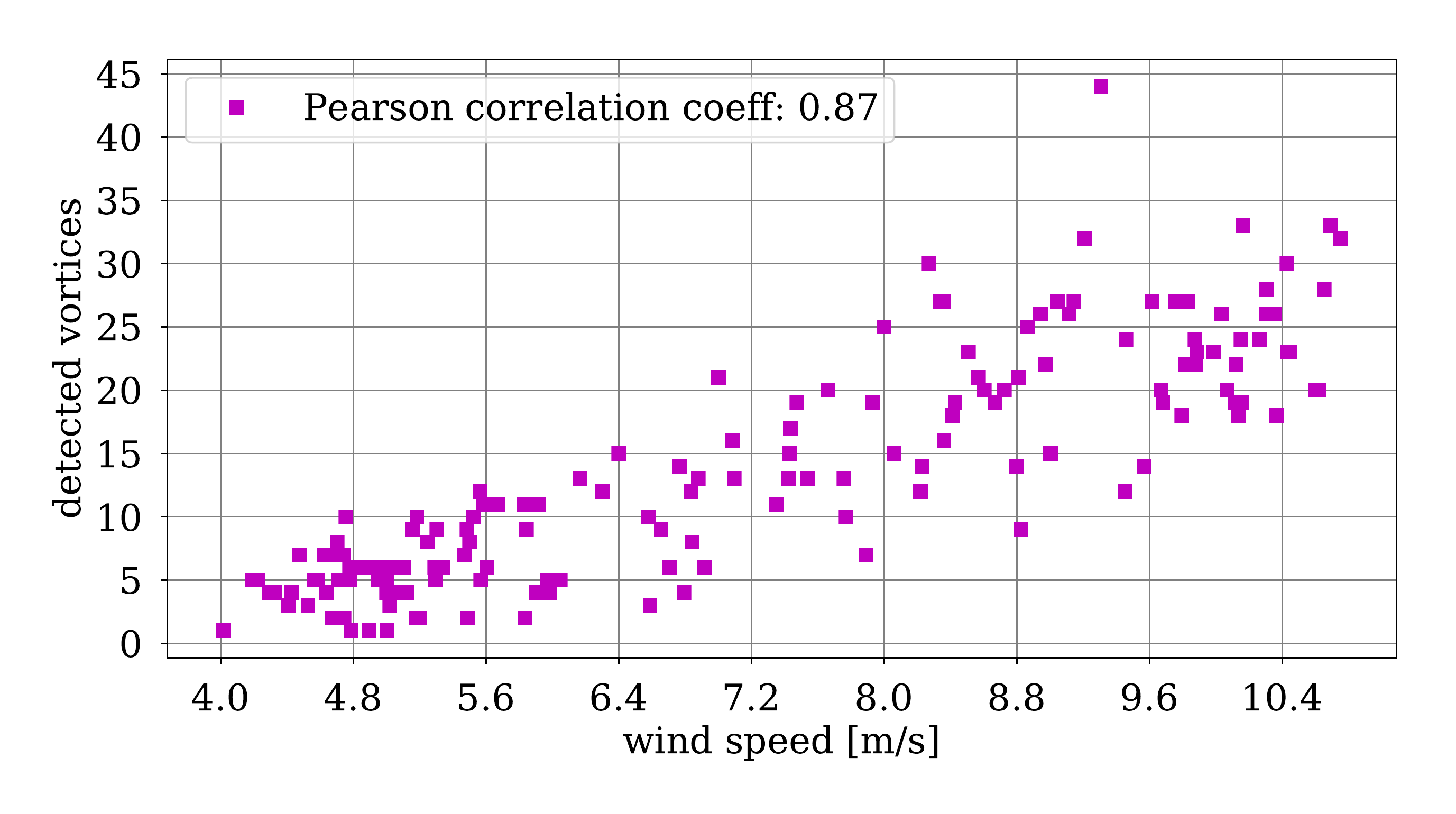}
\includegraphics[width=0.6\textwidth]{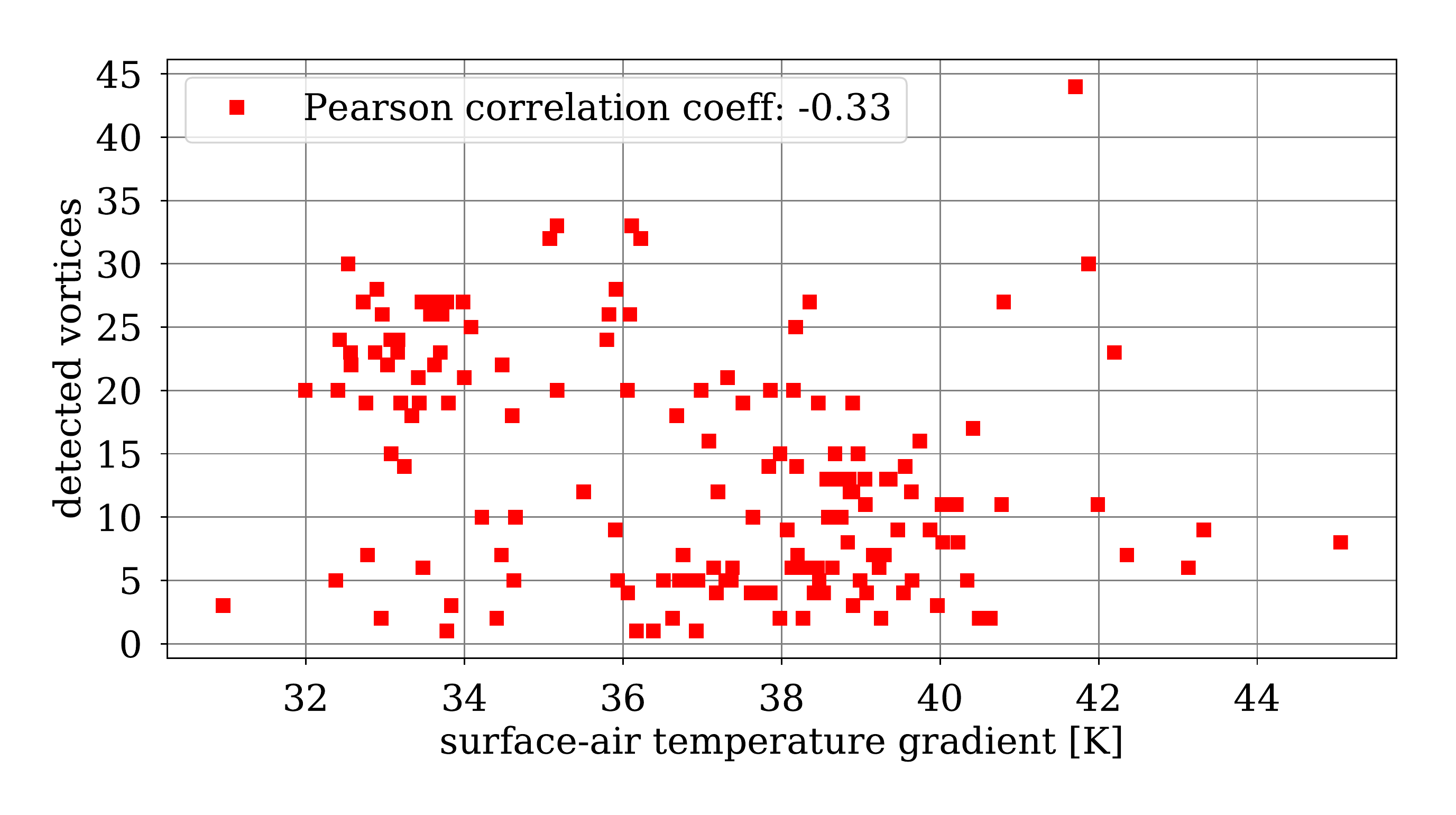}
\includegraphics[width=0.6\textwidth]{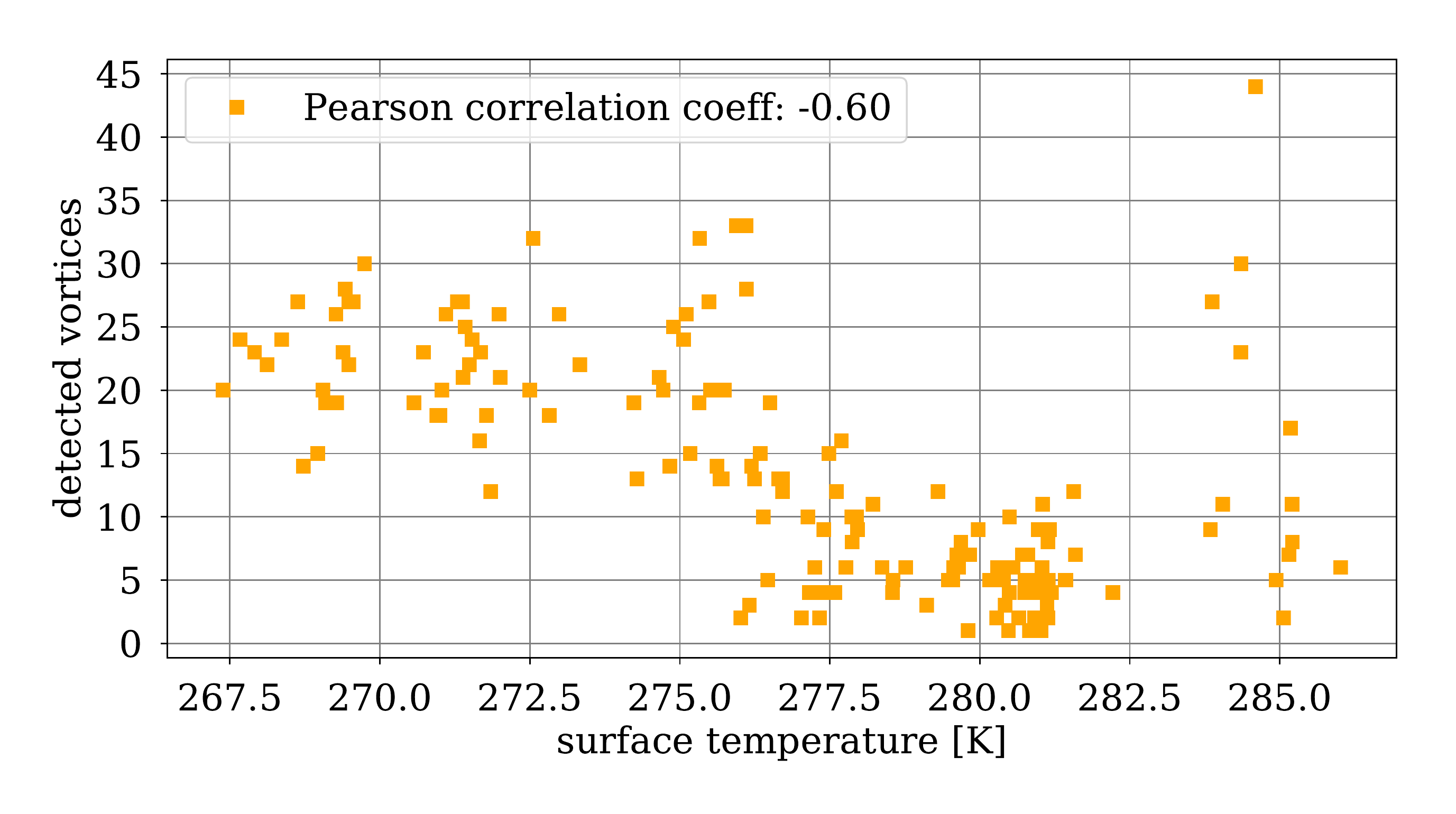}
\caption{ \textit{Counts of vortex-induced pressure drop are displayed as a function of environmental conditions (from top to bottom: wind speed, surface-to-atmosphere gradient, surface temperature). Each square represents a sol shown in Figure~\ref{fig:vortices}. The Pearson correlation coefficient is estimated in each case and indicated as a legend in each plot.} 
\label{fig:correlation}}
\end{center}
\end{figure}
} \new{\santiago}

This correlation between
the activity of
convective vortices 
and ambient wind speed
is degenerate.
Higher ambient wind speed
causes
larger sensible heat flux,
hence a putatively more active 
turbulence -- although on
Mars the radiative forcing
of the daytime PBL is
dominant.
However, 
vortices are also advected 
by the ambient wind
\cite{Balm:12,Reis:14}
hence move faster if
the ambient wind is large.
Thus, if we assume a similar vortex
formation rate at low
and high wind conditions,
the probability of encounter
by a fixed station such as InSight
would be larger in the high-wind case.
Large-Eddy Simulations are proposed
in section~\ref{sec:les} to further investigate this
question.
At the same time, it should be noted that shearing
may prevent the formation
of convective vortices if the ambient wind speed is too high
\cite{Kurg:11,Balm:12}.
This does not seem to
be the case at the InSight
landing site where,
even in the 
low-surface-temperature
and
high-wind-speed conditions
of the \textit{Aphelion season}
sequence, the number
of vortex encounters 
is very high.

\newcommand{\mars}{
\begin{figure}[h!]
\begin{center}
\includegraphics[width=0.99\textwidth]{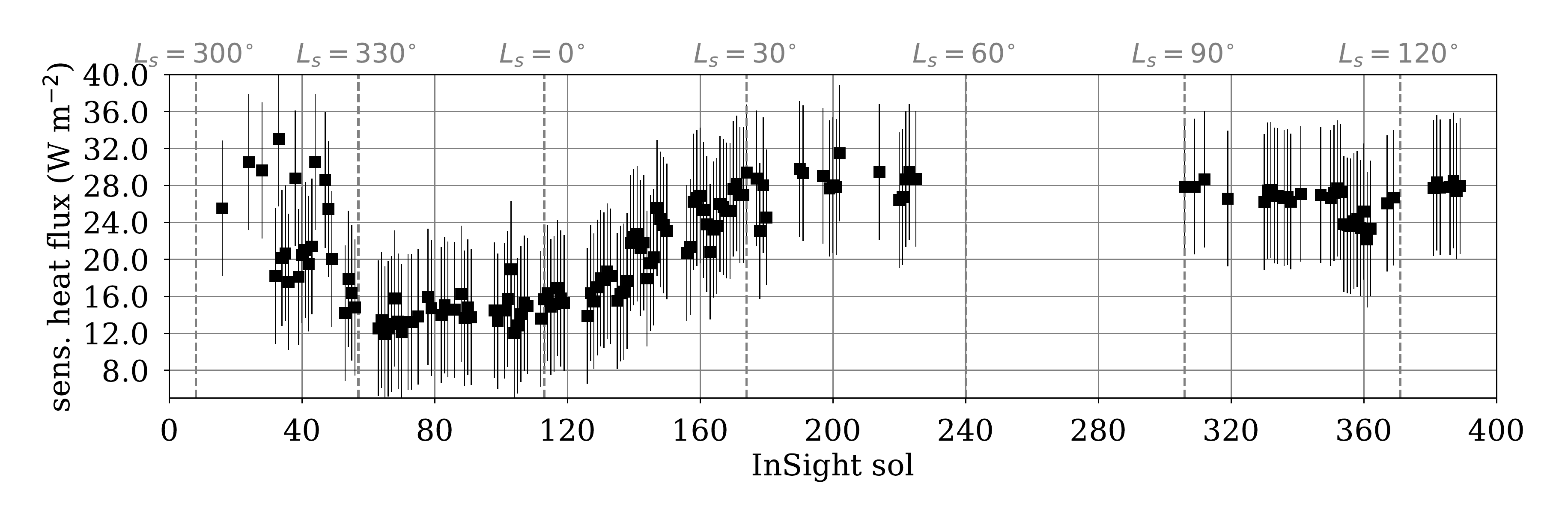}
\includegraphics[width=0.7\textwidth]{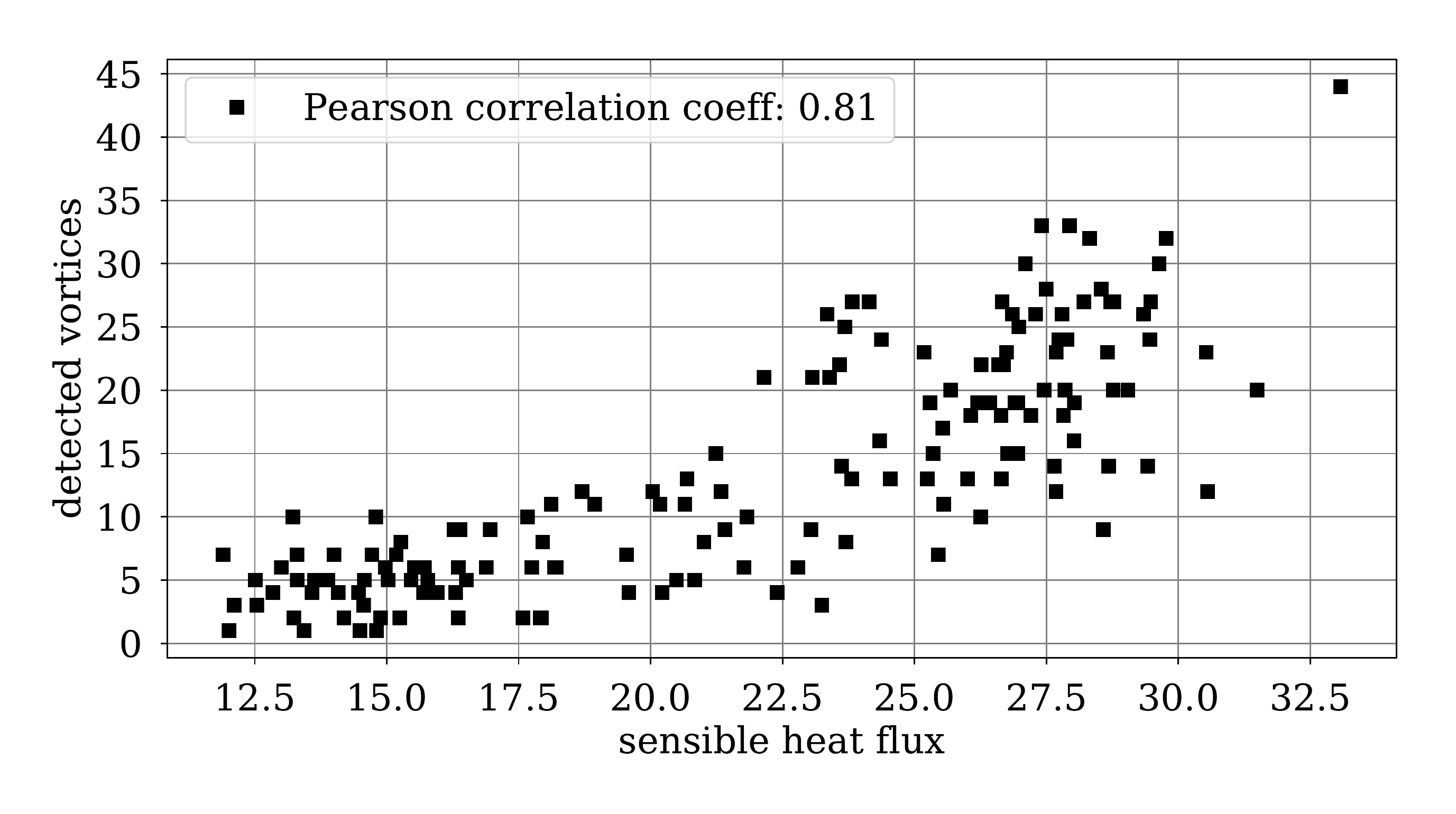}
\caption{ \textit{Sensible heat flux
computed from InSight observations
with a bulk formulation (see text)
is shown in the top panel in a similar
setting as Figure~\ref{fig:season}.
Its correlation to the number
of detected vortices is shown
in the bottom panel
in a similar setting as
Figure~\ref{fig:correlation}.} \label{fig:seasonSHF}}
\end{center}
\end{figure}
The seasonal variability
of sensible heat flux~$H_s$
deserves further comments;
this quantity can be calculated
from InSight observations
(Figure~\ref{fig:seasonSHF})
using the bulk formulation
$ H_s = \rho \, c_p \, u_* \, \theta_* $
where
friction velocity~$u_*$ 
and temperature scale~$\theta_*$
are computed
from equations 3 and 4 
in \citeA{Spig:18insight}
using the InSight observations
of respectively
ambient wind speed~$V$
and surface-atmosphere 
gradient~$T_s - T_a$,
while
atmospheric density~$\rho$
is computed from
pressure~$P$ and air temperature~$T_a$
observed by InSight,
and $c_p$ is specific
heat capacity on Mars.
Interestingly, 
as is shown
in Figure~\ref{fig:seasonSHF},
the seasonal
variability of the number
of vortex encounters
at the InSight landing site
is well correlated to
the observed sensible heat flux.
Drawing conclusions
on the driving mechanisms
for vortices
from this correlation
is tantalizing;
we consider, however, that
sensible heat flux is an ambiguous
diagnostic for two reasons.} 
\newcommand{\venus}{Computing sensible
heat flux mixes environmental
variables controlling
both the formation rate of vortices
(e.g., near-surface instability)
and the advection of vortices
(e.g., ambient wind speed).
Consequently, a valuable physical
interpretation of the correlation
between vortices and sensible
heat flux is difficult.} 
\new{\mars}
\begin{enumerate}
\item Contrary to Earth, 
in the low-density
martian atmosphere, 
the near-surface
instability that drives
the daytime turbulence
is mostly a result of 
radiative warming
through 
CO$_2$ 
(and, to lesser extent,
H$_2$O and dust)
absorption
of incoming surface 
infrared flux
(see \citeA{Habe:93pbl},
\citeA{Savi:99},
and \citeA{Spig:10bl} section 4).
Sensible heat flux still plays
a role in driving daytime turbulence
on Mars, but less so 
than radiative contributions (see section~\ref{sec:gustiseason}), 
in contrast to the terrestrial case.
The contribution of 
sensible heat flux
on the Martian PBL 
only becomes dominant
in extreme regional wind conditions encountered
over steep slopes \cite{Spig:11ti}.
\item \new{\venus}
\end{enumerate}

\subsubsection{Wind gustiness \label{sec:gustiseason}}

\begin{figure}[h!]
\begin{center}
\includegraphics[width=0.99\textwidth]{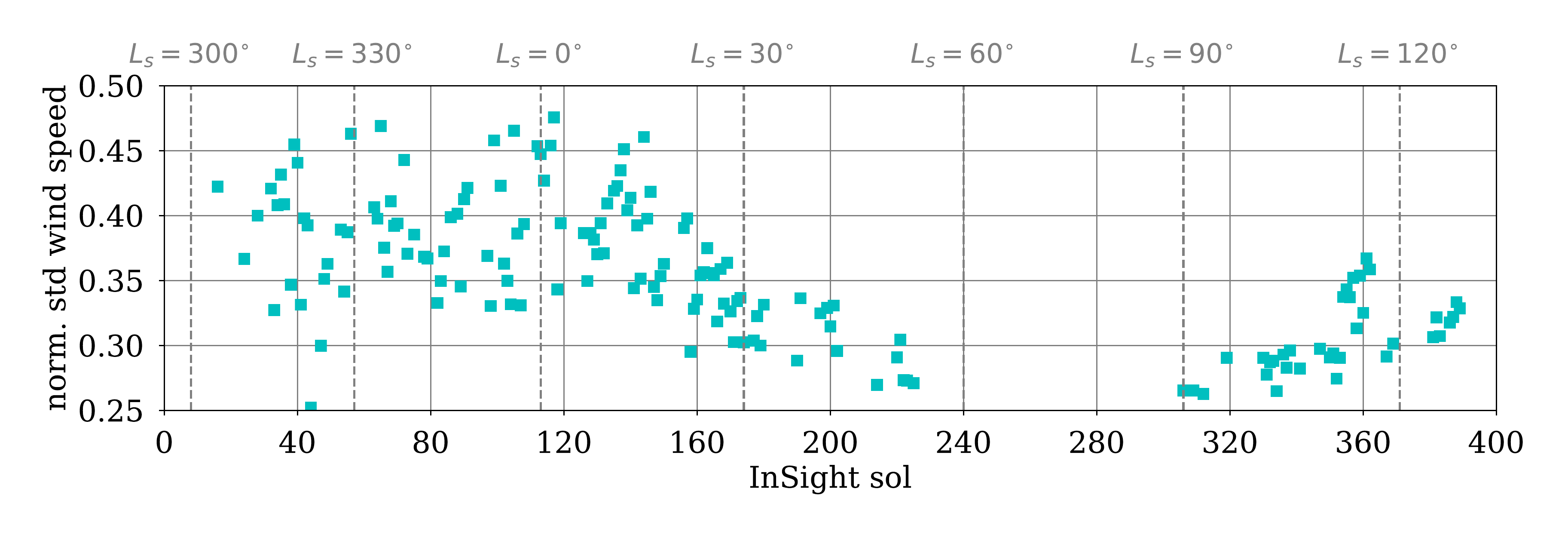}
\includegraphics[width=0.48\textwidth]{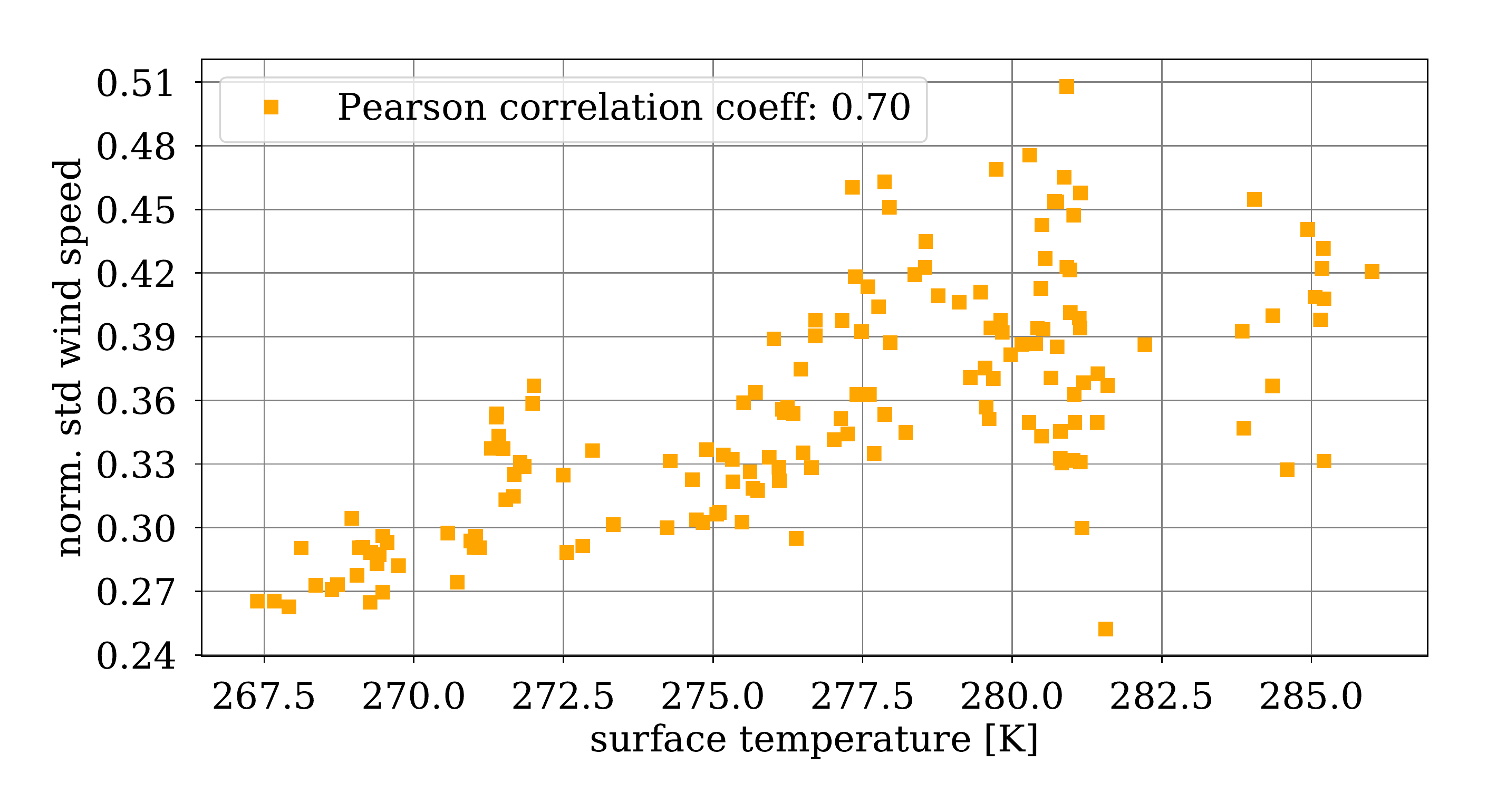}
\includegraphics[width=0.48\textwidth]{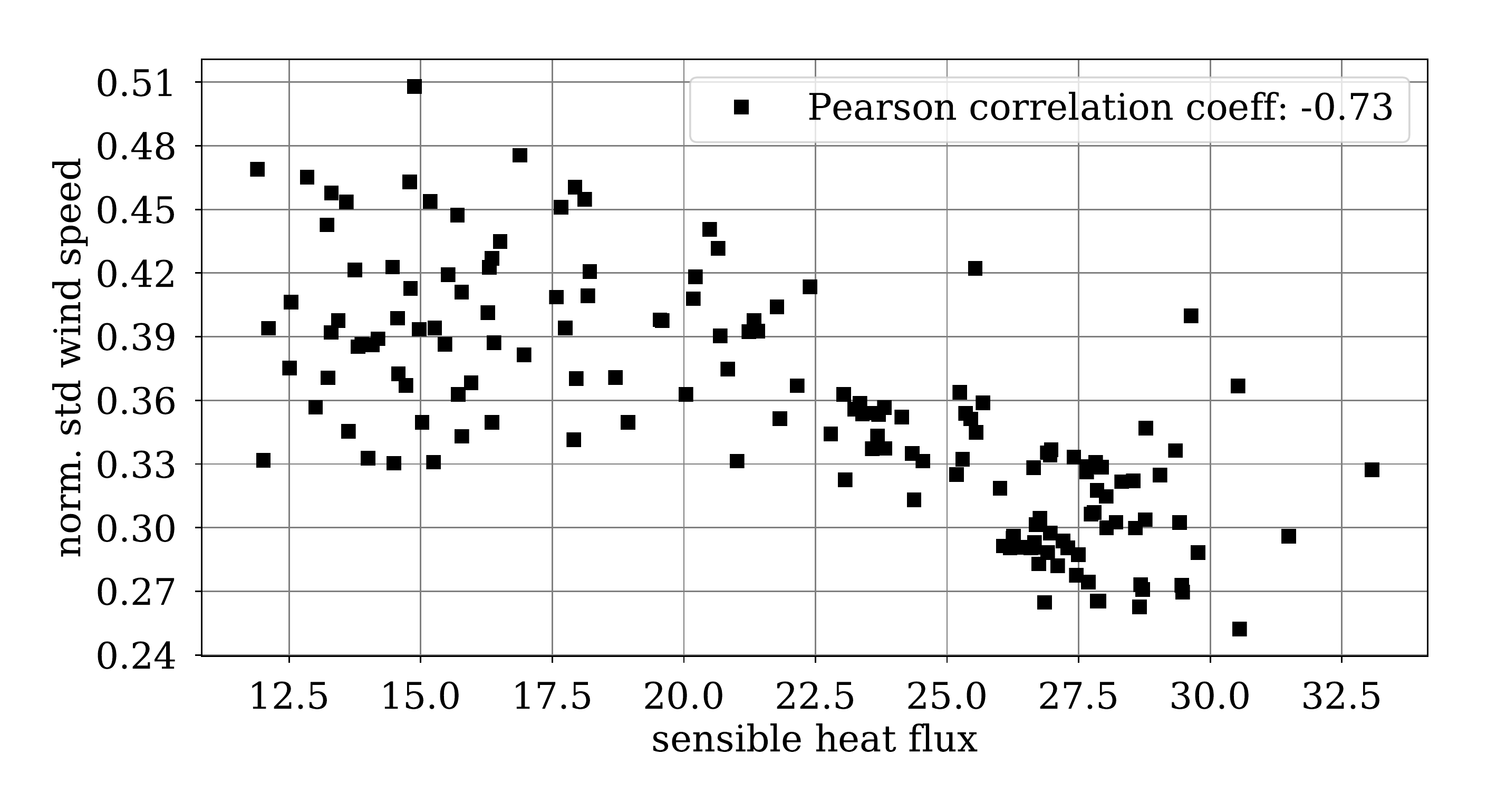} 
\caption{ \textit{The seasonal evolution
of normalized gustiness,
defined as the standard deviation
of wind speed (representing turbulence)
normalized with the mean wind speed
(i.e. the ambient wind speed)
is shown here in the top plot in the same fashion
as is done in Figure~\ref{fig:season}).
The same local-time interval
of 11:00-14:00 as Figure~\ref{fig:season}
is used to compute the
mean and standard deviation
of wind speed.
The correlation of normalized gustiness 
to surface temperature (bottom left panel)
and sensible heat flux (bottom right panel)
is shown in a similar setting as
Figure~\ref{fig:correlation}.
The correlation plot between normalized gustiness
and wind speed is very similar to the bottom right panel
(with a slightly-different Pearson coefficient of -0.78).
}
\label{fig:season_STW}
}
\end{center}
\end{figure}

Figure~\ref{fig:season_STW} (top) shows 
the normalized daytime gustiness, obtained
from the standard deviation of wind speed
divided by the mean wind speed
(computed over
intervals of
local times
11:00-14:00
LTST).
Interestingly,
in both the 
\textit{Early mission}
and the
\textit{Dust storm and spring}
sequences,
the gustiness remains roughly
constant at values
35-45\%.
There is no apparent influence
of the local dust storm
on this normalized gustiness:
the strong decrease in vortex encounters
noticed at sol 50 
in Figure~\ref{fig:vortices}
is not observed in the normalized
gustiness.
Gustiness and vortices are two
integral parts of daytime
convective turbulence in the PBL;
however, contrary to the vortex count,
the local normalized gustiness
is supposedly 
corrected of the
effect of advection
by the wind normalization
(a vortex count normalized by ambient wind
speed is sometimes used also,
see \citeA{Elle:10} and \citeA{Hols:10}).
The fact that, 
in the \textit{Dust storm and spring} sequence,
vortex count decreases,
while 
normalized gustiness does not,
suggests that the advection effect
is the dominant explanation
for the seasonal correlation
between vortex encounters
and ambient wind speed
at the InSight landing site.

What Figure~\ref{fig:season_STW} (top)
also indicates is a decrease
of gustiness in the 
\textit{Aphelion season} sequence,
from a value of 40\% to 25\%.
This decrease of gustiness
appears to
be associated with
the decrease of both the
surface-to-atmosphere gradient
and the surface temperature
shown in Figure~\ref{fig:season},
while the ambient wind speed
increases.
However, the seasonal evolution 
puts the
surface-to-atmosphere gradient
at the same
values at $L_s = 60^{\circ}$
as during the local dust storm
at $L_s = 330^{\circ}$;
this is not the case
for surface temperature
which reaches much lower
values at $L_s = 60^{\circ}$.
This, and the fact that
normalized gustiness has
not decreased during
the local dust storm
while it has decreased
significantly 
at $L_s = 60^{\circ}$,
indicates
that normalized
gustiness is primarily
sensitive to surface temperature.
\newcommand{\oslo}{Furthermore, 
Figure~\ref{fig:season_STW} 
(bottom) indicates 
a positive correlation 
(Pearson coefficient 0.7)
between daytime gustiness 
and surface temperature,
while a negative correlation
is observed between
daytime gustiness 
and sensible heat flux
(or wind speed).
This is in agreement 
with the martian daytime PBL turbulence
being mainly
driven by radiative contributions
rather than sensible heat flux, contrary to Earth
-- and further reinforces
our claim in section~\ref{sec:season_vortex_spec}
that sensible heat flux
could be well correlated
to vortex encounters,
without it being
the actual driver of
daytime turbulence on Mars.}\new{\oslo}
Lower daytime surface temperature
in the \textit{Aphelion season}
sequence implies lower infrared flux
from the surface
to the atmosphere,
hence lower radiative flux
absorbed by the CO$_2$
atmosphere overlying 
the Martian surface,
and as a result
less intense
convective turbulence
\cite{Savi:99,Spig:10bl}.
The
\textit{Aphelion season} sequence,
during which 
gustiness is lower than 
in the \textit{Early mission} sequence
while vortex encounters
are as numerous,
strongly suggests
that advection 
by ambient wind speed
is a key element
for explaining
sequences of
sustained
vortex encounters
at the InSight 
landing site.

The slight increase 
of normalized gustiness
from 25\% to 30\%
close to
$L_s = 120^{\circ}$
is also correlated
with the slow seasonal
increase of surface
temperature at
the end of norther summer.
What remains to be explained is why
the drop in daytime surface temperature
during the regional dust storm
is not associated with a drop
of normalized gustiness.
A possibility is that the
dust particles 
injected by the distant regional
dust storm and 
present in the PBL
at the InSight landing
site cause
an increase of infrared
absorption in the PBL
that would add up to
the CO$_2$ absorption
and compensate (approximately)
the decrease in energy input
coming from the surface
that received less sunlight
because of dust absorption
and scattering.

\section{Comparison with Large-Eddy Simulations (LES) \label{sec:les}}

\begin{table}[]
\centering
\begin{tabular}{c|cccccccc}
\hline\hline
season $L_s$ ($^{\circ}$) & 300 & 300 & 0 & 0 & 30 & 30 & 120 & 120 \\
\hline
ambient wind $V$ (m/s) & 10 & 20 & 10 & 20 & 10 & 20 & 10 & 20 \\
\hline\hline
\end{tabular}
\caption{The parameters explored by the eight
Large-Eddy Simulations carried out for this study
are provided in this table. Further details on
the other (common) modeling settings are provided
in section~\ref{sec:methodo_les}.
The ambient wind corresponds to conditions
in the free atmosphere not influenced
by friction and turbulence close
to the surface; at the height
of InSight measurements, the equivalent
ambient wind is about~$V/2$.}
\label{tab:les}
\end{table}

\begin{figure}[h!]
\begin{center}
\includegraphics[width=0.95\textwidth]{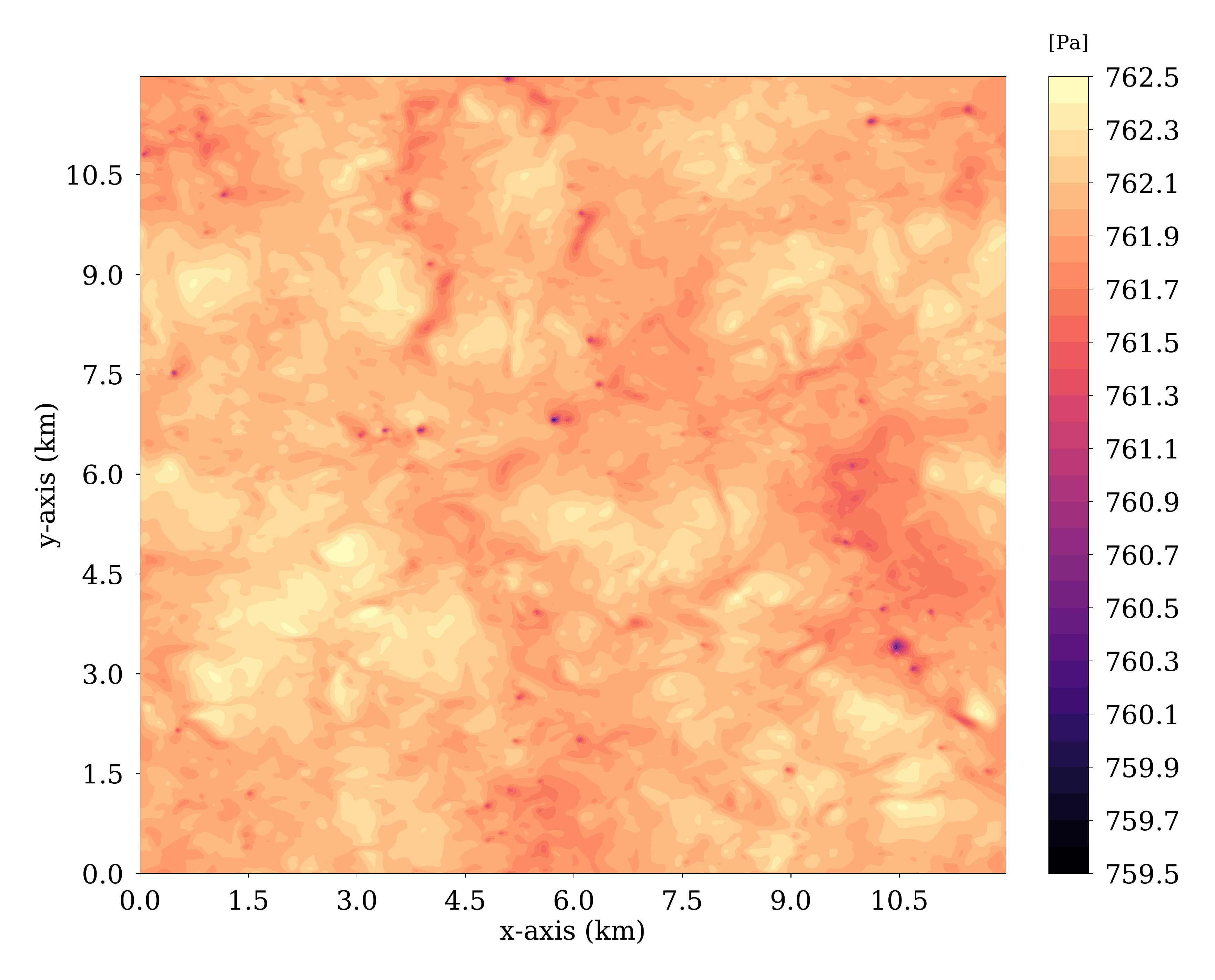}
\caption{ \textit{The typical surface pressure
field obtained in our 25-m-resolution 
Large-Eddy Simulations is displayed here 
on the whole 144 km$^2$ domain,
for the case~$L_s$=300$^{\circ}$ 
with ambient wind~$V$ = 10 m/s.
The center of convective cells 
can be seen as large areas of
``burgeoning'' maxima 
of surface pressure in yellow colors.
The convective vortices can be
seen as localized round-shaped
areas of pressure minima in violet colors.}  
\label{fig:les}}
\end{center}
\end{figure}

We performed eight LES runs
which all share the 
same simulation configuration 
described in 
section~\ref{sec:methodo_les}.
Table~\ref{tab:les}
summarizes the parameters chosen
for the exploration of
environmental conditions
encountered at the InSight
landing site during the
first 400 sols of operation.
What differs from one
simulation to the other
is the ambient wind speed,
and the season considered
for the LES radiative transfer
computations (and, accordingly, 
the initial
temperature profile).
This is designed
to explore the impact
of the seasonal variations
of environmental conditions
described in section~\ref{sec:envseason}.
A typical pressure field
predicted by LES
is shown in Figure~\ref{fig:les},
with localized
vortex-induced
pressure drops
forming at the 
intersection of
larger-scale
convective
cells
\cite{Kana:06,Toig:03,Mich:04,Spig:16ssr}.
It should be emphasized here
that the results we discuss
in this paper
with our 25-m LES 
will be in need to be confirmed
by future work
using higher-resolution LES
(typically 5 m, 
a factor of five better)
to better resolve the
population of 
small-radius vortices
\cite{Nish:16,Gier:19}.

\subsection{Vortex statistics \label{sec:lesvortexstat}}

To compare the vortex statistics
predicted by our LES runs with 
those obtained from InSight observations 
(see section~\ref{sec:statvortex}),
LES time series 
of pressure ``measurements''
\new{emulating} InSight's
are generated by
randomly picking up 
a given grid point 
in the LES domain
for each different sol
(this is a practical application
of the ergodic principle).
We generate by this procedure
576 different ``sols''
for each LES listed 
in Table~\ref{tab:les}.
Then, the exact same pressure-drop
detection algorithm as is used for the InSight
data, described in section~\ref{sec:methodovortex}, 
is applied to the LES time series
for each generated ``sol''.

\begin{figure}[h!]
\begin{center}
\includegraphics[width=0.49\textwidth]{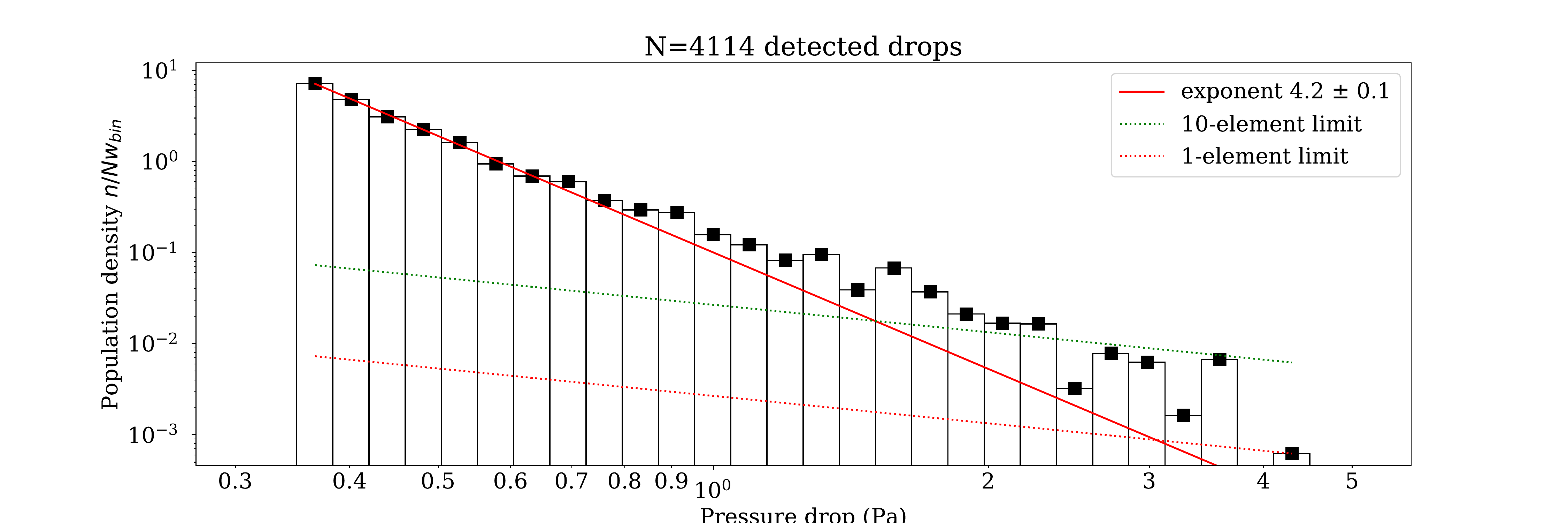}
\includegraphics[width=0.49\textwidth]{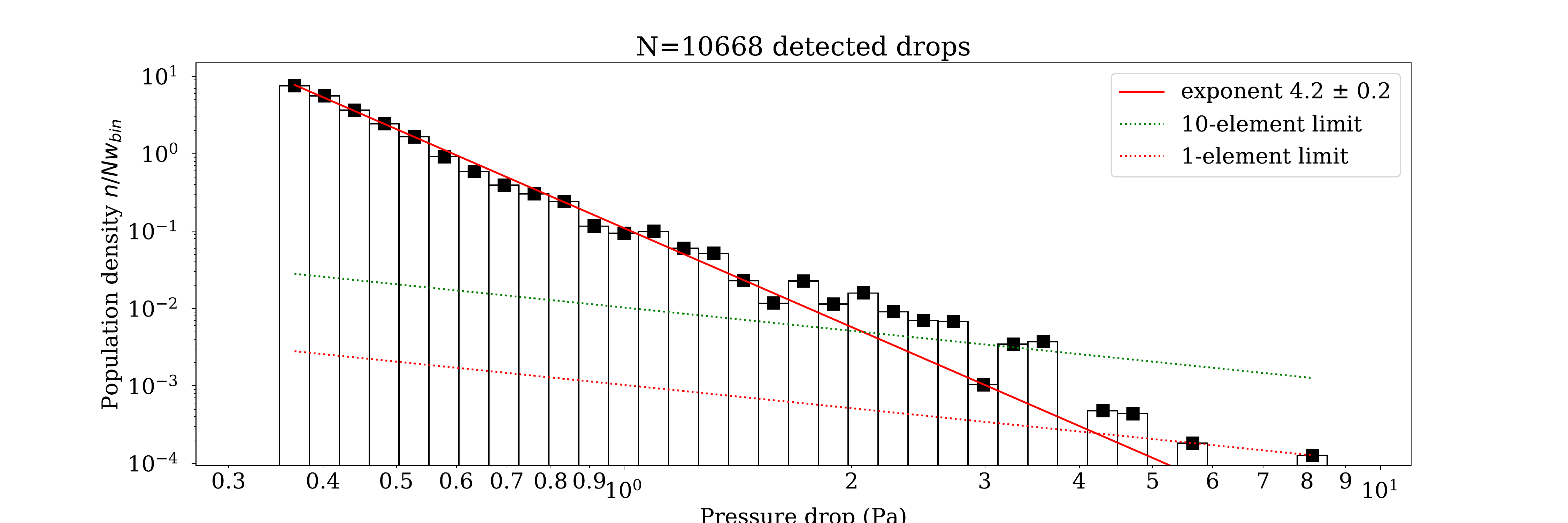}
\includegraphics[width=0.49\textwidth]{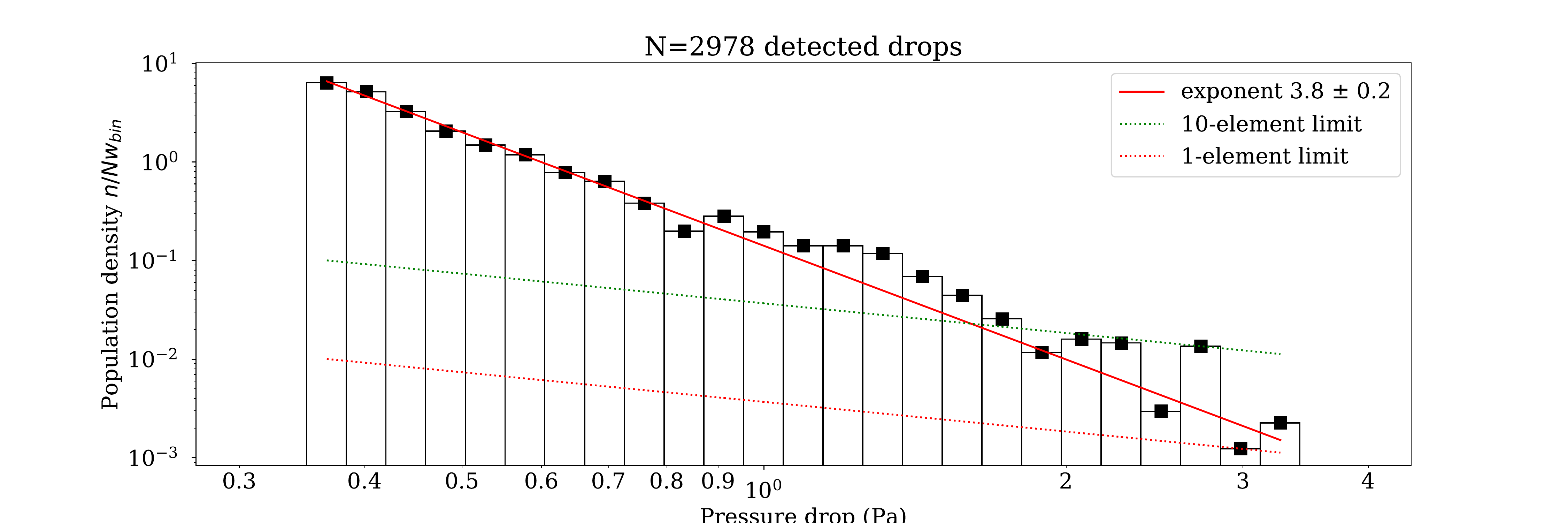}
\includegraphics[width=0.49\textwidth]{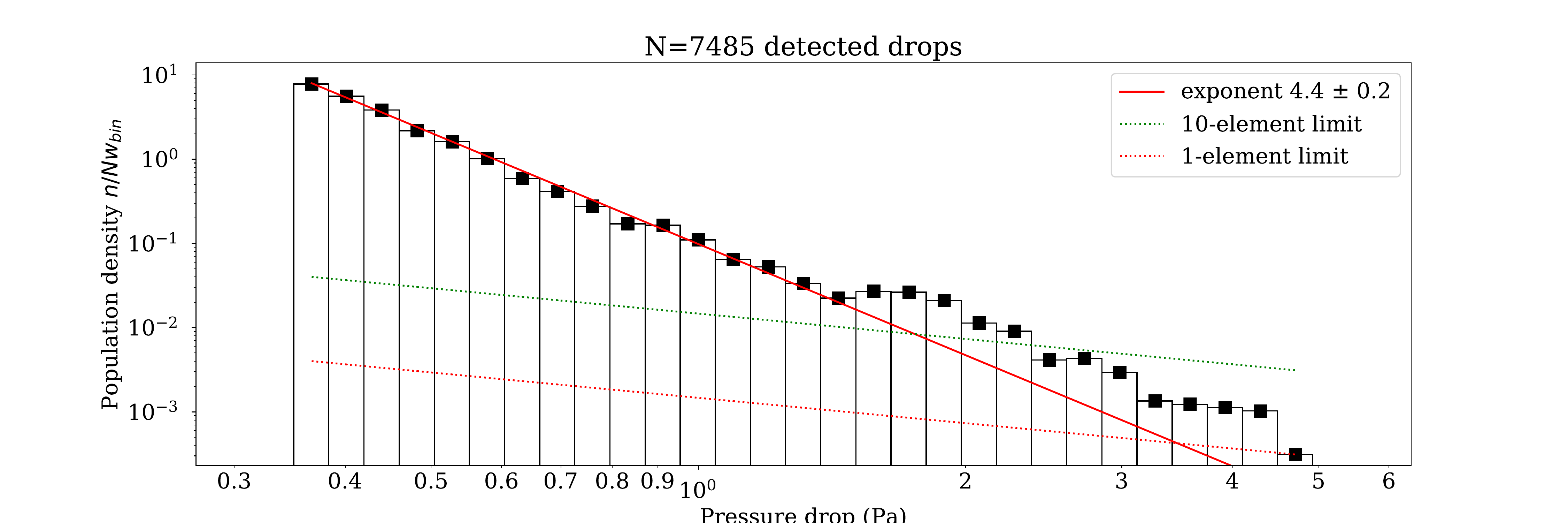}
\includegraphics[width=0.49\textwidth]{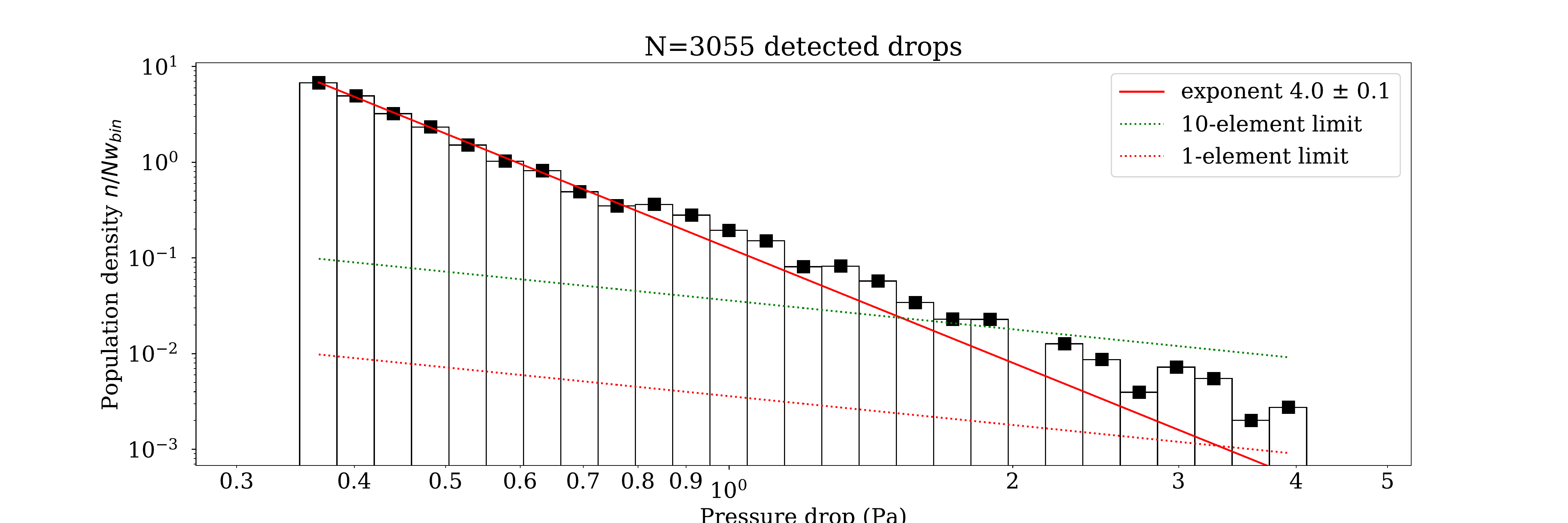}
\includegraphics[width=0.49\textwidth]{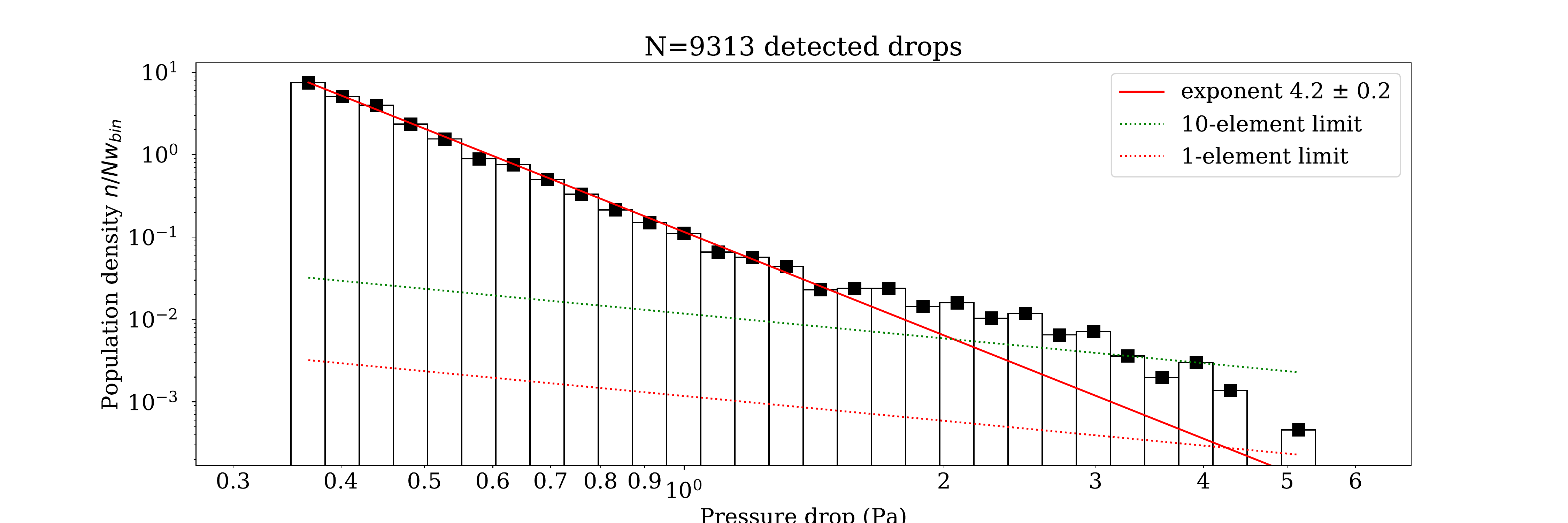}
\includegraphics[width=0.49\textwidth]{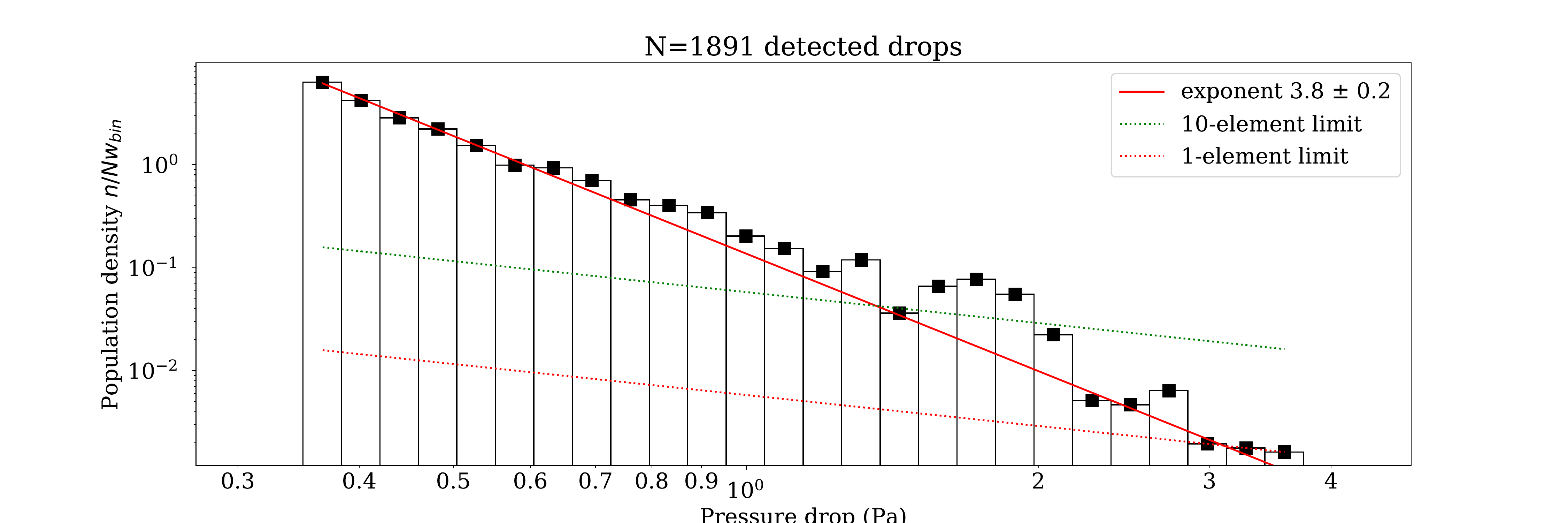}
\includegraphics[width=0.49\textwidth]{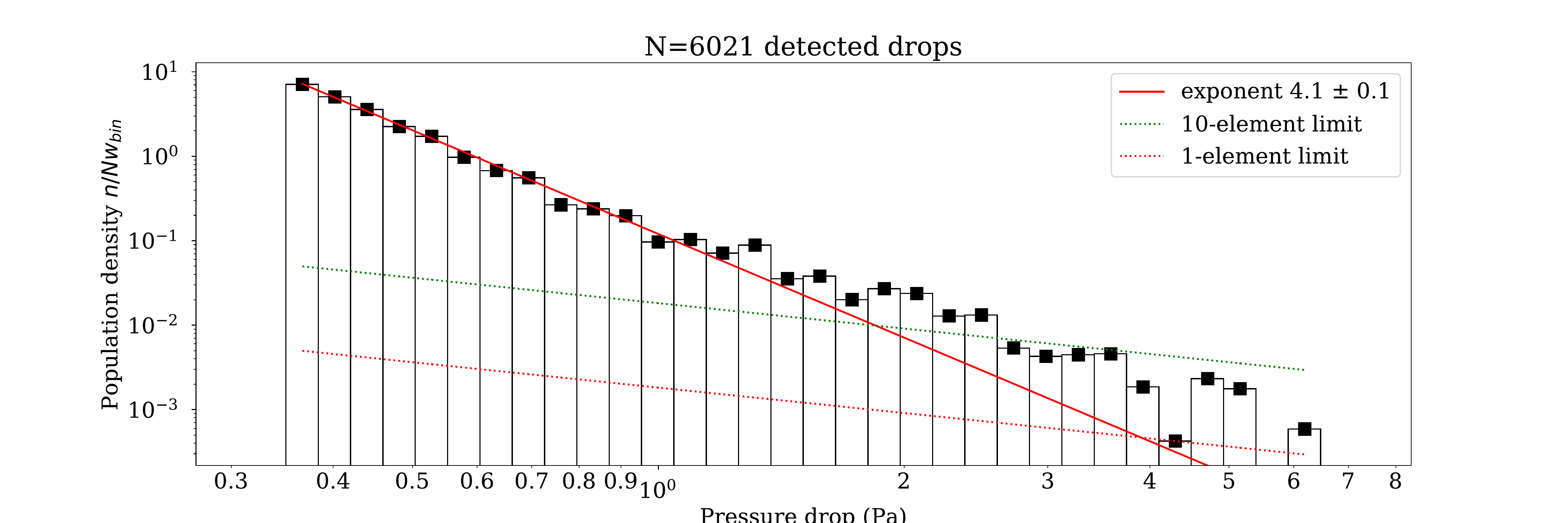}
\caption{ \textit{The normalized (differential)
distributions of pressure drops 
deeper than 0.35~Pa
detected in the
Large-Eddy Simulations are shown in similar
diagrams to those shown in
Figure~\ref{fig:vortices_distrib_fit}.
This is obtained by the same
detection/histogram method
as the one used for InSight observations.
A total of 576 time series
of pressure emulating different
``sols'' are included for each LES case.
The left and right plots are obtained for 
respectively LES with ambient wind
of 10 m/s and 20 m/s
and rows from top to bottom
corresponds to simulations
for 
$L_s = 300^{\circ}$,
$L_s = 0^{\circ}$,
$L_s = 30^{\circ}$,
$L_s = 120^{\circ}$.
The number of vortex encounters
detected in each case
is indicated in the title
of each plot.} \label{fig:histles}}
\end{center}
\end{figure}

A first result that can be discussed
is the distribution of pressure
drops obtained in the LES.
As is shown in Figure~\ref{fig:histles},
the differential distribution
of pressure drops caused by
convective vortices
are suitably represented
by power laws 
with exponents
\new{close to 4},
which is in agreement
with the results obtained
with Insight
(3.4 exponent; see 
Figure~\ref{fig:vortices_distrib_fit}).
We also note that,
at all seasons,
there is systematically
about 2 - 3 times
more vortex encounters
in the case with higher
ambient wind speed.
The exponent of the 
optimum power-law
distribution also
appears to change 
with ambient
wind speed
\new{(the larger the wind speed, 
the larger the exponent --
further work on LES will be
needed to confirm this point).}
How the power-law exponent
changes with season
is much less clear;
no particularly
clear trend 
can be drawn.
We also note that
the tendency 
found in InSight
observations
(section~\ref{sec:statdrop})
of the deepest
pressure drops
departing from a power
law with exponent 3 - 4
\new{appears}
to be reproduced 
by our LES runs.
Future studies 
with a more extended
period of time
covered by InSight
(e.g., two complete 
martian years),
thereby including
more of the deepest vortex
encounters,
will allow
this question to be fully addressed.

\begin{figure}[h!]
\begin{center}
\includegraphics[width=0.99\textwidth]{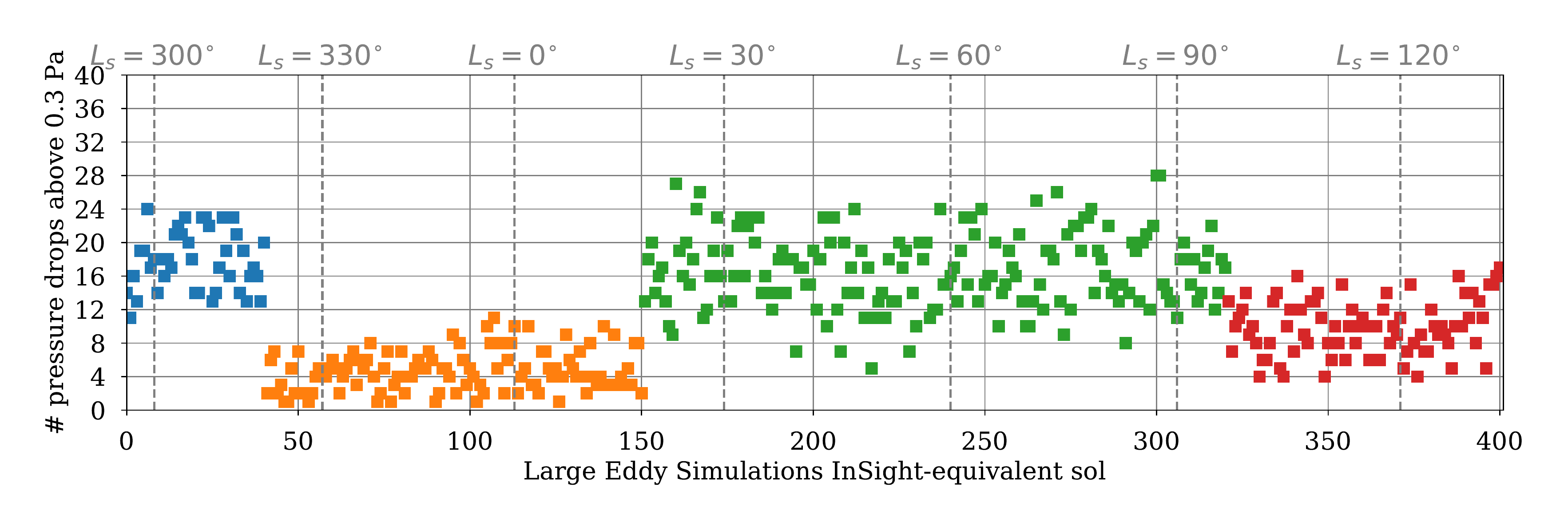}
\caption{ \textit{This seasonal plot emulates
what is shown in 
Figure~\ref{fig:vortices}
about the seasonal evolution
of InSight vortex encounters.
Four LES cases
corresponding
to the seasons
and the wind
conditions
experienced by
InSight are included:
$L_s=300^{\circ}$ and $V=20$~m/s
(blue squares, \textit{Early mission} sequence),
$L_s=0^{\circ}$ and $V=10$~m/s
(orange squares, \textit{Dust storm
and spring} sequence),
$L_s=30^{\circ}$ and $V=20$~m/s
(green squares, early \textit{Aphelion season} sequence),
$L_s=120^{\circ}$ and $V=20$~m/s
(red squares, late \textit{Aphelion season} sequence).
A random selection
amongst the
576 available
``sols'' for each
considered LES case
(i.e. color)
emulates the 
daily variability.}
\label{fig:seasonles}}
\end{center}
\end{figure}

Figure~\ref{fig:seasonles}
summarizes the 
\new{emulated}
``daily''
and seasonal
variability
of convective 
vortices 
obtained in LES.
This figure is
obtained by
assembling
the four LES cases
corresponding
to the seasons
and the wind
conditions
experienced by
InSight so far:
high-wind cases for
$L_s=300^{\circ}$
(\textit{Early mission} sequence),
$L_s=30^{\circ}$
(early \textit{Aphelion season} sequence),
$L_s=120^{\circ}$
(late \textit{Aphelion season} sequence)
and
low-wind case for
$L_s=0^{\circ}$
(\textit{Dust storm
and spring} sequence).
A random selection
amongst the
576 available
``sols'' for each
considered case
emulates the 
daily variability.

\begin{figure}[h!]
\begin{center}
\includegraphics[width=0.99\textwidth]{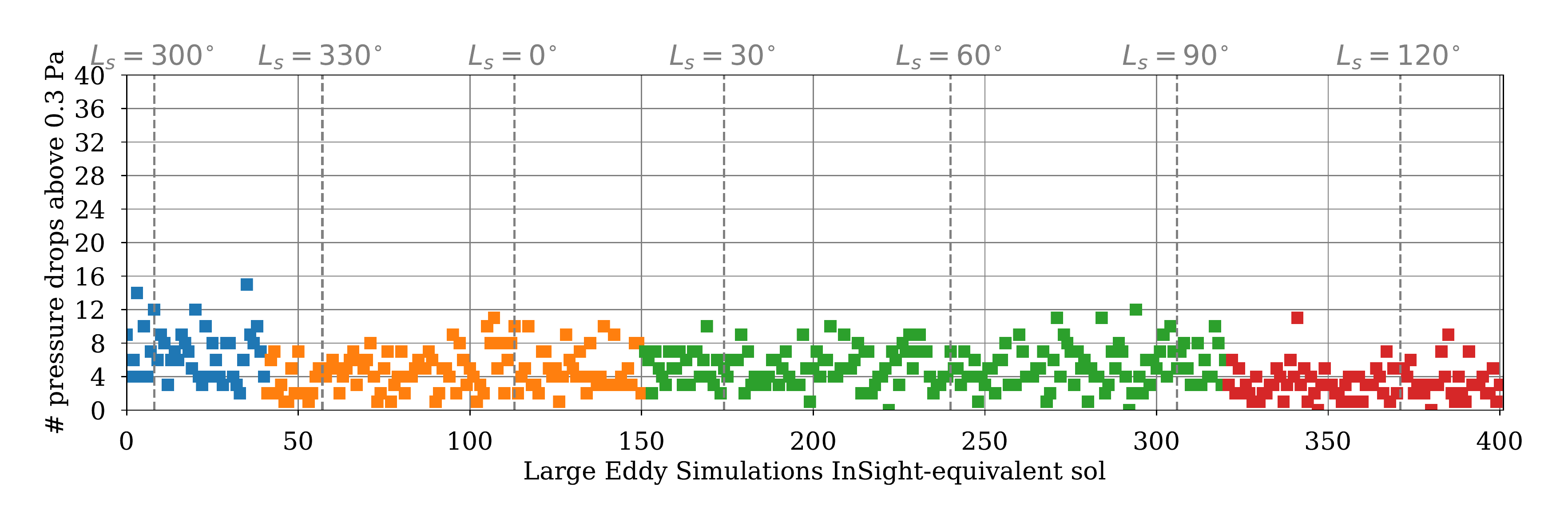}
\caption{ \textit{This Figure is 
constructed similarly to
Figure~\ref{fig:seasonles}, except
that only the LES cases with
an ambient wind of 10 m/s
are included:
$L_s=300^{\circ}$ and $V=10$~m/s
(blue squares, \textit{Early mission} sequence),
$L_s=0^{\circ}$ and $V=10$~m/s
(orange squares, \textit{Dust storm
and spring} sequence),
$L_s=30^{\circ}$ and $V=10$~m/s
(green squares, early \textit{Aphelion season} sequence),
$L_s=120^{\circ}$ and $V=10$~m/s
(red squares, late \textit{Aphelion season} sequence).}
\label{fig:seasonles_windsame}}
\end{center}
\end{figure}

A first remark is that,
for all the cases
displayed in
Figure~\ref{fig:seasonles},
the typical number of
detected vortex encounters
per sol is 
in agreement
between the InSight observations 
and the LES.
Secondly, the 
LES-reconstructed
``daily''
variability 
of vortex encounters 
within a given
sequence
can be quite large,
as is observed by InSight.
\newcommand{\nice}{This shows that the
daily variability
of vortex encounters
observed by InSight
can be described in great part
by the statistical nature of
turbulence.} \new{\nice}
Each sol
of InSight observations
would be an instance 
of InSight
being placed
at a different location
in the horizontal structure
of the PBL daytime turbulence
(exhibited for instance
in Figure~\ref{fig:les}).
Thirdly, the overall
seasonal variability of convective
vortices observed by InSight,
and the three 
above-mentioned sequences,
are well reproduced by
the set of LES runs.
A notable exception is
the decrease of 
vortex encounters
at~$L_s = 120^{\circ}$
onwards, 
that is predicted 
by the LES but not observed
by InSight
(Figure~\ref{fig:vortices}).

\subsection{Advection effects \label{sec:lesadvec}}

The results in section~\ref{sec:lesvortexstat} provide
confidence that LES
are valuable tool to help to
interpret the InSight vortex statistics.
Can we confirm with LES 
the conclusion suggested
in section~\ref{sec:season_vortex} 
that ambient wind speed seems to
be a dominant driver of
the seasonal variability 
of the number of vortex encounters?
Figure~\ref{fig:seasonles_windsame}
shows the LES-generated seasonal
plot of vortex variability
as in Figure~\ref{fig:seasonles}, except that
the LES 
runs are considered 
at the relevant
seasons but
with the choice of
same ambient wind
throughout (10 m/s).
This figure 
demonstrates that, if not
for the seasonal variability
in ambient wind speed,
the vortex encounter at
the InSight landing 
should have decreased
steadily, following 
the tendency
of surface temperature
(see Figure~\ref{fig:season}).
This is a similar
tendency as the one
drawn for normalized
gustiness in section~\ref{sec:gustiseason}:
the surface is 
colder and colder,
meaning the 
surface infrared flux 
is lower and lower,
thereby reducing the
major energy input
in the martian PBL
(i.e. absorption of
surface infrared flux by CO$_2$
particles in the lowest
part of the PBL)
hence the strength
of convective 
turbulent activity.

LES support the ambient wind
speed as the major influence 
on the seasonal
variability of 
vortex encounters
observed at the InSight 
landing site.
Now, 
as is discussed
in section~\ref{sec:season_vortex},
the influence of ambient wind
speed might be twofold.
On the one hand,
ambient wind speed influences
vortex activity
through the advection effect:
stronger ambient wind advects
more vortices to a given point,
making the encounters on 
a given sol more frequent.
On the other hand, 
ambient wind speed influences
vortex activity
through 
the formation rate:
stronger wind on Mars 
could make
the sensible heat flux term
less negligible
compared to the radiative
term on the PBL energy budget, 
hence leading to
stronger turbulence and
more vortices forming
-- provided that enhanced
shearing effects would not 
prevent the formation of vortices.
We can use the LES to distinguish
the two effects in a different
fashion than what is permitted by the InSight time series.
Instead of a vortex 
count performed
along the time dimension
to mimic InSight detections,
we performed vortex counting
on the whole 
horizontal LES domain
of 144 km$^2$,
identifying pressure minima
in the complete pressure field
as in Figure~\ref{fig:les}.
This allows formation-rate
effects to be emphasized, 
rather than advection effects.

\begin{figure}[p!]
\begin{center}
\includegraphics[width=0.49\textwidth]{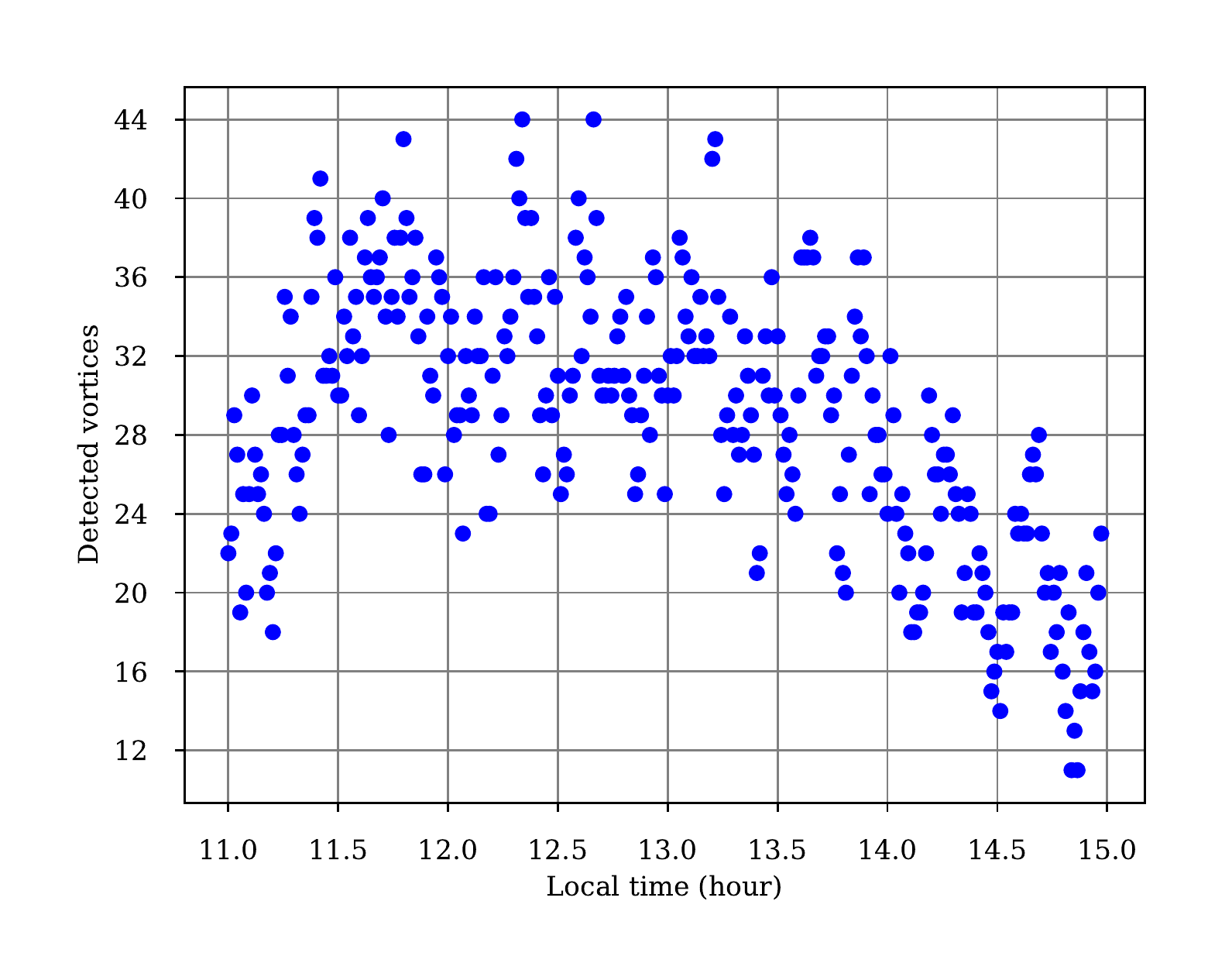}
\includegraphics[width=0.49\textwidth]{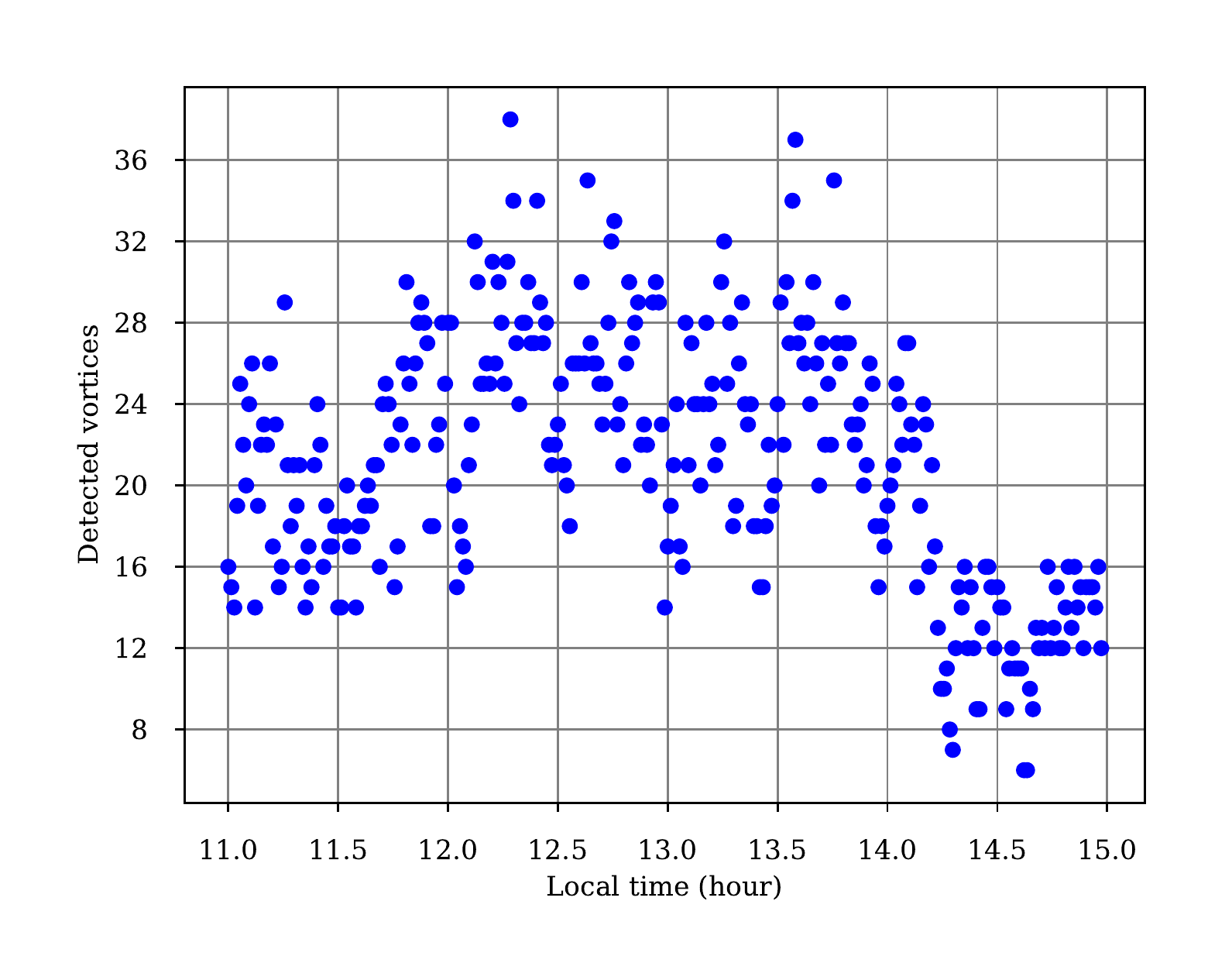}
\includegraphics[width=0.49\textwidth]{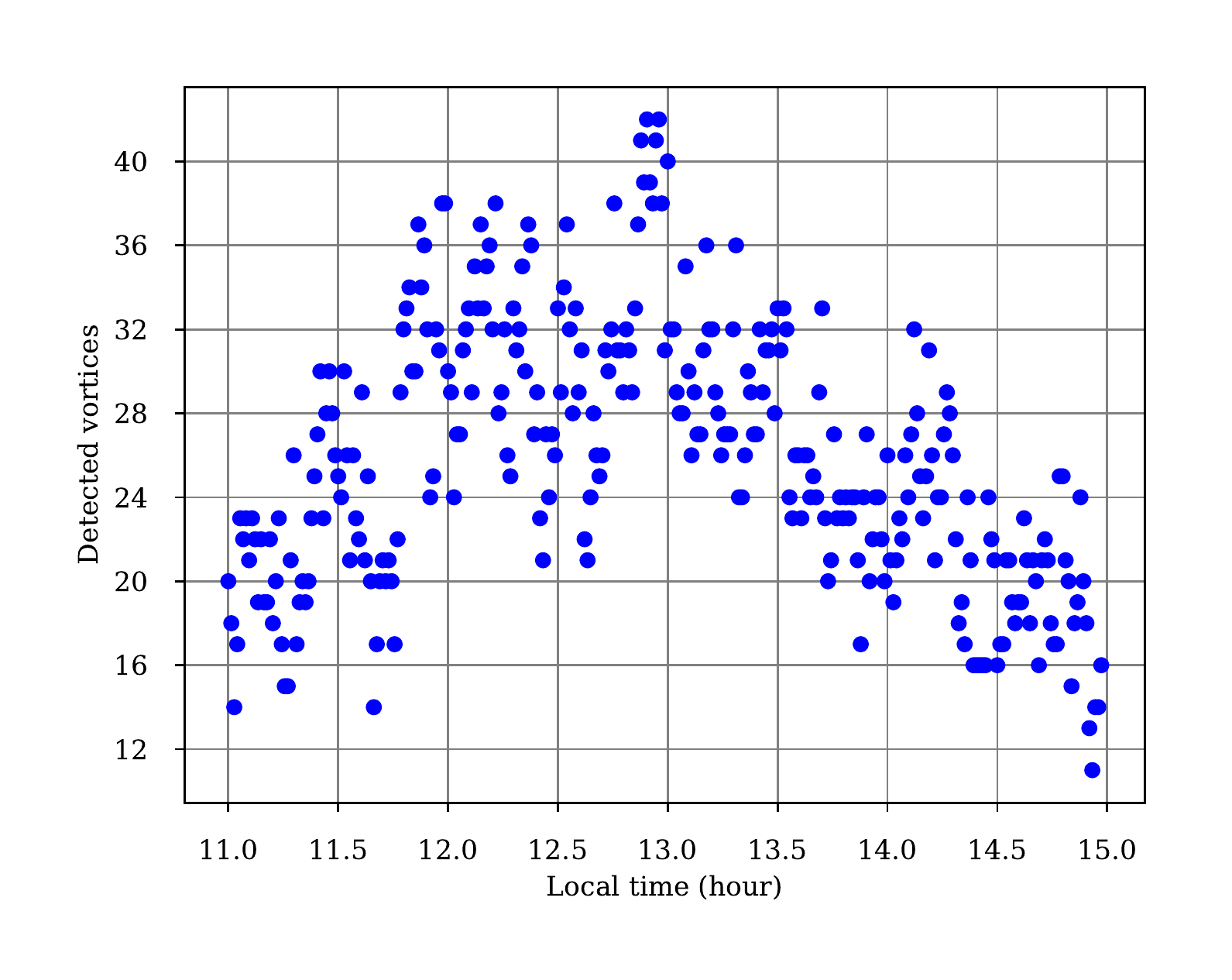}
\includegraphics[width=0.49\textwidth]{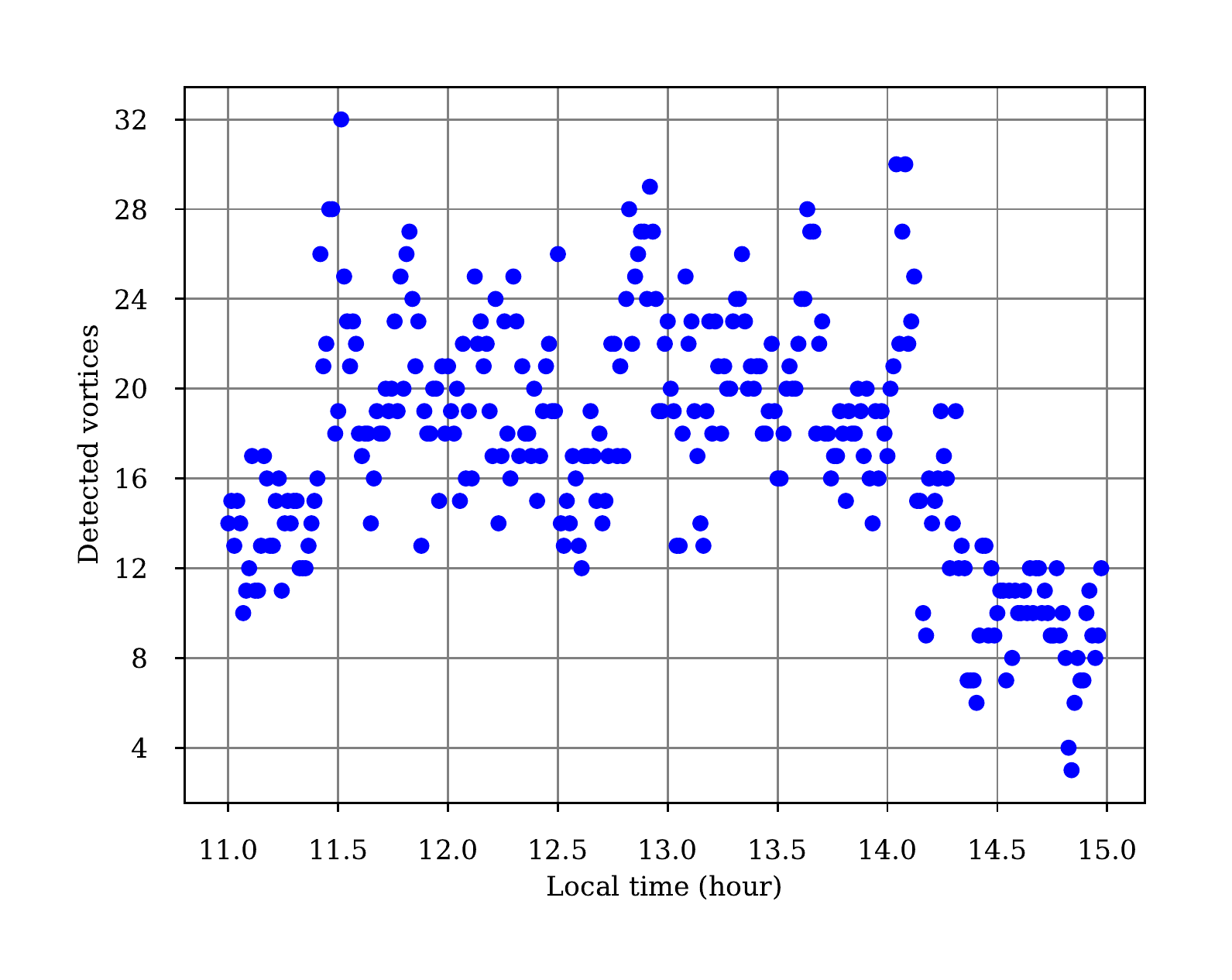}
\includegraphics[width=0.49\textwidth]{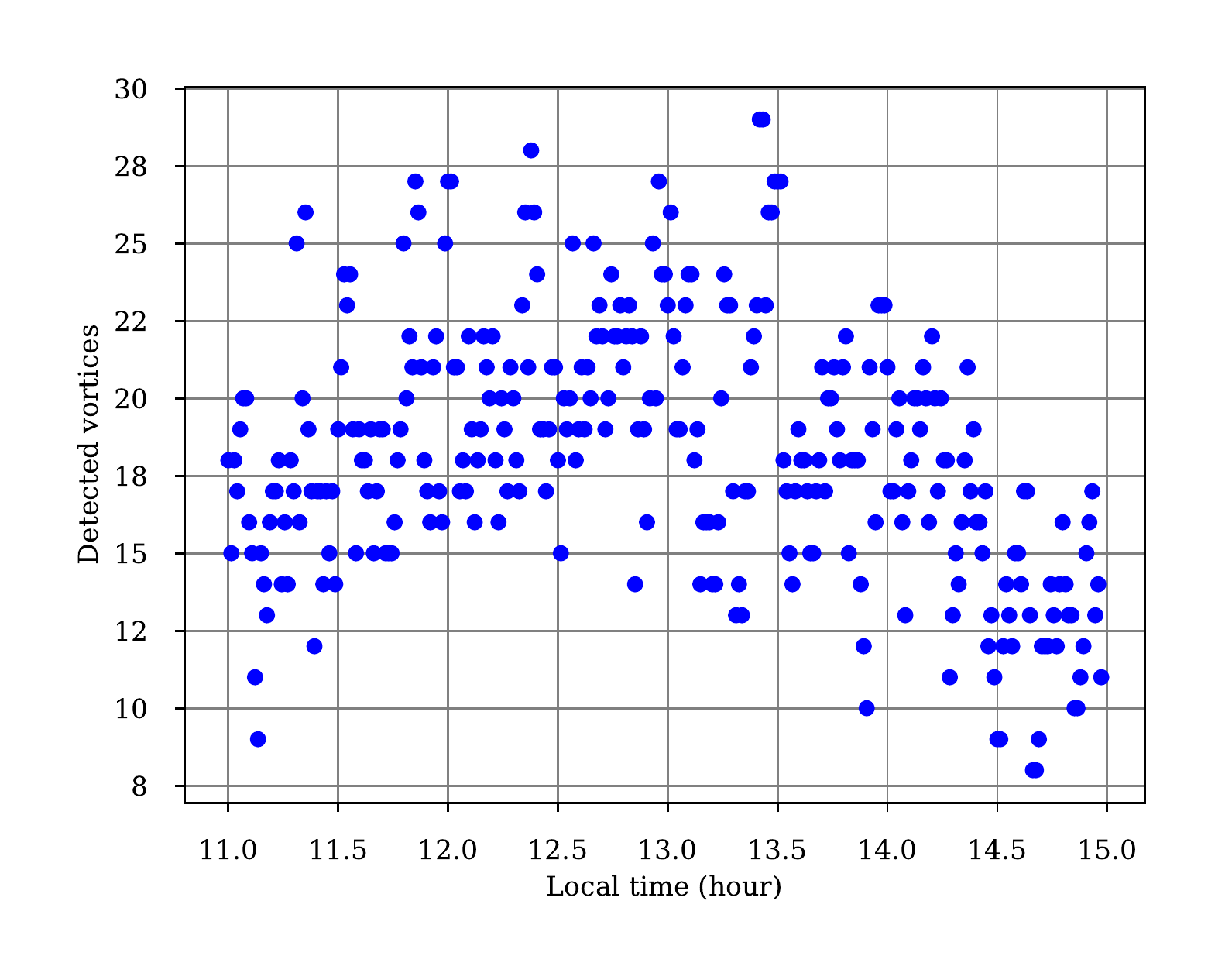}
\includegraphics[width=0.49\textwidth]{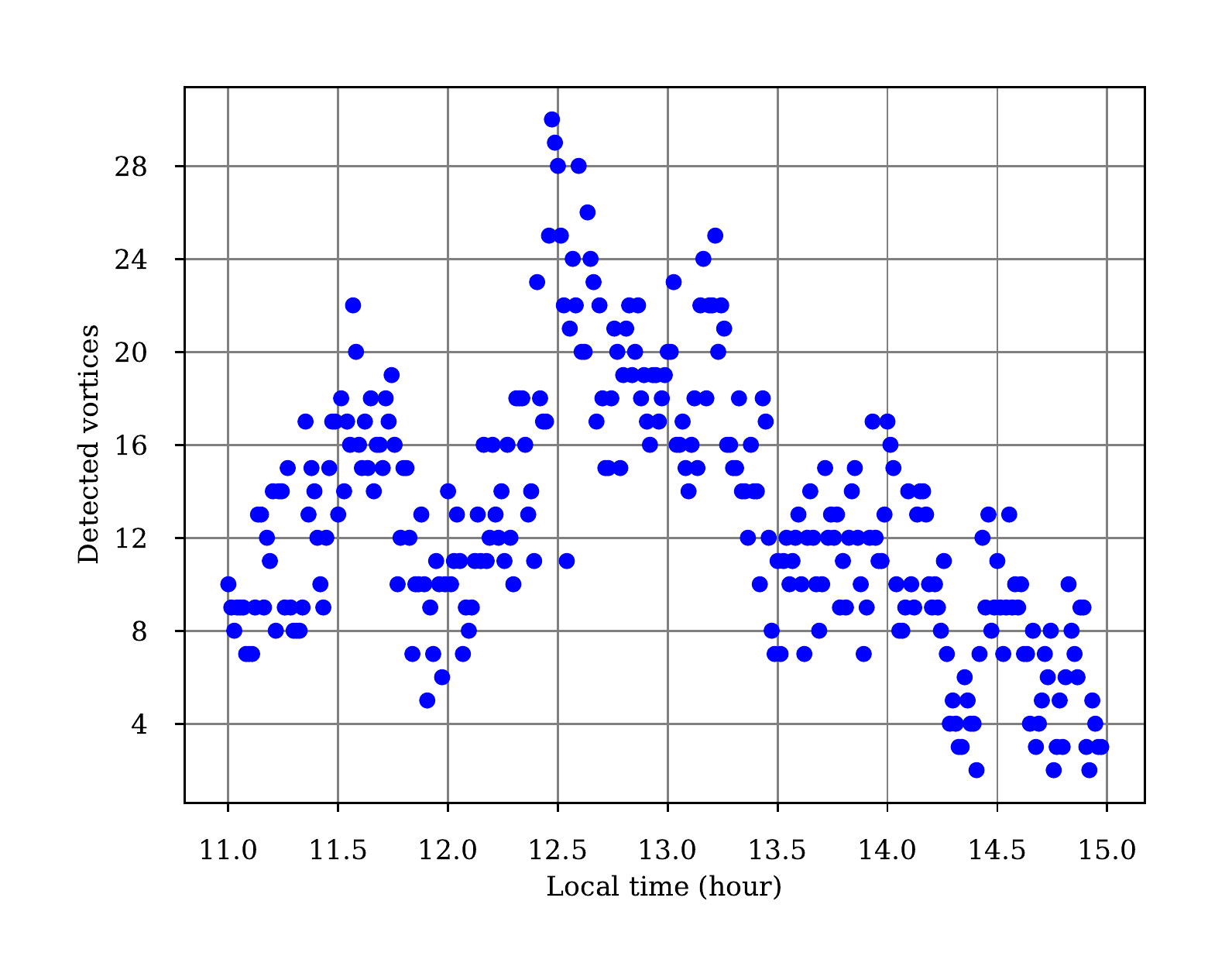}
\caption{ \textit{Pressure drops caused
by convective vortices are detected
here in the horizontal
surface pressure field
(as is shown in Figure~\ref{fig:les})
rather than InSight-equivalent
time series.
The plots show the number
of vortices detected
at different local times
in the most active period
for turbulence convection in LES.
The left and right plots
respectively refer to
LES runs with ambient
wind speed of 10 m/s and 20 m/s.
From top to bottom, the LES
cases for 
$L_s=300^{\circ}$,
$L_s=30^{\circ}$,
$L_s=120^{\circ}$
are considered.} 
\label{fig:rateformation}}
\end{center}
\end{figure}

The results are shown in
Figure~\ref{fig:rateformation}.
We found that in our LES, 
for all the four seasons
considered
and consistently
at all relevant local times,
fewer convective vortices
are forming when the background
wind is doubled from
10 to 20 m/s.
This could be considered 
as a counter-intuitive
result since large
ambient wind speed
enhances horizontal
vorticity known
to be a precursor
of convective vortex formation
\cite{Toig:03,Rafk:16}.
Yet, as is mentioned above
in the text,
this robust conclusion
that less vortices form
in larger-ambient-wind 
LES runs
could be explained
by shearing effects:
strong ambient
wind is deforming the convective
cells and adversely affecting
the formation of vortices;
in terrestrial field studies, 
windy days are well-known to be met
with far fewer, if any,
dust-devil vortex encounters
\cite{Balm:12,Kurg:11,Lore:16}.
Another potential 
line of explanation
is a possible 
lower longevity 
of convective vortices in the
high-wind case --
since vortices would be
more short-lived in
the high-wind case,
fewer of them would
be detected 
in the horizontal LES
pressure field.
This was also found
in the high-resolution 
terrestrial LES
by \citeA{Gier:19}:
low-wind conditions
favor more long-lasting vortices
than high-wind conditions.

The analysis of LES thus
strongly suggests
that the seasonal
variability of vortex encounters
observed at the InSight
landing site is
dominated by the seasonal variability
of wind speed, most
probably through an advection effect.

\subsection{Vortex tracks \label{sec:tracks}}

\newcommand{\delhi}{The upper middle 
and right panels
respectively show a binarized
version of the upper-left image
derived from LES results
and a Radon transform of this
binarized image to detect
linear tracks.
Those 
detection methods
used here 
on LES predictions
are exactly similar
to those developed
for HiRISE images
and detailed
in \citeA{Perr:20}.} 
\begin{figure}[h!]
\begin{center}
\includegraphics[width=0.99\textwidth]{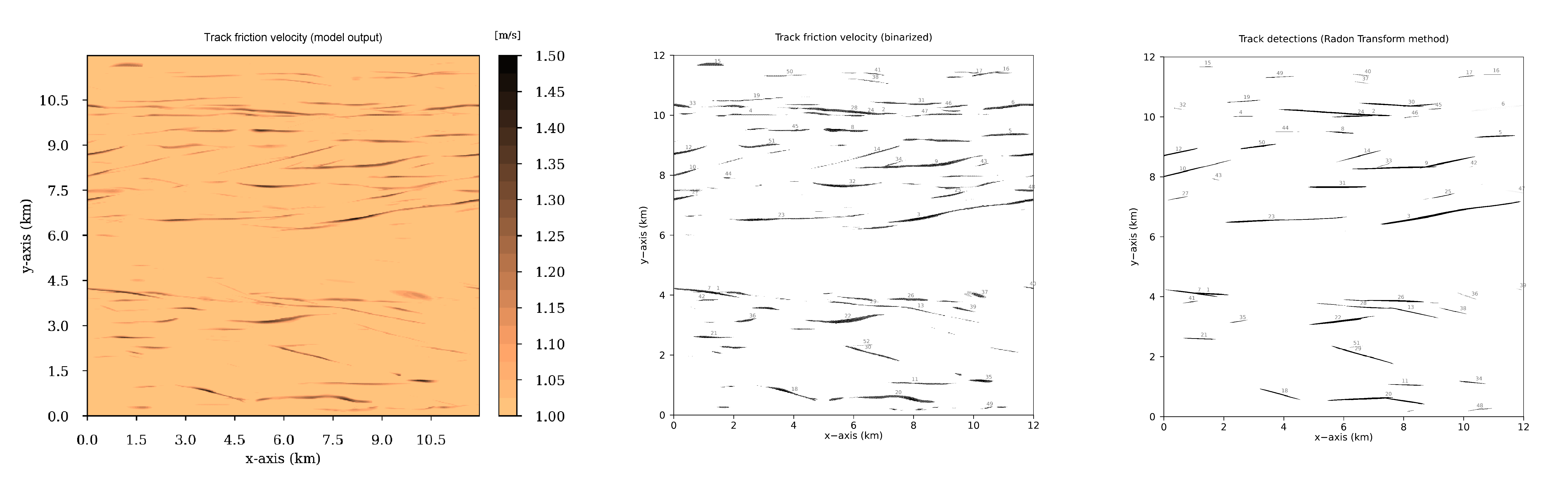}
\includegraphics[width=0.48\textwidth]{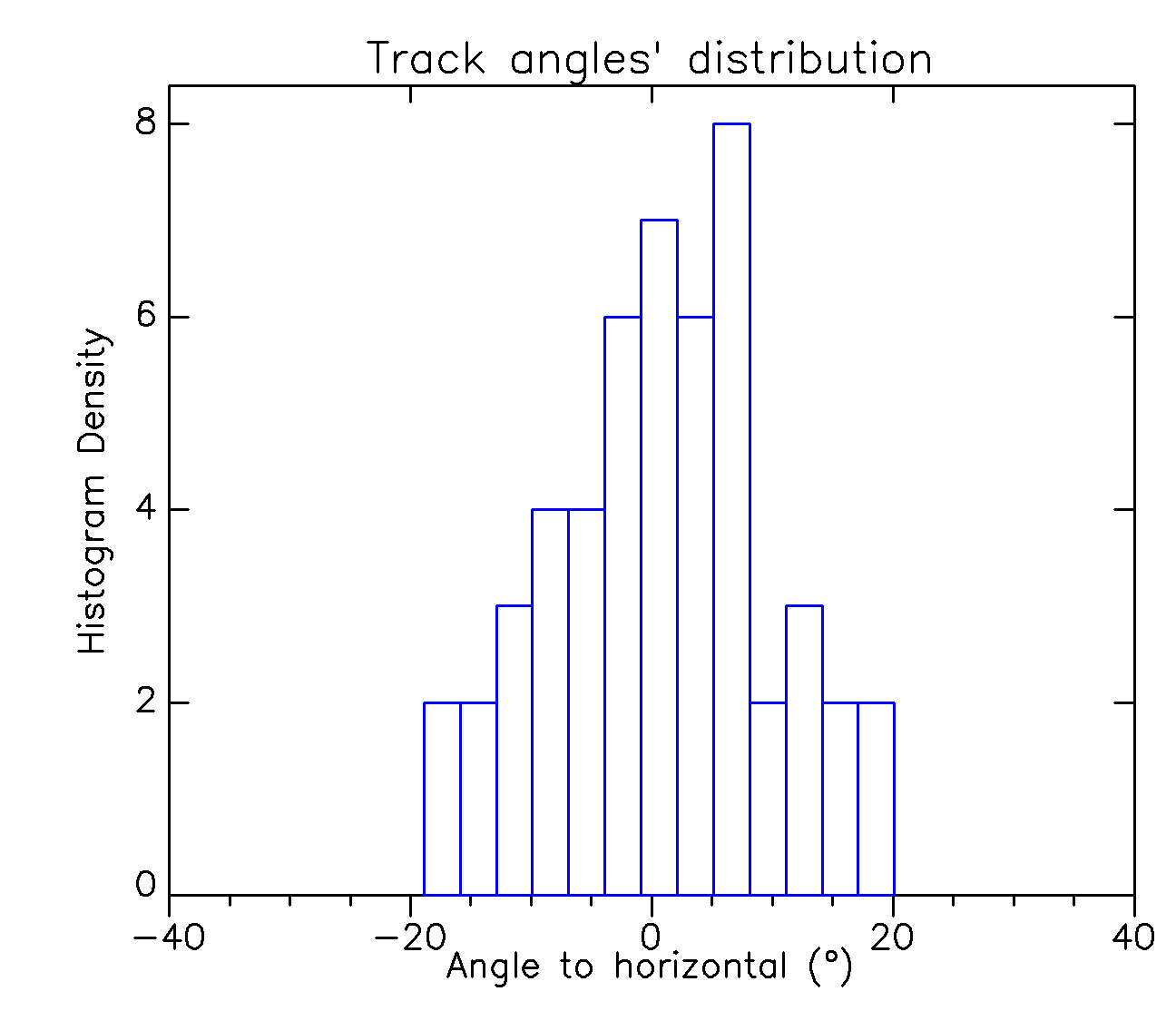}
\includegraphics[width=0.48\textwidth]{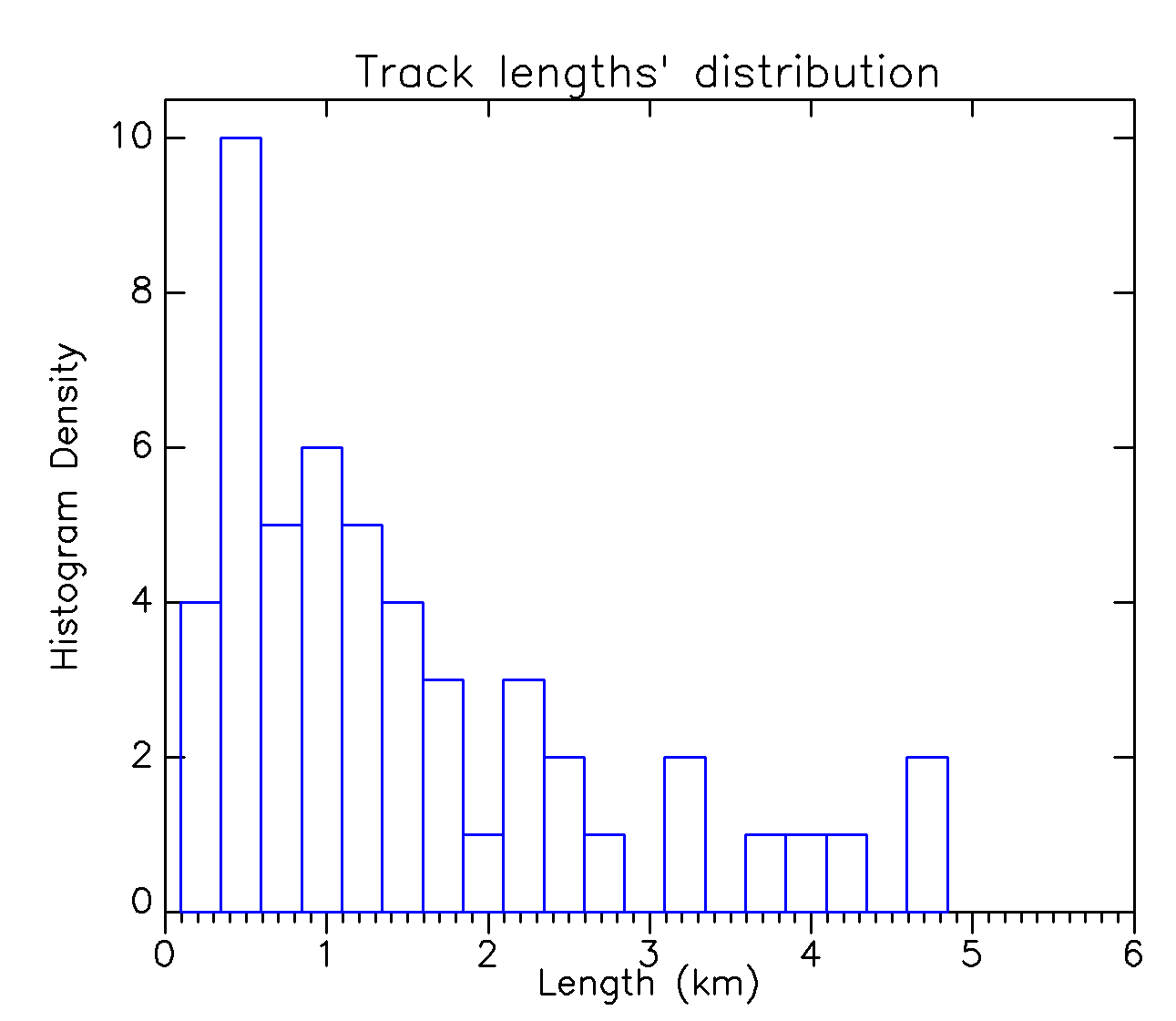}
\caption{ \textit{The upper left 
panel shows
a ``dark track'' spatial map
emulated from LES
by extracting the maximum
friction velocity
at each grid point of the
domain in the local time
interval~[12:00,13:00].
The same orange colour is used
for all values of 
friction velocities
below 1 m/s.
Conditions for the 
\textit{Early mission}
sequence are considered here
(i.e. LES run with $L_s = 300^{\circ}$
and ambient wind speed of~$20$~m/s).
\new{\delhi}
The histograms 
of the distribution
of track angles (left)
and lengths (right)
are displayed
at the bottom of the figure.}
\label{fig:tracks}}
\end{center}
\end{figure}

Although no visible dust devils have been detected by 
the InSight cameras \cite{Banf:20},
numerous fresh tracks
were detected from orbit
in the region of the
InSight landing site
\cite{Perr:20},
sometimes corresponding
to tracks identified
by InSight cameras
(\citeA{Bane:19nat}
and Charalambous et al., 
in revision for this issue).
Those dark tracks are
putatively formed
by convective vortices
able to lift enough
bright dust particles
from the surface
to make the underlying
darker material
apparent -- although 
those vortices
probably do not carry enough
dust particles in their
vortical structures
to be seen as
\emph{dust devils}
by the InSight cameras.

The formation of 
dark tracks seen
from orbit can be
emulated by
LES.
Assuming the above
formation mechanism,
tracks would correspond
to locations where
the wind stress
would
exceed a certain 
lifting/saltation
threshold value
(\citeA{Mich:06dd}; 
see also Baker et al. 
in revision for
this issue).
We show in 
Figure~\ref{fig:tracks}
a possible mapping
of tracks produced
by our LES integrations,
obtained by
calculating 
the maximum of
friction velocity
(see section 6.1 in
\citeA{Spig:18insight})
at each grid
point during 
an active daytime
one-hour
interval
of the LES 
simulation.
The ``contrast''
of the image is set
by defining the same
color for 
all friction
velocities below
1 m/s, 
this color 
acting as a proxy for
undisturbed 
martian soil
devoid of dark tracks.
The ``equivalent
orbital image''
of dark tracks obtained
from LES is then
analyzed with the
exact same semi-automated tracking
procedure explained
in \citeA{Perr:20}.

On Figure~\ref{fig:tracks},
a total of 51 tracks
are detected on
the 144 km$^2$ LES
domain within one
hour.
Considering 6 hours
of daytime vortex activity
(Figure~\ref{fig:vortices_lt})
and assuming for simplicity
a constant formation rate, 
this translates
to a maximum track formation
rate of 2 tracks
per sol per km$^2$.
This is clearly
much larger than
the minimum formation rate of 0.04 - 0.06 tracks
per sol per km$^2$
found by \citeA{Perr:20},
even considering
the very exceptional period 
of intense dust devil activity 
at the beginning of the InSight 
mission (i.e. 0.68 tracks per sol per km$^2$).
Actually, this ``threshold'' 
value of 1 m/s
is chosen to be
permissive to 
detect enough tracks
to form a reasonable
statistics to compare
to images in \citeA{Perr:20}
in the next paragraph.
A normal track formation rate, 
like in \citeA{Perr:20}
and the pre-landing
estimates by \citeA{Reis:16},
yields about 1 track
per hour for a LES
domain size of 144 km$^2$.
This means that only the 
darkest track in 
Figure~\ref{fig:tracks}, 
obtained for friction 
velocities of about 1.4-1.5 m/s
\newcommand{\moscou}{(200 values amongst
the $8{\times}10^{8}$
values output by LES
over the considered hour)}
\new{\moscou},
would correspond to
a realistic case of
the orbital images of
\citeA{Perr:20}, which
illustrates the stringent 
conditions for lifting
dust particles from the
surface in the vicinity of
the InSight landing site.
This is echoed by the scarcity
of surface change events
witnessed by InSight cameras,
which corresponds 
to the strongest
pressure drops 
and associated wind gusts
monitored by InSight APSS
(see Charalambous et al.
and Baker et al., 
in revision for this issue).

The typical 
track length
(ranging from
500~m to about
5 km)
and the
distribution
angle 
(ranging
20$^{\circ}$
apart from the
ambient wind
direction,
with a standard
deviation of 9$^{\circ}$)
are in good 
agreement with
the values
obtained with
HiRISE orbital
images with
the same 
method \cite{Perr:20}.
Furthermore, the low standard
deviation of tracks under fairly 
high ambient wind speed (10 m/s) 
confirm that the linearity of 
tracks increases with the ambient wind speed 
(see \citeA{Balm:12, Perr:20}
on Earth and on Mars, respectively).

\section{Convective cells and an estimate of the convective PBL depth \label{sec:ccell}}

The pressure sensor on board InSight is
more sensitive than any pressure sensor
sent on a lander to Mars \cite{Banf:18}.
This allows the signal of
convective cells to be detected as quasi-periodic
signals in the pressure 
time series, on longer
timescales than 
vortices and gusts
(see Figure~\ref{fig:firstpbl})
because their spatial
scales
are
larger
(see the brightest areas
in Figure~\ref{fig:les}).
Here we report a first analysis
on a typical case, but a more in-depth
analysis is warranted in the future.

\begin{figure}[h!]
\begin{center}
\includegraphics[width=0.48\textwidth]{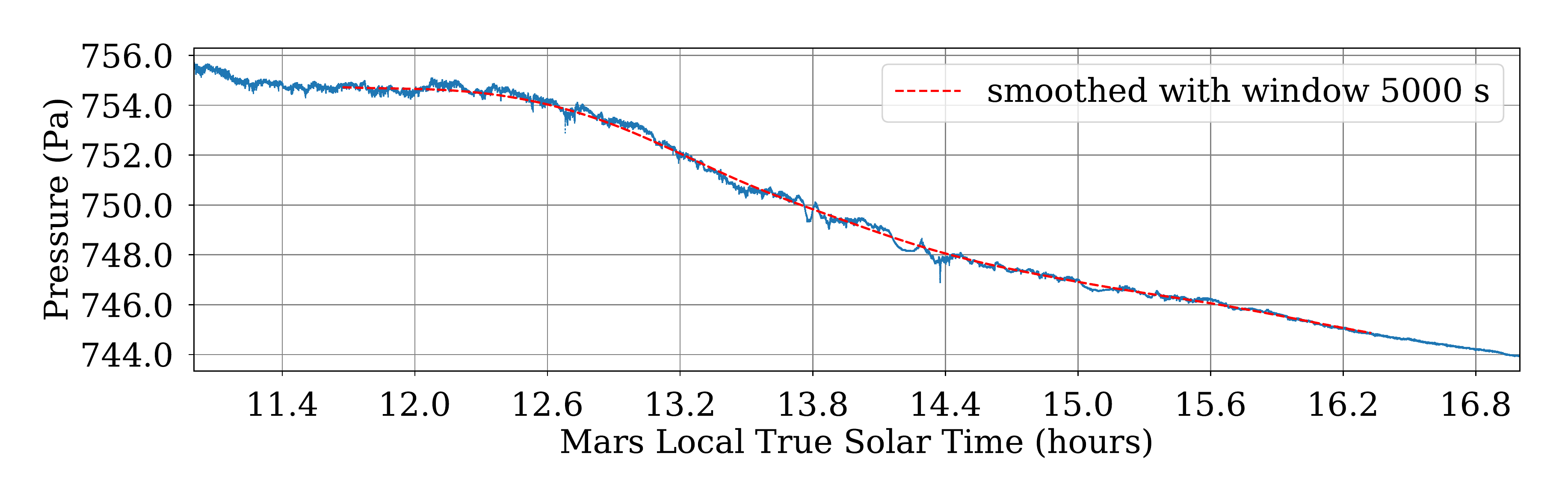}
\includegraphics[width=0.48\textwidth]{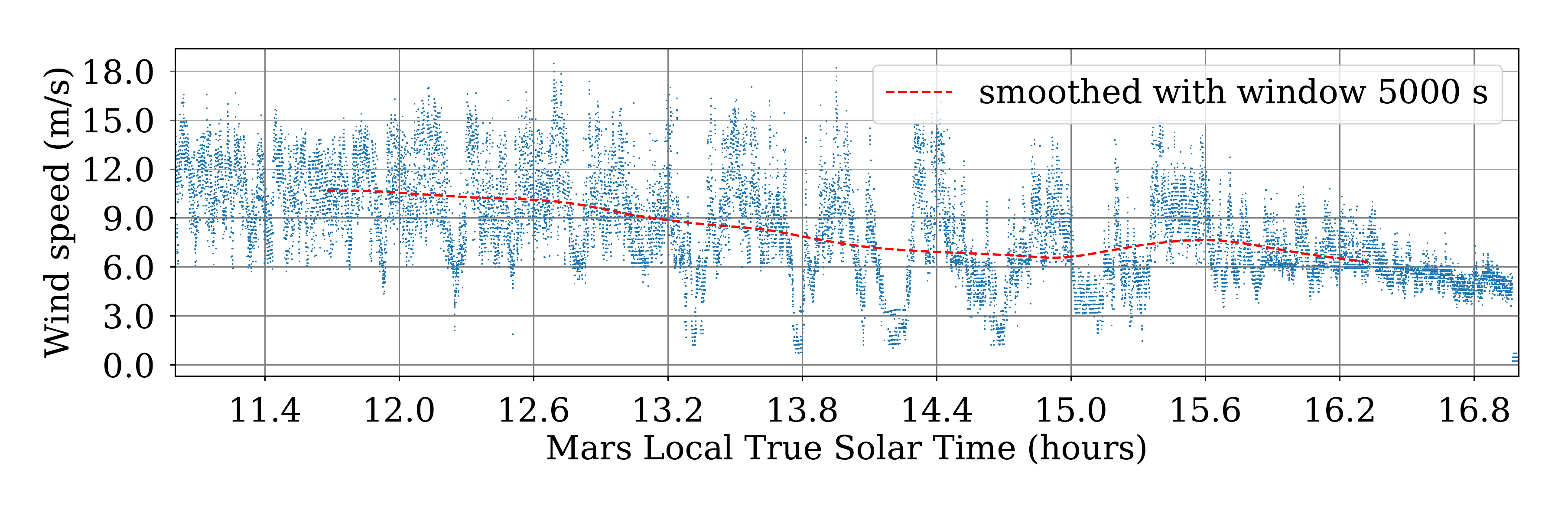}
\includegraphics[width=0.48\textwidth]{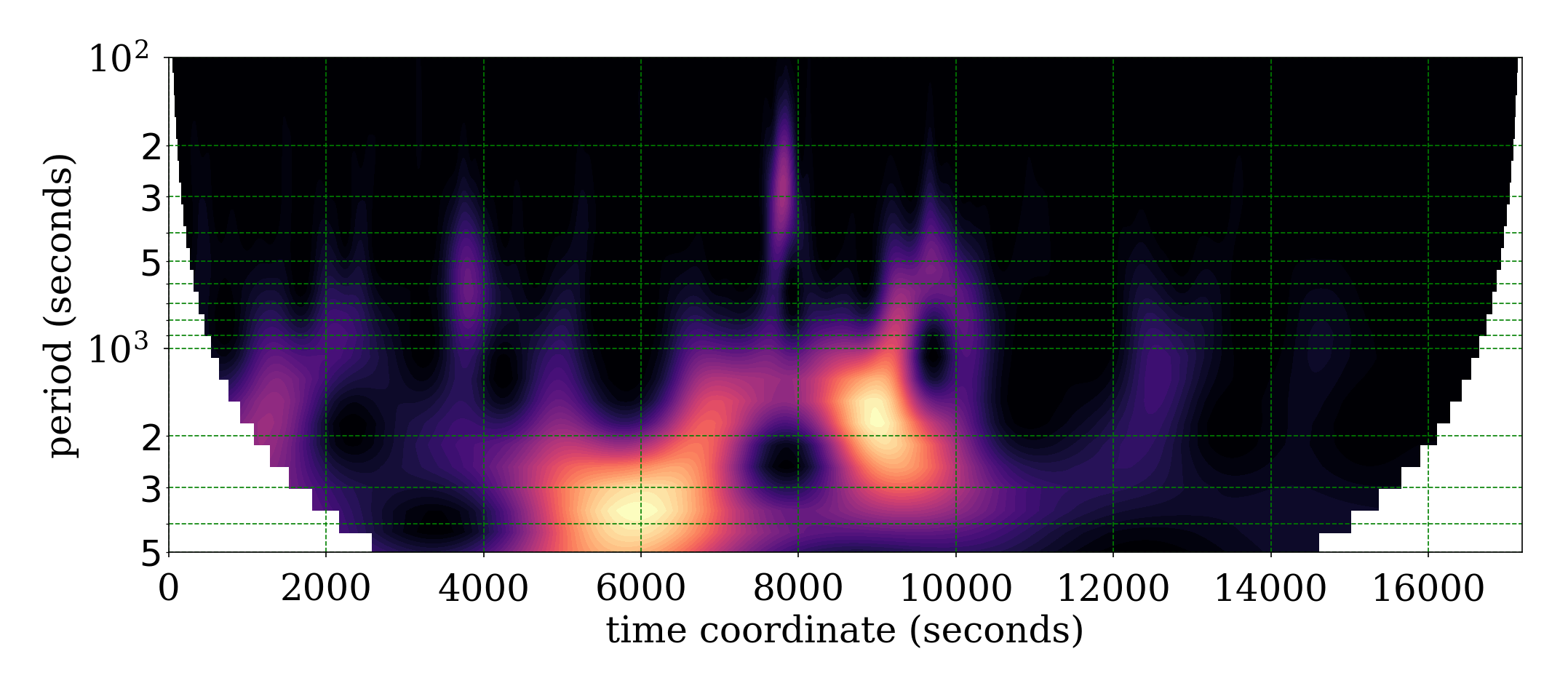}
\includegraphics[width=0.48\textwidth]{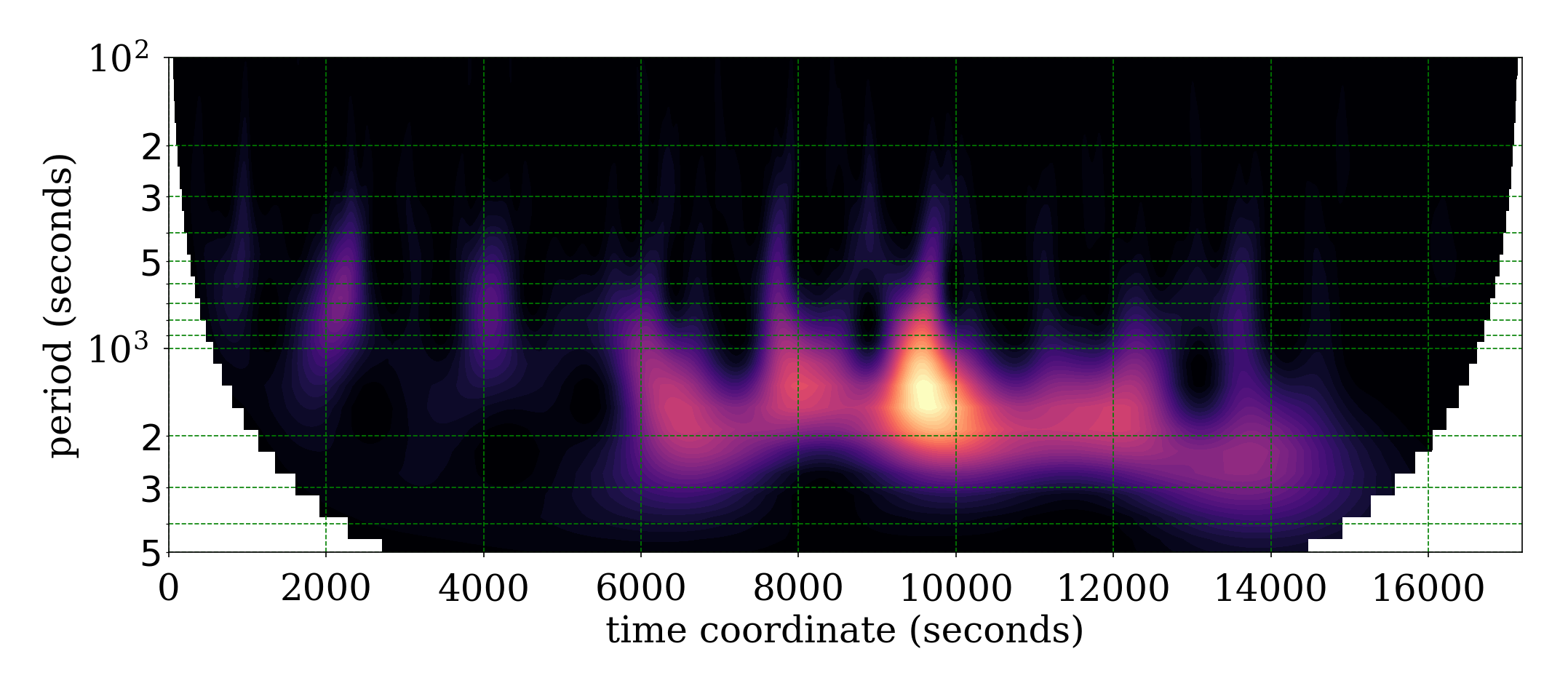}
\includegraphics[width=0.48\textwidth]{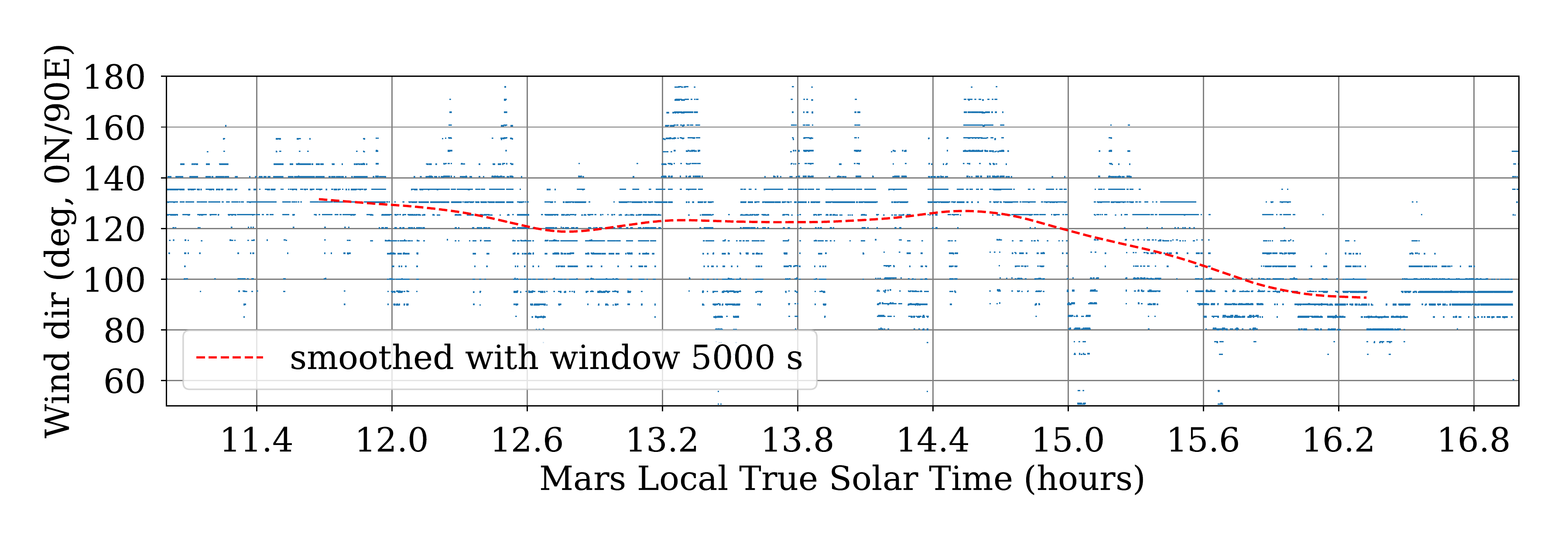}
\includegraphics[width=0.48\textwidth]{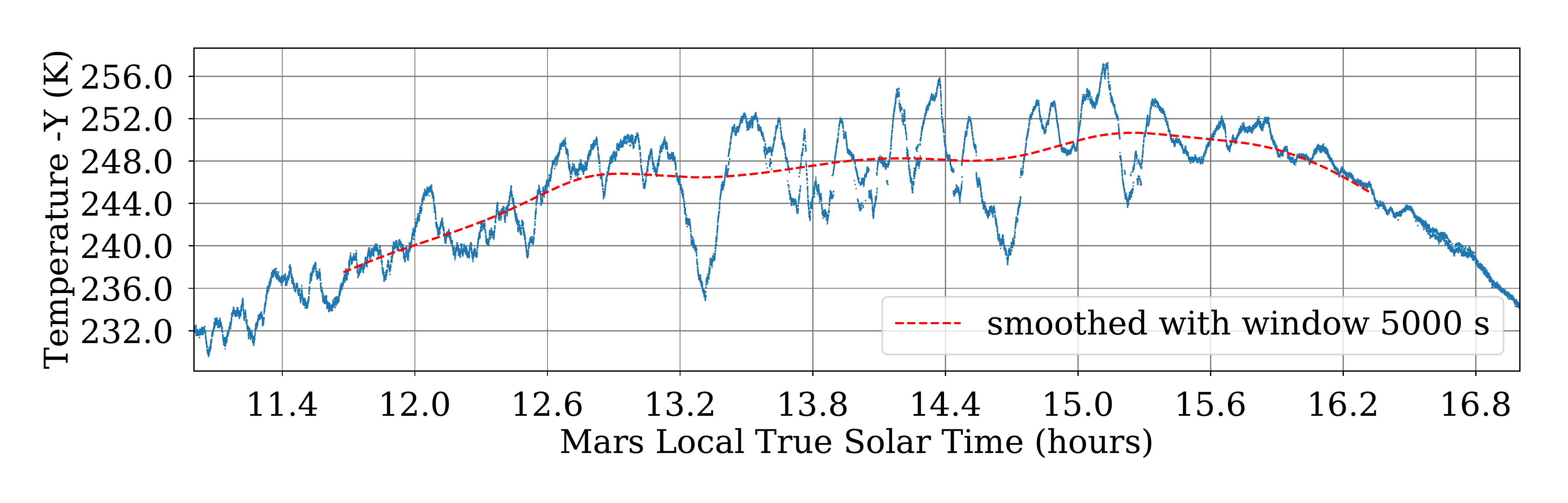}
\includegraphics[width=0.48\textwidth]{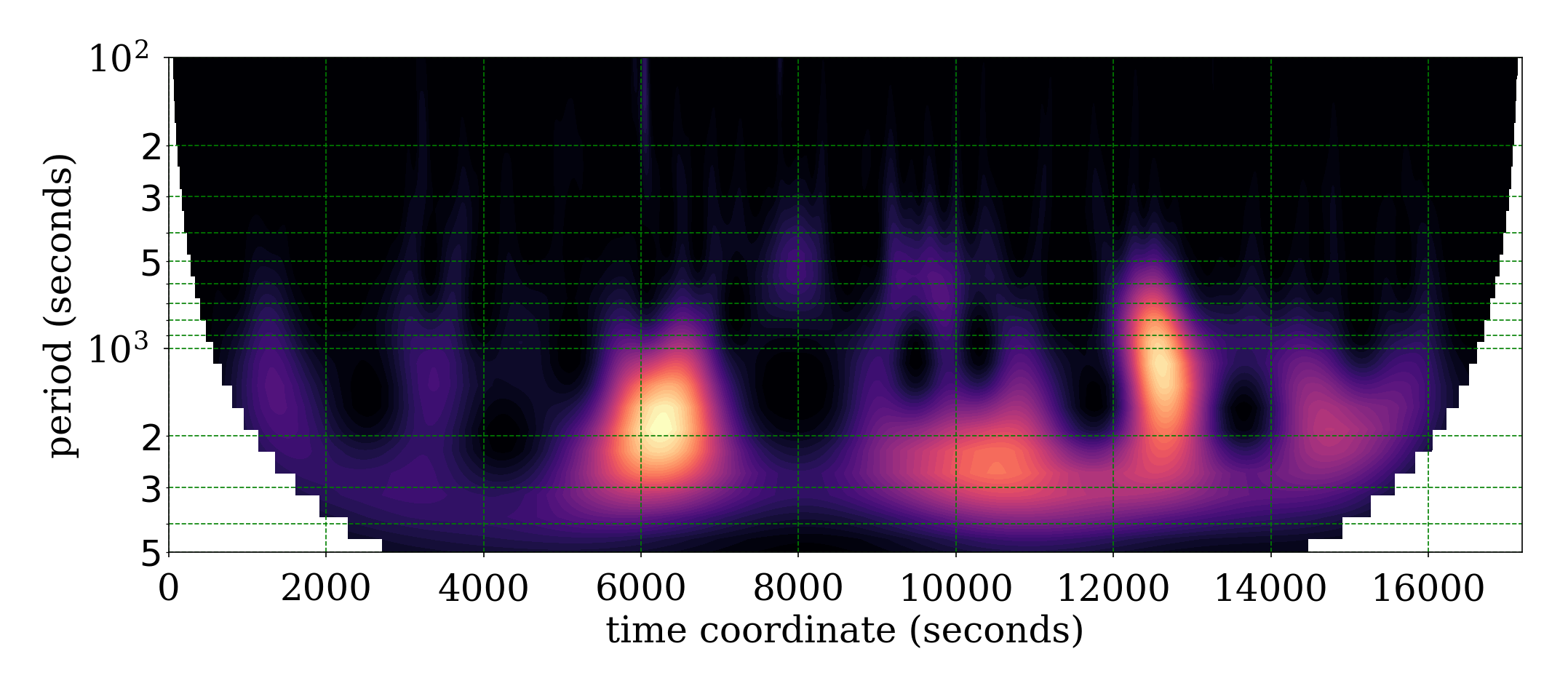}
\includegraphics[width=0.48\textwidth]{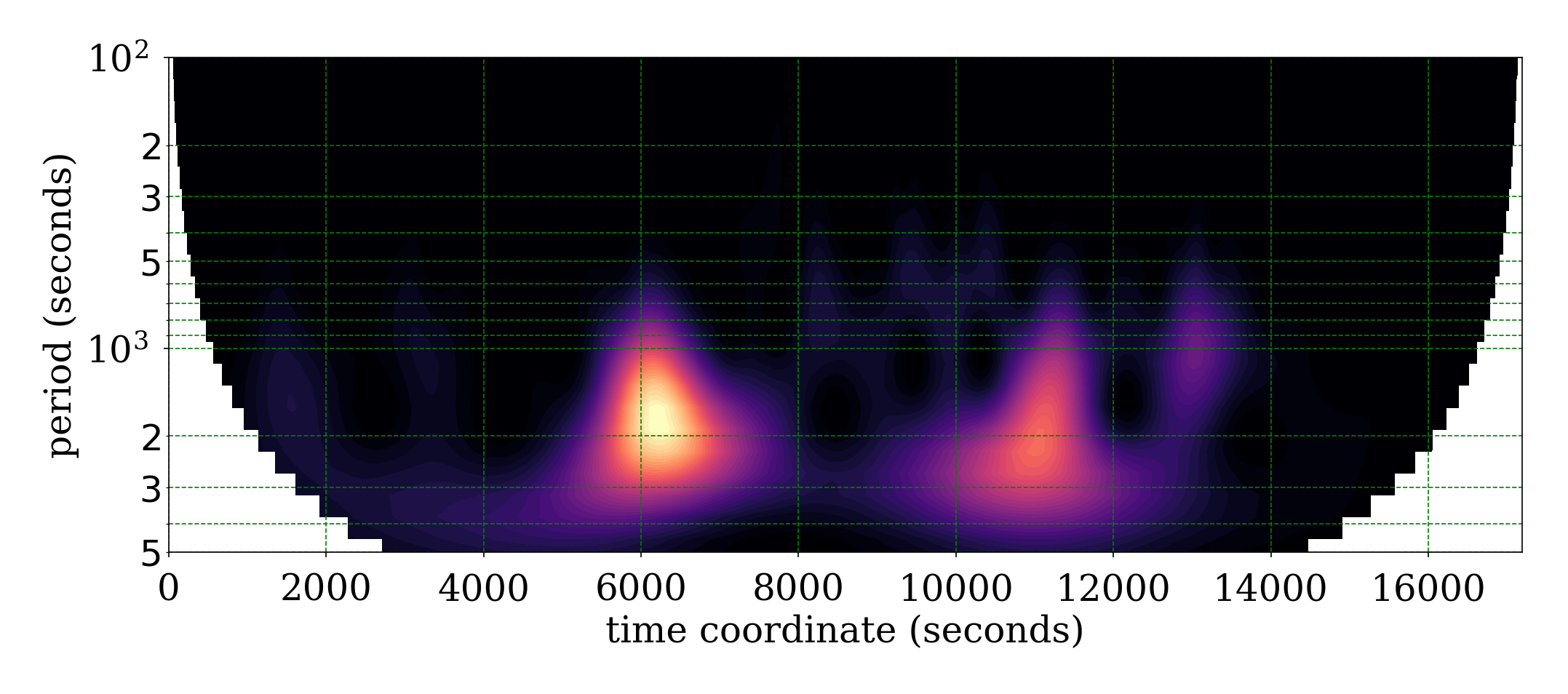}
\caption{ \textit{In this figure, the blue lines
show the observed daytime
signal by
the InSight APSS instruments
on sol 234 ($L_s = 57^{\circ}$):
pressure (upper left),
wind speed (upper right),
wind direction (lower left),
\new{air} temperature (lower right).
The red lines 
in each subpanel 
corresponds to a smoothing
average of the signal
with a Hanning window
of 5000 seconds.
The signal detrended
by this smoothing
average is 
analyzed with
wavelet transforms
to identify
quasi-periodic patterns.
Each displayed 
atmospheric time series
has its wavelet analysis
shown below the line plot
and encompassing
the same local time
interval.
Details on the wavelet analysis are provided 
in the \textit{Methods} section of \citeA{Banf:20}.
It is based on
the approach described
in \citeA{Torr:98}
and coded in Python
by Evgeniya Predybaylo.}
\label{fig:cells}}
\end{center}
\end{figure}

A particularly clear example
is provided in Figure~\ref{fig:cells}.
On all atmospheric measurements
(pressure,
wind speed,
wind direction,
atmospheric temperature),
quasi-periodic variations
with periods 1500-2000 seconds
are observed
(a similar signal
is also found
on the solar array data,
see \citeA{Lore:20}).
This is also observed
in the field 
on Earth \cite{Lore:12}
and predicted by
Large-Eddy Simulations
\cite{Spig:12gi,Spig:18insight}.

Quasi-periodic
variations 
of pressure,
wind,
temperature
are caused
by the advection of
convective cells by
the ambient wind.
In other words,
this signal in the InSight data
is another illustration
of the key role played
by advection of PBL
turbulent structures
by the ambient wind.
The width of convective cells
scales with the PBL mixing height 
\new{(\citeA{Will:79} found the
former is a factor $1.2$ of the latter)},
hence the quasi-periodic
signal makes it possible
to estimate the PBL mixing height.

In the case considered
in Figure~\ref{fig:cells},
the ambient wind is about 
7-8 m~s$^{-1}$ when the quasi-periodic
oscillations appear,
which means 
the width of the convective cells
(hence, the PBL depth)
ranges 
from 10.5 km
to 16 km,
assuming 
direct advection 
of the
cell-induced
pressure signatures
by the ambient
wind speed.
It is important to note here
that, in a LES case where the
PBL depth is known, 
\citeA{Spig:12gi}
found that multiplying the
quasi-period 
with the ambient wind
yields an estimate of PBL depth
that is about twice
the real value of the PBL depth
deduced from studying 
the vertical mixing
depth in LES
(e.g., \citeA{Spig:10bl}).
We conclude that the PBL depth
in the case of the InSight observations
ranges from 5 km to 8 km,
which is typical of 
active Martian 
daytime PBL conditions
\cite{Till:94,Hins:08,Fent:15}.
However, this range is also
large and corresponds to
the typical regional
variability of the PBL depth 
on Mars \cite{Hins:08,Spig:10bl,Hins:19}.
Furthermore, the typical PBL depth
obtained for the LES with InSight conditions
is about 5-6~km, 
i.e. corresponding
to the lower range obtained
by the above estimate.

This makes the estimate
of PBL depth by 
the quasi-periodic signal
probably valid only to
an order of magnitude.
As a result, 
by this method,
it was
difficult 
to obtain the variability of
the PBL depth with local time
and season.
Furthermore, the case
displayed in Figure~\ref{fig:cells}
is one of the most
favorable: while
quasi-periodic
signals are very often
detected in the daytime
measurements by InSight,
clearly determining
their period has resulted
to be challenging -- especially given
the challenges posed
by InSight temperature measurements (see section~\ref{sec:obs}).
A systematic exploration
is warranted as future work
and considered out of the
scope of the present paper.

\section{Conclusions}

\newcommand{\bastia}{although the power-law 
fit for the vortices
with the deepest pressure drops
appears to differ
from this whole-population fit} 

The conclusions of our study
may be summarized as follows.
\begin{enumerate}
\item High-sensitivity continuous 
pressure, wind, temperature measurements
by InSight exhibit signatures of gusts,
convective cells and vortices,
associated with daytime 
Planetary Boundary Layer (PBL)
turbulence.
InSight measurements can be 
fruitfully compared
to turbulence-resolving
Large-Eddy Simulations (LES).
\item 
Simultaneous quasi-periodic
variations of pressure,
temperature, and winds,
with periods 1000 to 2000 seconds,
are attributed to
the advection of
convective cells
by the ambient wind.
The typical daytime PBL 
mixing depth
obtained from this
signal is in the
range 5-8~km.
\item The InSight landing
site is particularly prone
to vortex encounters.
More than 6000 pressure drops
deeper than 0.35~Pa
are detected in the first
400 sols of InSight operations.
\item The differential
distribution of observed
vortex-induced pressure drops
at the InSight landing
can be well represented
by a power-law with
a 3.4~$\pm$~0.3 exponent,
\new{\bastia}.
\item The equivalent
distribution in LES
is %in agreement with 
\new{close to}
InSight observations,
exhibiting exponents
around 4.
\item With the help
of LES, the 
variability of vortex
encounters 
from one sol to
the other
at the InSight landing site
can be
explained 
\new{in great part}
by 
the
statistical
nature of daytime
PBL turbulence.
\item \label{conc:wind}
On a seasonal basis,
the vortex encounters at
the InSight landing site
are much more correlated
to the ambient wind
speed than with the
surface temperature
and
surface-to-atmosphere 
temperature gradients.
This conclusion is 
supported by LES.
\item \label{conc:gusti} 
Normalized wind gustiness
(i.e., standard deviation
of wind speed over mean wind
speed) is positively correlated 
to surface temperature
rather than sensible heat flux,
confirming the radiative
control of the daytime martian PBL.
\item \label{conc:les} An analysis
of vortex population
in the horizontal pressure field
of the LES runs indicates
that fewer convective
vortices are forming
when the background wind
is doubled from 10 to 20 m/s.
\item Conclusions
\#\ref{conc:wind}
\#\ref{conc:gusti}
\#\ref{conc:les}
led us to conclude that
the long-term
seasonal variability
of vortex encounters at the InSight landing
site is mainly controlled by the
advection of convective vortices
by the ambient wind speed.
\item Typical tracks
followed by vortices forming
in the LES show
a similar distribution
in direction and length
as orbital imagery
of the InSight region;
to match the rate of
track formation, only
the strongest vortex-induced 
wind gusts predicted
by LES (close
to friction velocities
1.5~m~s$^{-1}$)
has to lead
to bright dust particles
being moved 
on the surface.
\end{enumerate}

The meteorological measurements
by the instruments on board InSight
make a uniquely rich dataset
to study the daytime PBL dynamics,
as is already demonstrated
by the first 400 sols of InSight operations.
It is not possible
to fully unleash here,
in one study,
the potential 
of the InSight measurements
to study atmospheric turbulence.
Some conclusions reached in this
paper will have to be revisited
once a more extended period
of time has been
monitored by InSight
\newcommand{\nairobi}{(covering in
particular northern fall,
the season at which
the annual maximum 
of surface temperature
is reached at the near-equator
landing site of 
InSight)} 
\new{\nairobi}.
We also emphasize here that
further comparisons between
turbulence-resolving models 
and \emph{in situ}
high-frequency 
continuous measurements
at the surface of Mars
will allow the broadening
of knowledge on
PBL turbulence both
on Mars and elsewhere.

\acknowledgments
All co-authors acknowledge NASA, CNES and its partner agencies and institutions (UKSA, SSO, DLR, JPL, IPGP-CNRS, ETHZ, IC, MPS-MPG) and the flight operations team at JPL, CAB, SISMOC, MSDS, IRIS-DMC and PDS for providing InSight data. 
The members of the InSight engineering and operations teams 
are warmly acknowledged for their dedication
and hard work in
allowing InSight
to perform the numerous
measurements used in this paper.
The first author 
and all French co-authors 
acknowledge support from the 
Centre National d'{\'E}tudes Spatiales (CNES).
Additional funding support was provided by Agence Nationale de la Recherche (ANR-19-CE31-0008-08 MAGIS).
The modeling part of this work was granted 
access to the High-Performance Computing (HPC) 
resources of 
Centre Informatique National 
de l’Enseignement Supérieur
(CINES) under the allocation 
A006-0110391 made by 
Grand Equipement National 
de Calcul Intensif (GENCI).
A portion of this research was carried out at the Jet Propulsion Laboratory, California Institute of Technology, under a contract with the National Aeronautics and Space Administration. Additional work was supported by NASA's InSight Participating Scientist Program.
Spanish co-authors acknowledge funding by the Centro de Desarrollo Tecnológico e Industrial (CDTI), Ministerio de Economía y Competitividad, and the Instituto Nacional de Técnica Aeroespacial (INTA).
We thank John Clinton and
the Mars Quake Service
for monitoring of the
atmospheric events in addition
to the seismic events.

The datasets produced to obtain the Figures in this study,
along with the Python codes used for the data analysis,
are available in the citable online
archive \url{https://doi.org/10.14768/2ddaba56-cf61-4d5b-83b7-e4a079ed836b}
hosted at the Institut Pierre-Simon Laplace (IPSL) datacenter.
All InSight data 
used in this study 
are publicly available 
in the Planetary
Data System.
Data from the APSS pressure sensor 
and the temperature and wind (TWINS) sensor
referenced in this paper is 
publicly available 
from the PDS 
(Planetary Data System)
Atmospheres node. 
The direct link to the InSight data archive 
at the PDS Atmospheres node is: \url{https://atmos.nmsu.edu/data_and_services/atmospheres_data/INSIGHT/insight.html}.
Surface brightness temperature
measured by the HP$^3$ radiometer
is publicly available from
the PDS Geosciences node:
\url{https://pds-geosciences.wustl.edu/missions/insight/hp3rad.htm}
(DOI reference \url{https://doi.org/10.17189/1517568}).
%%
%Most of the Python codes
%used to create
%the figures
%are available at
%\url{https://github.com/aymeric-spiga/insight-PBL-JGRplanets}.
%%
%The list of all convective vortices
%detected in the first 400 sols of 
%InSight observations is also included
%in this repository at
%\url{https://raw.githubusercontent.com/aymeric-spiga/insight-PBL-JGRplanets/master/WORKFLOW_drops/outputjgr/alldrop_ordered.txt}.
%%
The code for the Large-Eddy Simulations
carried out for this study
is managed on an online repository 
at LMD (access granted upon request);
this paper uses the revision 1937 
of the code.
This paper is 
InSight Contribution Number 115.

%%%%%%%%%%%%%%%%%%%%%%%%
%bibliographystyle{apalike} 
%bibliography{newfred}

%%%%%%%%%%%%%%%%%%%%%%%%
\end{document}